\documentclass[journal]{IEEEtran}
\usepackage{xcolor,soul,framed} 
\colorlet{shadecolor}{yellow}
\usepackage[pdftex]{graphicx}
\graphicspath{{../pdf/}{../jpeg/}}
\DeclareGraphicsExtensions{.pdf,.jpeg,.png}
\usepackage[cmex10]{amsmath}
\usepackage[cmex10]{amsmath}
\usepackage{csquotes}
\usepackage[T1]{fontenc} 
\usepackage{textcomp} 
\usepackage{array}
\usepackage{mdwmath}
\usepackage{mdwtab}
\usepackage{eqparbox}
\usepackage{url}
\usepackage{hyperref}
\usepackage{caption}
\DeclareCaptionFont{custom}{\fontsize{8.4pt}{10pt}\selectfont}
\usepackage{subcaption}
\usepackage{booktabs}
\usepackage{pifont}
\usepackage{verbatim}
\usepackage{float}
\usepackage{hyperref}
\newcommand{\mydash}{\rule[0.5ex]{0.4em}{1.0pt}\,}
\usepackage{orcidlink}
\usepackage{ulem}
\usepackage{tcolorbox}

\usepackage{multirow}
\usepackage{geometry}
\usepackage{multicol}
\usepackage{booktabs}       

\begin{document}

\newcommand{\orcidiconAZB}{\href{https://orcid.org/0009-0003-0514-9090}{\includegraphics[scale=0.1]{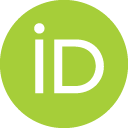}}}

\newcommand{\orcidiconOBA}{\href{https://orcid.org/0000-0003-2523-3858}{\includegraphics[scale=0.1]{photo/orcid.png}}}

\bstctlcite{IEEEexample:BSTcontrol}
    \title{Sustainable and Precision Agriculture with the Internet of Everything (IoE)}
  \author{Adil Z.~Babar \orcidiconAZB,~\IEEEmembership{Student Member,~IEEE} and \"Ozg\"ur B.~Akan  \orcidiconOBA,~\IEEEmembership{Fellow,~IEEE}\\

  \thanks{The authors are with the Center for neXt-generation Communications (CXC), Department of Electrical and Electronics Engineering, Ko\c{c} University, Istanbul 34450, Turkey  (e-mail: \{ababar23, akan\}@ku.edu.tr).
  
  O. B. Akan is also with the Internet of Everything (IoE) Group, Electrical Engineering Division, Department of Engineering, University of Cambridge, Cambridge CB3 0FA, UK (email: oba21@cam.ac.uk).
  
    This work was supported in part by the AXA Research Fund (AXA Chair
    for Internet of Everything at Ko\c{c} University)}
 }


\maketitle


\begin{abstract}

Agriculture faces critical challenges from population growth, resource scarcity, and climate change, driving a shift toward advanced, technology-integrated farming. Mechanization has transformed agriculture, enhancing sustainability and crop productivity. Now, technologies like artificial intelligence (AI), robotics, biotechnology, blockchain, and Internet of Things (IoT) are advancing precision agriculture. The concept of Internet of Everything (IoE) has gained traction due to its holistic approach towards integrating various IoT specializations, called \enquote{IoXs} where X referring to a specific domain. This paper explores the transformative role of the IoE in agriculture, expanding beyond traditional IoT applications to integrate niche subdomains like molecular communication (MC), the Internet of Nano Things (IoNT), Internet of Bio-Nano Things (IoBNT), designer phages and Internet of Fungus (IoF). Our study provides a detailed review of how these IoE subdomains, in conjunction with 6G, blockchain, and machine learning (ML), can enhance precision farming in areas like crop monitoring, resource management, and disease control. Unlike prior IoT-centric reviews, this work uniquely focuses on IoE's potential to advance agriculture at molecular and biological scales, achieving more precise resource utilization and resilience. Key contributions include an exploration of these technologies' applicability, associated challenges, and recommendations for future research directions within precision agriculture.

\end{abstract}


%
\IEEEpeerreviewmaketitle



\section{Introduction}

\IEEEPARstart{A} {griculture} is crucial for human sustenance and a  foundational pillar of national economies, stimulating agribusiness, job creation, and generating foreign exchange through exports. Its importance is historically established, with milestones such as the first agricultural revolution around 10,000 B.C. \cite{Bowles-doi:10.1086/701789} and the British Revolution in the 17\textsuperscript{th} –19\textsuperscript{th} centuries \cite{Clark-2002}, which led to significant productivity increases during the latter period. Later, the mid-20\textsuperscript{th} century Green Revolution further boosted agricultural yields through research, development, and technology transfer \cite{Ameen-2018}. Today, Agriculture 4.0  integrates information and communication technologies (ICT) with traditional agricultural practices, setting the stage for a new phase of agricultural evolution focused on sustainability and efficiency \cite{Sundmaeker-2016}. Fig. 1 illustrates this agricultural evolution, highlighting advancements from crop rotation and high-yield crops to modern precision technologies.

\begin{figure}
  \begin{center}
  \includegraphics[width=3.5in]{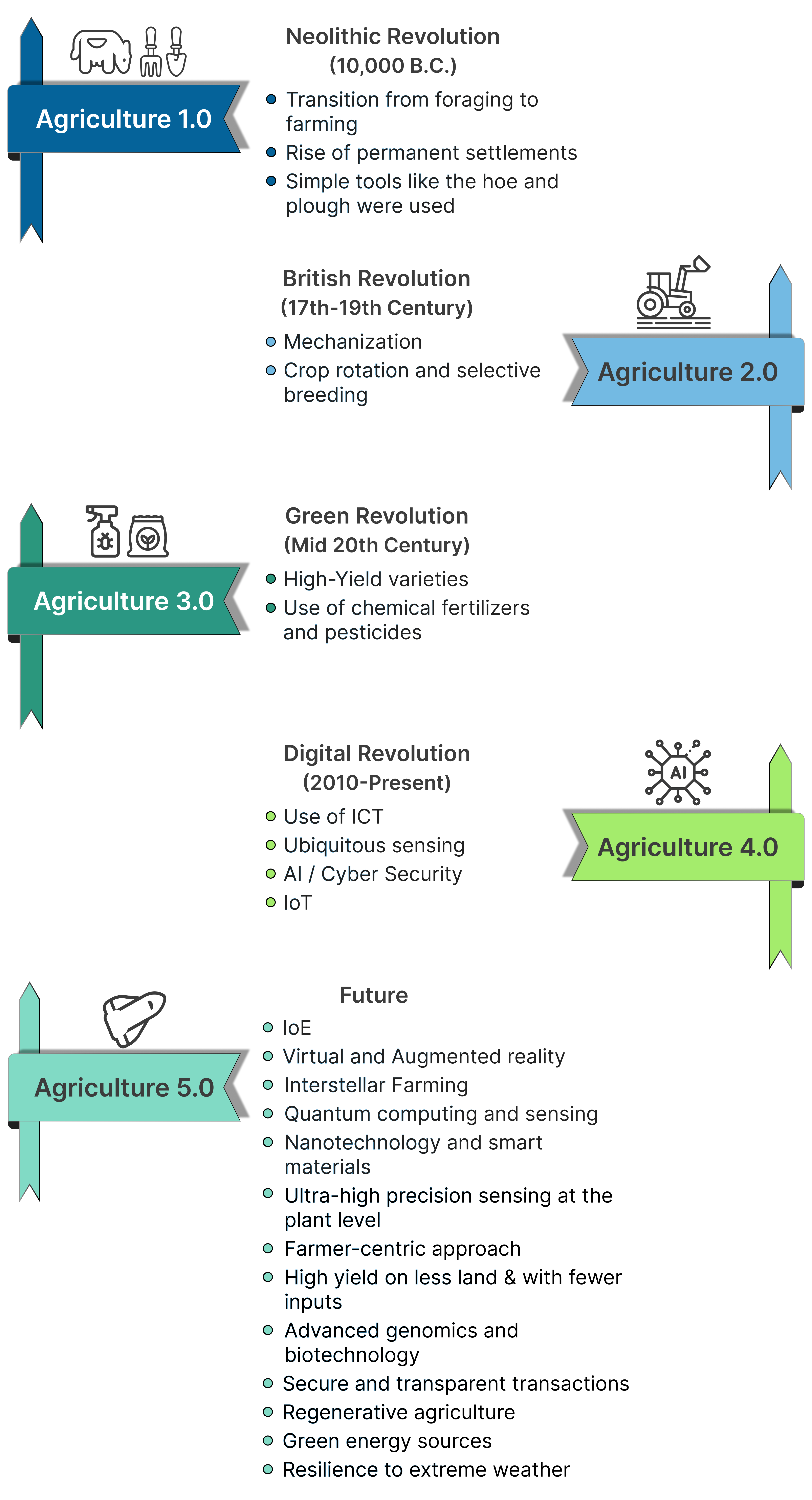}\\
 \caption{History of Agricultural Revolution }\label{Figure 1.}
  \end{center}
\end{figure}

Technological advancements in wireless communication have significantly transformed agriculture by enabling extensive coverage, increased capacity, high data rates, and lower latency. Emerging technologies, such as Wi-Fi 7 and 5G, address challenges related to bandwidth, efficiency and speed, supporting applications in smart agriculture and urban development \cite{Attaran-2021}. However, digital inequality in rural areas remains a significant barrier to adoption, highlighting the need for improved infrastructure, network resilience, and technical skill development \cite{PHILIP2017386}.

The Internet of Things (IoT) has emerged as a formidable catalyst in the advancement of technology, allowing for real-time data processing, analytics, and advanced connectivity, which are particularly valuable in agriculture for applications like precision farming, livestock monitoring, and greenhouse management \cite{WANG2021107174}, \cite{Farooq-2020}. By creating a network of interconnected devices and data-driven decision-making processes, IoT advances the goals of Agriculture 4.0. This transformation towards a more efficient and accessible world is effectively depicted in Fig. 2. This figure includes a block diagram and details the various applications areas of IoT technology.

\begin{figure}
  \centering
  \begin{subfigure}[b]{0.45\textwidth}
     \includegraphics[width=\textwidth]{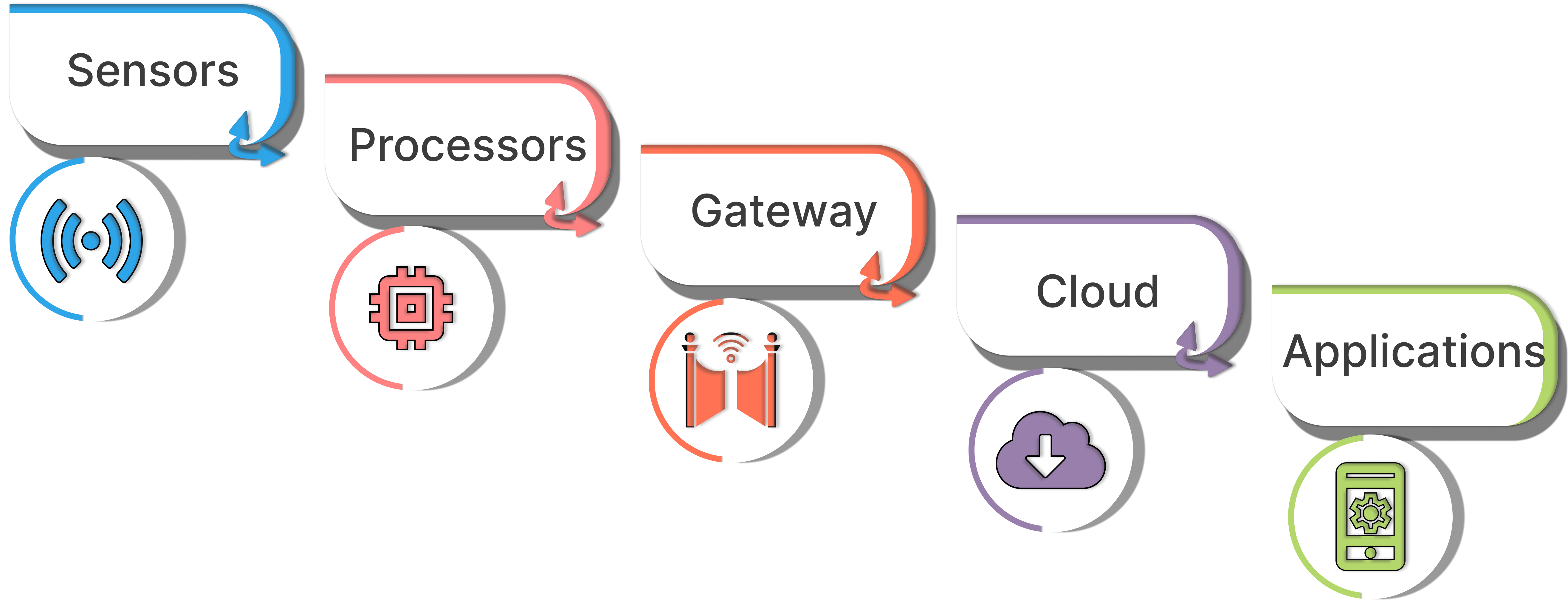}
    \caption{}
    \label{IoT2a}
  \end{subfigure}
  \hfill
  \begin{subfigure}[b]{0.45\textwidth}
    \includegraphics[width=\textwidth]{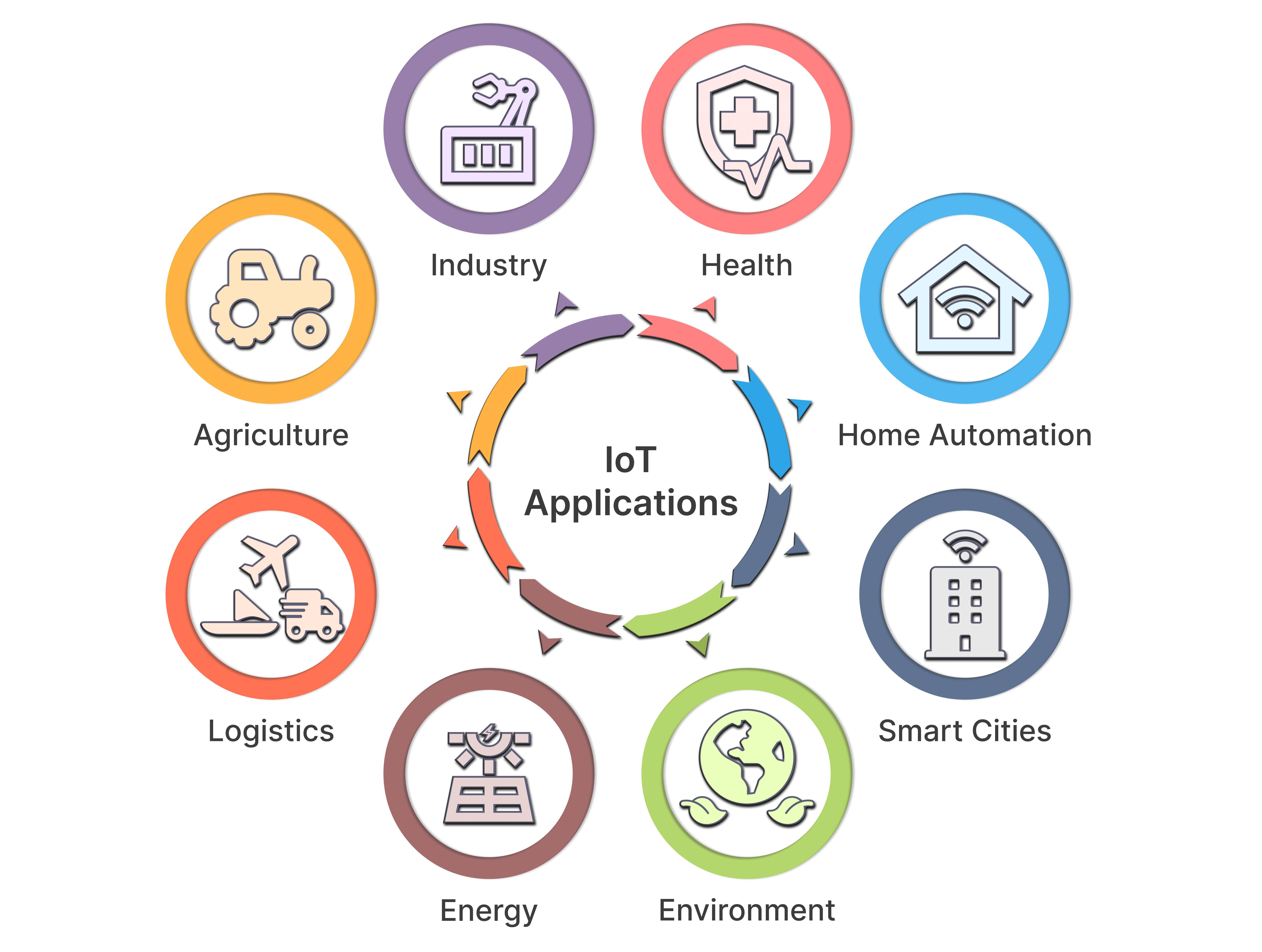}
    \caption{}
    \label{App2b}
  \end{subfigure}
\caption{Internet of Things (IoT) (a) Basic block diagram (b) Application domains (Created with figma.com)}
\label{IoTApp}
\end{figure}

Expanding on IoT, the Internet of Everything (IoE) connects diverse entities-such as sensors, data, and living organisms-into a unified network. This approach enables scalability, intelligence, and cross-domain functionality with potential applications like molecular-level interactions with plants and animals. The IoE Framework, as introduced in \cite{akan-2023-IoE}, specializes IoT into specific domains, or IoXs, such as the Internet of Agricultural Things (IoAT), the Internet of Energy (IoEn), or the Internet of Vehicles (IoV), etc., fostering collaboration across domains. In smart farming, IoE’s potential lies in integrating IoXs and emerging technologies to optimize resource management, monitoring, and decision-making, supporting a more sustainable and productive agricultural future.

This study aims to investigate the transformative potential of the IoE in enhancing precision agriculture. While numerous survey and review articles have explored IoT’s role in smart and precision agriculture, they primarily focus on conventional IoT systems and wireless communication. To the best of our knowledge, this study is the first to  thoroughly investigate the integration of both unconventional IoE subdomains- such as molecular communication (MC), the Internet of Bio-Nano Things (IoBNT), designer phages, \& the Internet of Fungus (IoF)- and emerging technologies like 6G and machine learning (ML) etc. within agricultural applications. This work identifies the unique challenges posed by each technology and presents a forward-looking perspective on potential research directions to harness these innovations for sustainable and efficient agriculture.

The paper begins by introducing the concept of critical role of agriculture in human sustenance and economic development, tracing its evolution through various technological eras. It then highlights the importance of technologies, especially IoT, in agricultural domain. Additionally, the transformative impact of IoE on the agriculture sector was examined. The  remainder of the paper is organized as follows : Section II outlines the research methodology, encompassing the framework, motivation, and research questions guiding the study of IoE applications in agriculture. Section III presents an overview of IoAT, detailing various application areas, challenges and future research opportunities. Section IV focuses on the application of nanotechnology in enhancing smart agriculture. Section V investigates harnessing phages to improve animal and plant health. Section VI explores the impact of IoF on contemporary agriculture. Section VII \& VIII delve into the Internet of Energy Harvesting Things and the integration of the Internet of Vehicles, Drones, and Space Things in achieving sustainable and efficient farming practices.  Sections IX through XI discuss the transformative potential of 6G, blockchain technology and ML in advancing agricultural practices. Section XII provides discussion and analysis of critical factors for successful IoE adoption in agriculture. Finally, the paper concludes with Section XIII, summarizing the key findings and insights.
To assist readers, the organization of this article is illustrated in Fig. 3, with acronyms listed in Table I.

\begin{table}[h!]
 \captionsetup{justification=centering}
 \captionsetup{labelsep=newline}
 \renewcommand{\arraystretch}{1.5}  
 \setlength{\tabcolsep}{6pt}        
 \caption{List of Acronyms}\label{table1a}
\centering
\resizebox{0.5\textwidth}{!}{  
\begin{tabular}{p{3cm} p{5.5cm}}  
 \toprule
\textbf{Acronym} & \textbf{Description}   \\
\midrule
IoAT & Internet of Agricultural Things  \\ 
ICT & Information and Communication Technologies \\ 
IoE & Internet of Everything \\
AI & Artificial Intelligence \\
IoT & Internet of Things \\
IoBNT & Internet of Bio-Nano Things \\
IoNT & Internet of Nano Things \\
IoD & Internet of Drones \\
IoV & Internet of Vehicles \\
IoST & Internet of Space Things\\
ML & Machine Learning \\
6G & Sixth Generation (Wireless Technology)\\
IIoT & Industrial Internet of Things\\
IoEn & Internet of Energy \\
PCA & Principal Component Analysis\\
SVD & Singular Value Decomposition \\
MDP & Markov Decision Process \\
DQN & Deep Q Network \\
SVM & Support Vector Machine \\
MC & Molecular Communication \\
V2V & Vehicle-to-Vehicle Communication \\
V2X & Vehicle-to-Everything Communication \\
VANET & Vehicle Ad-hoc Networks \\
IRS & Intelligent Reflecting Surface \\
NFV & Network Function Virtualization \\
CSA & Climate-Smart Agriculture \\
THz & Terahertz \\
RIS & Reconfigurable Intelligent Surface \\
UAV & Unmanned Aerial Vehicle \\
GPS & Global Positioning System \\
LWP & Leaf Water Potential \\
PAR & Photosynthetically Active Radiation \\
VRT & Variable Rate Technology \\
NP & Nanoparticles \\
IoS & Internet of Sensors \\
VOCs & Volatile Organic Compounds \\
RFID & Radio-frequency Identification \\
\bottomrule
\end{tabular}}
\end{table}

\begin{figure*}
\centering
  \includegraphics[width=\textwidth, height=0.95\textheight, keepaspectratio]{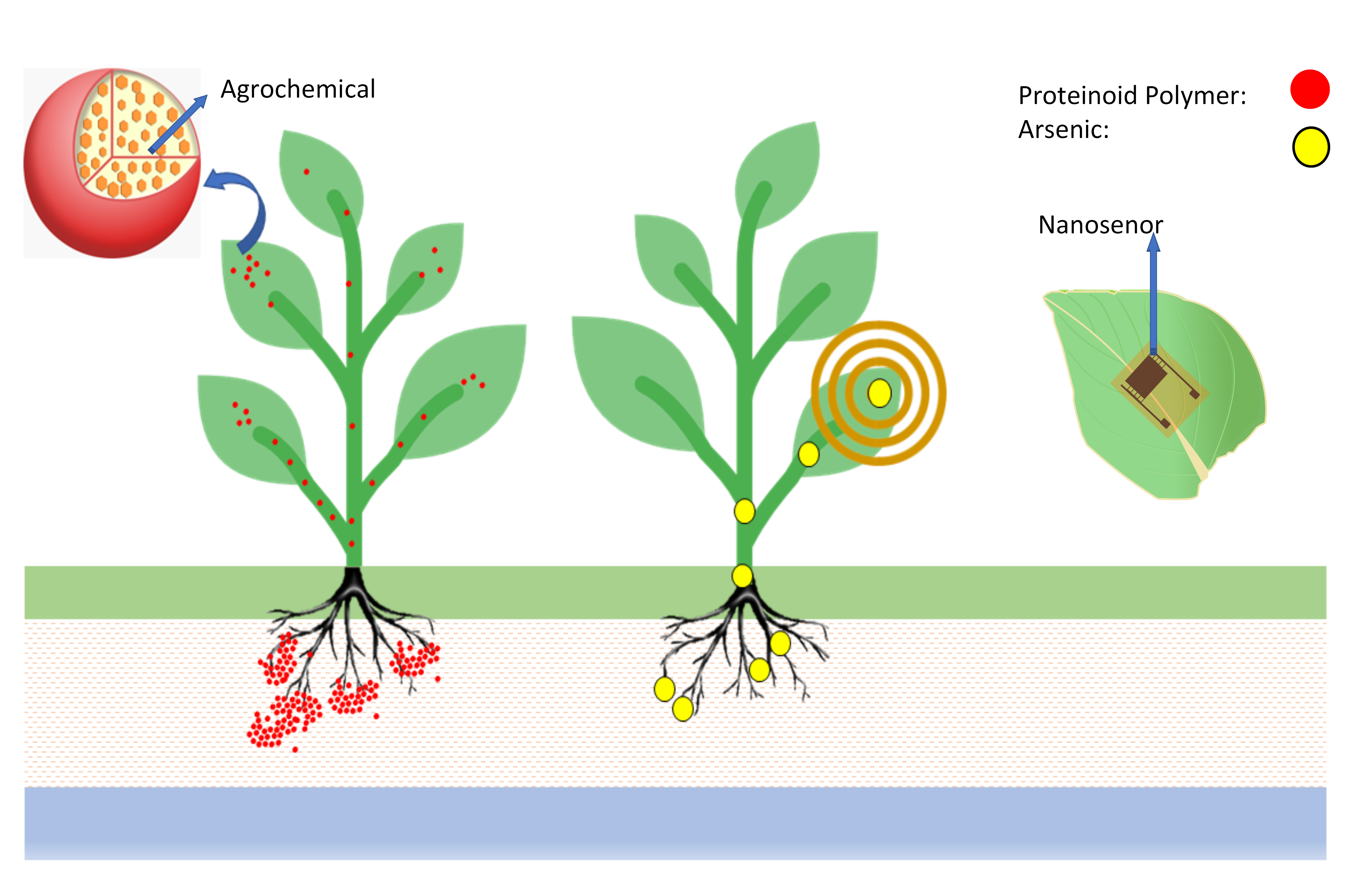}
  \caption{Organization / Structure of this article} 
  \label{Fig3-arch}
\end{figure*}

\section{Research Methodology}
A well-structured research methodology is key to ensuring the rigor and reliability of any study. It provides a systematic framework that guides the research process, facilitating precise data collection, in-depth analysis, and insightful interpretation to effectively address specific research questions.This section outlines the formulation of research motivation, the development of relevant \& specific research questions, a search string strategy, screening of search results, and the steps of data extraction and evaluation, as depicted in Fig.4.

\begin{figure*}
  \begin{center}
  \includegraphics[width=6.5in, height=1.5in]{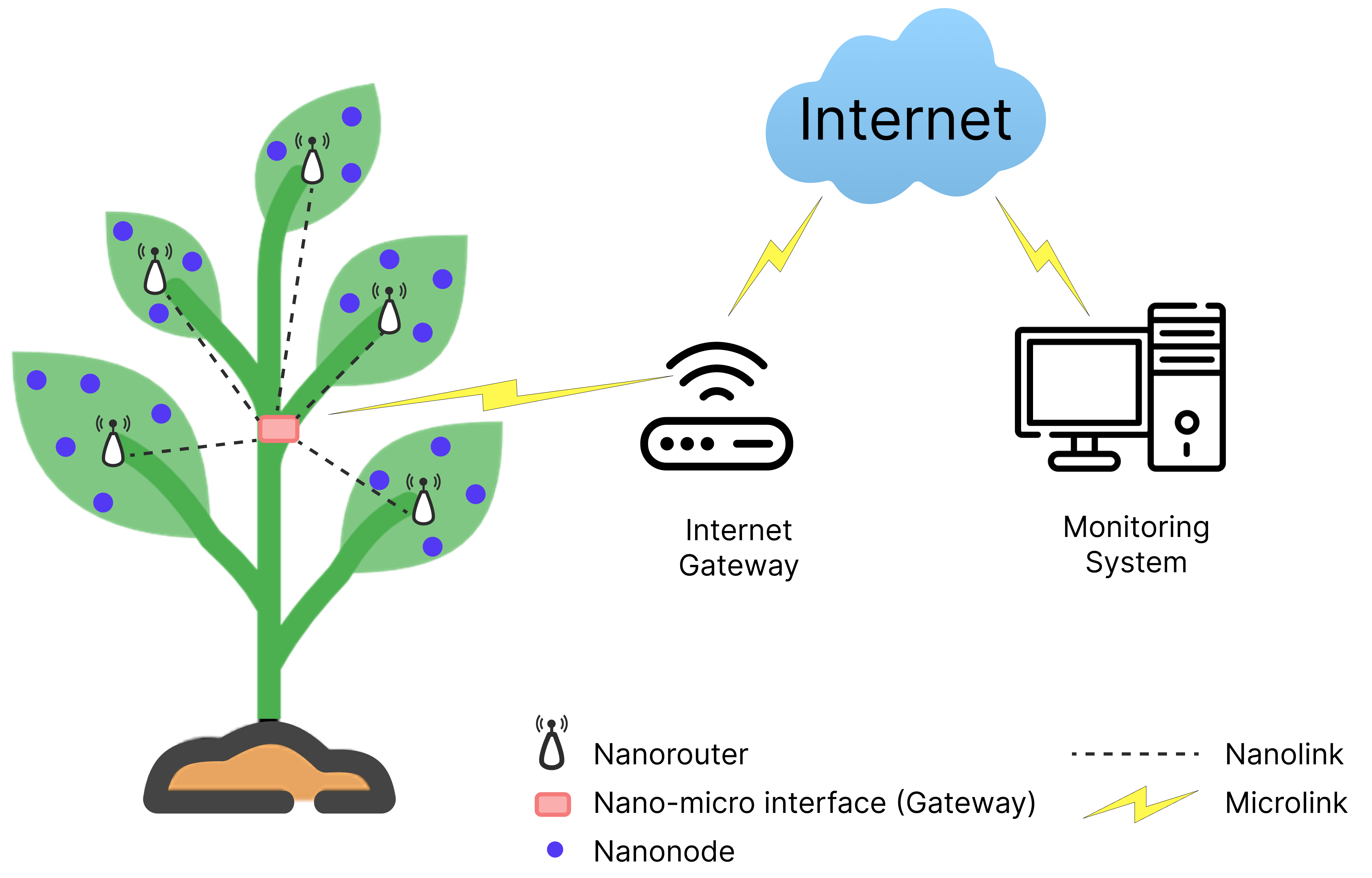}\\ 
  \caption{Fundamental stages of systematic literature review}\label{Fi}
  \end{center}
\end{figure*}

\subsection {Motivation} 
The motivation for conducting this study is grounded in several compelling reasons, which are enumerated below:

\begin{enumerate}
    \item  The rapid advancement of technology is transforming various sectors, including agriculture. This shift calls for moving beyond traditional agricultural practices to enhance productivity and efficiency. Consequently, it is essential to provide an overview of the application areas where ICT can drive smart and precision agriculture.
    \item The integration of IoT as a core component in realizing IoAT introduces multiple challenges that demand effective management or complete resolution to optimize the efficacy and productivity of agriculture.
    \item Existing literature reviews and survey articles typically focus on conventional IoT and wireless communication advancements in the agricultural domain. However, there is an urgent need to uniquely explore niche, uncharted technologies and their potential transformative impacts on agriculture.
\end{enumerate}

The main objective of this review paper is to addresses the critical gap in understanding how these various technologies in the IoE domain can advance precision agricultural.

\subsection {Research Questions}
Defining specific research questions (RQs) is essential when conducting a literature review, as it provides a clear direction and structured roadmap to facilitate the selection of relevant articles and data for analysis. For this study, the following RQs have been established to guide the research process effectively.

\begin{itemize}
    \item \textbf{RQ1}: \textit{What is the importance of ICT in precision agriculture?}
    \item \textbf{RQ2}: \textit{What are the application areas of IoT in smart agriculture?}
    \item \textbf{RQ3}: \textit{What are the current and emerging trends of IoT in smart agriculture?}
    \item \textbf{RQ4}: \textit{What are the key challenges specific to IoT in this sector?}
    \item \textbf{RQ5}: \textit{What novel IoXs can be employed in precision agriculture?}
    \item \textbf{RQ6}: \textit{What are the limitations of these IoXs?}
    \item \textbf{RQ7}: \textit{What are the potential research direction of these IoXs?}
    \item \textbf{RQ8}: \textit{What are the critical factors influencing the successful adoption and integration of IoE in agriculture?}
\end{itemize}

\subsection {Search Strings}

To gather the most relevant and recent studies aligned with the RQs and objectives of this paper, specific search terms and structured search strategies were employed. Precise keywords were essential in narrowing the search and focusing on relevant literature. Table II lists the search engines, digital libraries and databases, while Table III presents key words/ string searches used in this research. It is worth noting that only articles published in English were considered.

\subsection {Related Works}
Over the past decade, substantial research has advanced the field of smart agriculture, particularly through the integration of IoT technologies for enhanced productivity and sustainability. Table IV presents a curated selection of key studies from the last five years, chosen based on their relevance to the convergence of IoT with emerging IoE subdomains, as explored in this work. Criteria for selection include the studies’ focus on technical challenges and solutions related to IoT-driven precision agriculture, their in-depth analysis of innovative  applications in smart farming, and the extent to which they identify future research directions. This focused approach enables a deeper examination of the transformative potential and practical applications of IoE technologies within agriculture, addressing both opportunities and challenges unique to the field.

\begin{table}[h!]
\centering
\caption{Search Engines \& Digital Libraries}
\begin{tabular}{@{}p{2.5cm} p{4.7cm}@{}}  
\toprule
\textbf{Source} & \textbf{URL} \\ 
\midrule
IEEE & \url{https://ieeexplore.ieee.org/} \\ 
FRONTIERS & \url{https://www.frontiersin.org/} \\ 
ELSEVIER & \url{https://www.elsevier.com/} \\ 
SPRINGER & \url{https://www.springer.com/gp} \\ 
MDPI & \url{https://www.mdpi.com/} \\ 
NATURE & \url{https://www.nature.com/} \\ 
Scopus & \url{https://www.scopus.com/} \\ 
Google Scholar & \url{https://scholar.google.com/} \\ 
Research Gate & \url{https://www.researchgate.net/} \\ 
\bottomrule
\end{tabular}
\end{table}

\begin{table}[h!]
\centering
\caption{List of Search Strings}
\begin{tabular}{@{} p{2.5cm} p{4.7cm} @{}}
\toprule
\textbf{Databases} & \textbf{Search String/ Keywords} \\ \hline
\multirow{10}{*}{Sources in Table II} & \scriptsize\textit{(Internet of Everything OR IoE) AND (Precision Agri OR Smart Agri) OR (Internet of Things OR IoT) AND (Precision Agri OR Smart Agri) OR (IoT agricultural survey) OR (IoE agricultural survey) OR (IoE based agriculture applications OR precision agriculture applications OR Machine learning models OR ML for smart agriculture applications) OR (Blockchain OR BC smart agriculture) AND (IoE OR Internet of Everything)} \\ \hline
\end{tabular}
\end{table}

\begin{table*}
 \captionsetup{justification=centering}
 \captionsetup{labelsep=newline}
 \renewcommand{\arraystretch}{1.5} 
 \setlength{\tabcolsep}{2pt} 
 \caption{Related Works - Precision / Smart Agriculture}\label{table1}
\centering
\scriptsize 
\resizebox{\textwidth}{!}{
\begin{tabular}{ p{2cm} | p{3cm} | p{4.7cm} | p{4.7cm} }
\hline\hline
\textbf{Reference} & \textbf{Key Enabling Technology } & \textbf{Research Objective} & \textbf{Application Domain }   \\
\hline

\cite{ghazal2024computer} 2024
& Computer vision, AI, ML, imaging.
& Analyzes computer vision techniques across crop life cycles for precision farming, focusing on automation and decision-making.
& Crop monitoring, disease detection, yield prediction.
 \\\hline

 \cite{jararweh2023smart} 2023
& IoT, blockchain, AI, ML, robotics
& Provides an overview of smart agriculture fundamentals, tech enablers, and potential future developments.
& Crop management, food security, sustainable practices.
 \\\hline

 \cite{prakash2023advancements} 2023
& IoT, wireless communication, sensors, mechanization
& Examines IoT and sensor integration to improve automation and control in agricultural practices.
& Crop health monitoring, irrigation management, yield improvement.
 \\\hline

 \cite{abbasi2022digitization} 2022
& IoT, robotics, WSN, AI.
& Systematically reviews Agriculture 4.0 technologies, adoption rates, and key barriers to digitization.
& Crop management, in-field monitoring, automation.
 \\\hline

 \cite{Misra-2022} 2022
& IoT, big data, AI, blockchain, gene sequencing.
& Reviews IoT and AI-driven data solutions for enhancing monitoring, food safety, and traceability in agri-food systems.
& Crop monitoring, food quality, supply chain traceability.
 \\\hline

 \cite{Qazi_IoT_challenges_2022} 2022
& IoT, AI, UAVs, WSNs
& Provides a critical review of IoT and AI advancements, deployment challenges, and future trends in smart agriculture.
& Smart irrigation, pest detection, crop monitoring.
 \\\hline

 \cite{sinha2022recent} 2022
& IoT, UAVs, remote sensing, data analytics
& Discusses IoT frameworks, components, security issues, and future trends for automating agricultural management.
& Automated irrigation, soil monitoring, pest detection
 \\\hline

 \cite{shaikh2022recent} 2022
& IoT, WSN, agbots, drones, AI.
& Reviews IoT-enabled sensor technologies and their integration with advanced computing for effective smart agriculture.
& Soil monitoring, pest detection, climate adaptation.
 \\\hline

 \cite{shaikh2022machine} 2022
& ML, IoT, robotics, drones, sensors.
& Investigates ML applications to enhance automation, crop health monitoring, and resource efficiency in agriculture.
& Precision farming, disease detection, irrigation, yield prediction.
 \\\hline

 \cite{Rehman-2022-agronomy12010127} 2022
& WSN, IoT, automation
& Explores IoT applications and challenges in monitoring and control for smart agriculture.
& Climate monitoring, soil fertility, pest detection, irrigation.
 \\\hline

 \cite{Yang-2021-6G} 2021
& IoT, AI, blockchain, cybersecurity, edge computing.
& Examines smart agriculture's development modes, enabling technologies, and security/privacy challenges.
& Precision farming, facility agriculture, CSA.
 \\\hline

 \cite{friha2021internet} 2021
& IoT, UAVs, SDN, NFV, cloud computing
& Provides a comprehensive overview of emerging IoT technologies and their applications for advancing smart agriculture.
& Smart monitoring, disease management, supply chain, water management.
 \\\hline

 \cite{mizik2021climate} 2021
& IoT, CSA practices, data analytics.
& Analyzes CSA adoption on small-scale farms, focusing on productivity, resilience, and GHG reduction.
& Small-scale farming, climate resilience, productivity gains.
 \\\hline

 \cite{Godwin-2021-AI-agri-102} 2021
& IoT, cloud computing, ML, AI, UAVs.
& Examines smart farming technologies, including IoT and AI, to address agriculture's productivity and climate impact challenges.
& Crop and animal production, post-harvest management, climate resilience.
 \\\hline

 \cite{Hassan-2021} 2021
& IoT, AI, drones, imaging (multispectral, hyperspectral)  
& To review and synthesize advanced control and monitoring strategies in smart agriculture for automation, crop health, and precision farming.
& Stress monitoring, irrigation management, disease detection, yield prediction.
 \\\hline

 \cite{pathan2020artificial} 2020
& AI, ML, WSN, imaging, robotics
& Investigates AI-driven cognition tools to address crop health, pest control, and resource optimization in agriculture.
& Precision farming, disease detection, crop phenotyping.
 \\\hline

 \cite{saiz2020smart} 2020
& IoT, robotics, AI, big data, VRT
& Reviews advancements in data management for smart farming, emphasizing automation and sustainability.
& Crop data management, precision farming, sustainable practices.
 \\\hline

 \cite{kim2019unmanned} 2019
& UAVs, AI, big data, IoT, robotics.
& Reviews UAV technology for agricultural tasks, emphasizing platform types, control systems, and future applications.
& Crop mapping, pesticide spraying, crop monitoring.
 \\\hline

 \cite{jha2019comprehensive} 2019
& AI, IoT, ML, wireless comms
& To assess AI-driven automation techniques addressing issues like crop diseases, irrigation, and pest control in smart farming.
& Disease detection, irrigation optimization, yield enhancement.
 \\ \hline

 \cite{o2019edge} 2019
& Edge computing, fog computing, IoT
& Examines edge computing's role in overcoming connectivity challenges for scalable smart agriculture applications.
& Livestock health, crop production, farm security.
 \\ \hline
This Review 
& IoAT, MC, IoNT, IoBNT, designer phages, IoF, ML, 6G, IoD, IoV, IoEn, Blockchain 
& Investigates the integration of IoE in agriculture, focusing on novel subdomains and highlighting challenges and opportunities for enhancing precision agriculture.
& Precision resource management, pest control, nutrient exchange, disease detection.
 \\ \hline


\bottomrule
\end{tabular}}
\end{table*}


\subsection {Contributions}
This study provides a pioneering perspective on IoE-enabled precision agriculture by investigating the integration of unconventional IoE technologies in agricultural systems. It delivers a forward-looking analysis of precision farming, emphasizing advanced technologies that push the boundaries of agricultural innovation. By identifying key research directions, this study positions itself as a foundational reference for advancing IoE in agriculture, distinguishing it from previous research focused solely on IoT. The primary contributions of this article are outlined as follows.

\begin{enumerate}

    \item \textit{Identification of IoT Application Areas in Smart Agriculture:} This study identifies the primary areas where IoT technologies can be applied, such as crop monitoring, disease control, livestock management, and soil nutrient management. By systematically examining these areas, the study reveals the diversity of IoT in addressing diverse agricultural challenges and enhancing operational efficiency.

    \item  \textit{Exploration of Novel IoE Subdomains in Agriculture:} Our study is the first to systematically investigate how unconventional IoE subdomains can be applied within precision agriculture. Unlike traditional IoT, these technologies operate at molecular and biological scales, allowing for advanced precision in resource management, pest control, and disease detection. Key innovations compared to existing literature review / survey papers are as follows.
    \begin{itemize}
        \item Focus on Molecular and Bio-Nano Communication in agriculture.
        \item Introduction of designer phages for sustainable disease control.
        \item Importance of the IoF for enhanced nutrient and information exchange.
        
    \end{itemize}

    \item \textit{Identification of Technical Challenges and Research Opportunities:} 
   This study identifies the core technical and operational challenges associated with implementing IoE at scale in agriculture, emphasizing sustainable solutions. It also presents key future research directions to guide the effective integration of these technologies into agricultural systems.

\end{enumerate}


\section{Internet of Agricultural Things}
In the domain of the Internet of Agricultural Things (IoAT), the integration of IoT in agriculture marks a new frontier in enhancing farming practices. IoAT plays a vital role in the IoE ecosystem, catering to the agricultural sector's distinct needs. In this section we provide an overview of IoAT, outline various application areas and identify both challenges and open research directions.

\subsection {Overview}
The food and agricultural industries are adopting Agriculture 4.0, a digital revolution targeting various challenges in agriculture, such as deforestation, depletion of natural resources, and climate change \cite{mitra-2022-everything}. In this era, many farmers are turning to IoT-powered information technology solutions \cite{TALAVERA-2017-283}. A prominent strategy in this shift is using IoT, specifically IoAT, to automate agricultural processes \cite{Salam-2020}, enhancing efficiency through better data collection, streamlined automation, and continuous monitoring  \cite{Akkas-2017}. As the demand on individuals' time grows, such automation systems offer a time-saving approach, setting a foundation for the future of farming \cite{Kumar-2020}.

Smart agriculture outperforms traditional farming by conserving water, improving the usage of fertilizers and pesticides for safer produce, increasing the efficiency of crop yields, lowering costs of operations, making farming viable in urban and desert areas, decreasing greenhouse emissions, mitigating soil erosion, and providing real-time data to farmers. This advancement paves the way for further exploration into various fields where such innovative technologies can be effectively implemented.

The integration of IoT as a fundamental element in the realization of IoAT presents several challenges and issues that must be addressed. These challenges require either effective management or complete resolution to augment the efficacy and productivity of IoAT applications. Key challenges related to IoT in the agricultural sector that demand attention include:
\begin{itemize}
    \item \textbf{Energy efficiency}: It's  vital  as multitude of sensors, devices, and machines consume considerable amount of energy, leading to increased costs and adverse environmental impacts, such as larger carbon footprint. Therefore, optimizing energy efficiency within this domain is a  critical research area \cite{Mao-2021-Energy-efficiencey-IoT}.
\end{itemize}
\begin{itemize}
    \item \textbf{Power management}: A major constraint for IoT devices like sensors and actuators is the battery life. There is a need for energy efficient designs to optimize power consumption, including usage of low power wireless communication protocols such as Bluetooth Low Energy (BLE) or LoRaWAN, and the development of predictive maintenance algorithms for early detection of battery degradation. Additionally, exploring energy harvesting techniques offers a promising approach to effectively manage and extend the battery life.
\end{itemize}
\begin{itemize}
    \item \textbf{Data management}: IoT devices generate huge amount of data which traditional computing methods cannot handle. Consequently, there is a need for scalability of analytics algorithms and processing of data \enquote{on-the-fly}. One contributing factor is that the sensors often lack sufficient storage to retain and process all collected data at its own.
    \end{itemize}
\begin{itemize}
    \item \textbf{Data reliability}: Ensuring data reliability in  IoT systems is vital. Issues such as \enquote{fail-dirty} phenomenon \cite{Moore-2019-fail-dirty}, where a sensor malfunctions and sends erroneous data can have detrimental consequences. This particular issue is difficult to diagnose as the sensor appears to be in normal operating condition. Addressing this issue requires strategies like  redundancy and the use of ML algorithms.
\end{itemize}
\begin{itemize}
    \item \textbf{Interoperability \& standardization}: The agricultural sector uses a wide range of IoT devices from different vendors resulting in interoperability issues. These heterogeneous devices and software operating as sub-systems within IoAT framework can potentially affect energy efficiency and also cause interference. In this regard, standardization across IoT applications, software, devices, infrastructure and processes becomes crucial for efficient agricultural operations. 
    \end{itemize}
\begin{itemize}
    \item \textbf{Privacy}: Protection of privacy emerges as a major challenge, as these devices and data nodes generate enormous data. To maintain trust and security within IoAT, it is imperative to safeguard this sensitive and confidential data and information.
\end{itemize}    
\begin{itemize}
    \item \textbf{Security}: While IoT improves agricultural efficiency, it also broadens the risk of cyber-attacks and data breaches. As the number of embedded devices in IoT increases, so does the potential attack surface for the hackers.
\end{itemize}
\begin{itemize}
    \item \textbf{Bandwidth}: Although individual IoT devices require minimal bandwidth, the cumulative data flow among numerous interconnected devices necessitates a higher bandwidth capacity. As agriculture moves towards  revolution 5.0, emerging technologies like 6G, terahertz communication will be crucial to tackle this challenge.
   \end{itemize}
\begin{itemize}
    \item \textbf{Simplified deployment}: The deployment and maintenance of sensors, actuators, devices and equipment within IoAT must be user friendly, since they are to be operated by farmers who may have limited technical knowledge and expertise. Therefore, simplifying the management and operation of these IoT devices for farmers is important.
\end{itemize}
\begin{itemize}
    \item \textbf{Usability vs component cost}: Balancing usability vs component cost  is critical in IoAT to proficiently address the needs of end-user \enquote{farmers}. It is essential to achieve a balance that satisfies the farmers' requirements while keeping the costs of IoT devices low.
\end{itemize}
\begin{itemize}
    \item \textbf{Robust devices}: Developing durable and resilient IoT hardware for agriculture is vital. They should withstand severe and harsh environmental conditions such as extreme temperature, humidity, and precipitation.
\end{itemize}

\subsection {Categorization of Application Areas}
It is imperative to categorize and understand the different application areas of IoAT within smart agriculture. This categorization will provide useful insights into how particular IoT solutions are revolutionizing farming practices. Furthermore, this investigation will be complemented by a brief examination of some related research work and studies.

\begin{enumerate}
\item \textit{Smart Water Management}:
Smart water management empowers farmers to increase crop productivity, improve quality, conserve vital water resources, and minimize their environmental impact. The Smart Water Management Platform (SWAMP) \cite{Kamienski-2019-water} was developed to improve precision agriculture, addressing key challenges such as information modelling, adaptability and system complexity.
Additionally, in \cite{Krishnan-2022-water} AI and deep learning (DL) techniques are integrated with IoT to devise a smart water management system that optimizes sustainable usage of natural water sources. 
Complementing these advancements, \cite{García-2020-Irrigration} presents an exhaustive literature review  for smart water management in precision agriculture. The authors focused their research based on current IoT solutions including sensors, actuators, nodes, wireless technologies etc., for precision agriculture.

Although limitations and challenges specific to IoT in the agricultural sector have been discussed previously, following are some open directions for further research that particularly focus on water management and analysis:
\begin{itemize}
    \item Development of advanced water quality sensors, coupled with  devising intelligent strategies to cope with changing water parameters like pH, nutrients, turbidity etc., that could severely affect the crops.
\end{itemize}
\begin{itemize}
    \item Investigating adaptive deployment strategies for sensors to be re-positioned beyond fixed or stationary locations/ setups. It could improve both accuracy and geographical reach of water quality data collection.
\end{itemize}

\item \textit{Smart Monitoring}:
Smart monitoring in IoAT utilizes sensors, analytical data processing, and ML to  monitor diverse aspects like livestock, crops, farmland, and environmental conditions (temperature, humidity, soil moisture, and light intensity) to improve crop growth. The concept of unmanned Farms enabling remote 24/7 monitoring and control, is explored in \cite{Wang-2021-agriculture11020145}.
In \cite{Bhargava-2016} an Edge Mining (EM) approach is discussed for dairy cattle heat stress prediction. The use of UAVs with Raspberry Pi for fire detection is demonstrated in \cite{kalatzis-2018-edge}. Authors in \cite{Lavanya-2019} discusses the development of an IoT-based prototype for soil nutrient monitoring \& analysis, focusing on Nitrogen-Phosphorus-Potassium levels using a colorimetric method with LDRs and LEDs.
Furthermore, \cite{Rehman-2022-agronomy12010127} provides a comprehensive review on the utilization of IoT-based technologies for monitoring and managing diverse factors such as climate conditions, soil fertility, irrigation, and pest detection.

A few suggested areas for future research are listed below.
\begin{itemize}
    \item Development of customized IoT monitoring solutions to meet the specific requirements of diverse regions, crops, soils and environmental conditions. 
\end{itemize}
\begin{itemize}
    \item In \cite{ASTILL-2020-Poutry-environment-105291} technical challenges related to environmental sensors were highlighted, which included inconsistent responses from multiple sensors deployed for measuring carbon dioxide in a livestock facility. This required frequent calibration of sensors. Therefore, further research is required in the development of self-calibrating sensors that can maintain accuracy over time.
\end{itemize}
\begin{itemize}
    \item Developing innovative strategies \& techniques for effective integration and interpretation of multi-modal sensor data in crops / livestock monitoring. This involves addressing the challenges associated with multi-modal fusion for the improvement of accuracy of DL models. 
\end{itemize}
\begin{itemize}
    \item A novel direction could be to investigate intelligent IoT to analyse and interpret the livestock voices during periods of distress or drastic environmental changes.
\end{itemize}
\begin{itemize}
    \item Research on reduction of noise in sensor-gathered data is required to improve system performance and reliability. This noise could arise from numerous internal and external contributors which can compromise the integrity of data.
\end{itemize}

\item \textit{Disease Management}:
A major concern in agriculture is the extensive damage inflicted on crops by diseases, attributable to factors like bacteria, viruses, and unpredictable weather. IoAT has played a pivotal role in addressing this challenge efficiently. One notable initiative involves the deployment of a Convolutional Neural Network (CNN) model, as outlined in \cite{Udutalapally-2021-disease}. This model analyzes images of crops, captured by solar-powered sensor nodes, to predict and diagnose crop diseases. Moreover, the application of CNNs in agriculture is further exemplified in \cite{joshi-2022-ricebios} where the focus lies on the image classification of rice crops as either healthy or stressed. In \cite{Oliver-2018} a versatile  monitoring framework rooted in IoT technology, was implemented and tested in a viticulture setting. The prime focus was on tracking various weather and soil parameters to proactively manage vineyard diseases. Furthermore, \cite{Orchi-2022-agriculture1-2010009} presents a contemporary survey on the use of IoAT for disease detection in crops.

Following are some areas that require further analysis and study.
\begin{itemize}
    \item Manufacturing bio-compatible and environmental friendly sensors, ensuring no harm to the agricultural ecosystem. 
\end{itemize}
\begin{itemize}
    \item Utilizing advanced ML algorithms and extensive data analysis to achieve accurate and precise detection of existing and emerging diseases.
\end{itemize}

\item \textit{Smart Harvesting}:
Smart harvesting, achievable through IoAT, allows farmers to automate their harvesting processes, thus reducing labor costs and improving time efficiency. Wireless sensors enhance crop monitoring accuracy, facilitating the integration of intelligent tools across the entire crop cultivation process, from initial sowing to harvest \cite{Ayaz-2018-harvest}. Integration of IoT and wireless sensors gathers vital data like temperature, humidity, soil moisture, and plant development patterns. Utilizing cutting-edge technologies \cite{chamara-2022-harvest-ag}, this data can then be analyzed to forecast future harvests and provide real-time insights into agricultural yields.
Furthermore, the concept of employing agricultural robots for fruit picking, contributing to the advancement of smart agriculture, has been introduced in \cite{V-Suma-2021}.

The open challenges that needs investigation include:
 \begin{itemize}
    \item Efficient mechanisms / tools / robots to reduce trees, plants and crops damage during harvesting.
\end{itemize}
\begin{itemize}
    \item Devising vision system that can effectively identify obstacles and fruits with occlusions.
\end{itemize}

\item \textit{Supply Chain Management}:
IoAT has the potential to improve supply chain management in agriculture through  provisioning of real-time data on inventory, supply chain transparency, temperature monitoring, predictive maintenance, and fleet management etc.
The synergy of technologies such as cyber-physical systems (CPS), IoT, AI, cloud computing, and big data  significantly boosts the efficiency of supply chains, particularly in agriculture  \cite{MOSTEIRO-SANCHEZ-2020-Supply367}. 
Research by \cite{yadav-2020-selection} focuses on selecting of third-party logistics services for IoT-based agriculture, while \cite{yadav-2020-supply-analysing} addresses the challenges in adopting IoT in agricultural supply chain management. Additionally, \cite{Aljabhan-2022-SupplyIoT} presents a comprehensive analysis of IoT's integration with logistics and supply chain management, benefiting agriculture as well.

Further investigation and research is required in the following areas.
\begin{itemize}
    \item Addressing the issue of operational uncertainty caused by erratic crop yields coupled with  external factors like environmental conditions, farmers expertise, network infrastructure, transportation systems and imbalances between supply \& demand.
\end{itemize}
\begin{itemize}
    \item Enhancing traceability within agriculture supply chain which often lacks transparency and thus difficult to track the origin of produce. Blockchain technology can verify its provenance, assist in automated payments and also ensure security of the vital data.
\end{itemize}
\begin{itemize}
    \item Improving critical aspects like quality of food products, waste management and ensuring optimal temperature for storage through the application of IoT technology.
\end{itemize}
\begin{itemize}
    \item Development of forecasting models that could address the perishability  dynamics of produce in order to implement effective inventory policies.
\end{itemize}

\end{enumerate}



%

\section {Advancing smart Agricultural through Nanotechnology and Nano Communication}

IoT is mainly about connecting tangible devices to the Internet. In contrast, IoE broadens this concept by integrating computing devices, inanimate and living entities, people, data, and operational processes into the network. Meanwhile, Internet of Nano things (IoNT) takes this idea even further, venturing into the realm of nano-scale technology for more advanced connectivity \cite{Miraz-2018-IoNT}. This section explores the contributions of nanotechnology, molecular communication (MC), IoNT and Internet of Bio-Nano Things (IoBNT) in the realization of smart agriculture, with a summary of applications, challenges and future research directions presented in Table V.

\subsection{Impact of Nanomaterials \& Nanoparticles in Agriculture}
Nanotechnology, from its inception, has offered advanced and efficient solutions spanning a variety of sectors, including agriculture, biomedical, industrial, and military applications \cite{nayyar-2017-Nano}. In agriculture, it offers solutions to pressing challenges like ever increasing plant diseases and threats to crop productivity. A wide range of nanoparticles (NP) utilizing gold, copper, iron, silica and silicon have been developed to support this technology. Moreover, research into porous silica  \cite{Martinez-2014-poroussilica-nano} has highlighted its biocompatibility and biodegradability, thereby underscoring its wide-ranging potential applications.

Nanomaterials have been utilized in the creation of biosensors as \enquote{sensing materials} within the realm of crop biotechnology. They are instrumental in detecting and quantifying various agricultural aspects, including plant metabolic fluxes, residues of pesticides in food, and a range of pathogens – bacterial, viral, and fungal \cite{DUHAN-2017-Nano-11}. The same study also reports that certain nanomaterials may exert adverse effects on plants, leading to phenomena such as root shunting, diminished germination rates, and reduced transpiration.

More recently, nanotechnology in plant science has been used for delivery of agrochemicals or biomolecules, showcasing its ability to enhance crop productivity, disease treatments, and expedited disease detection. Among the innovative developments in nanotechnology and nanomaterials, proteinoid polymers, pioneered by Fox and his team \cite{Sidney-1960-proteinoid}  exhibits groundbreaking breakthrough. A type of polymer made from natural amino acids connected through thermal step-growth polymerization. These protein-like particles have garnered significant attention for their potential biomedical applications, particularly in drug delivery systems and controlled release mechanisms, due to their safety, biodegradability, biocompatibility, and immunological inertness \cite{serra-2015-proteinoid-advances}. 

In \cite{sasson-2020-proteinoid-engineering} authors investigated the promising yet relatively uncharted application of proteinoid nanoparticles (NPs) in agricultural biotechnology, aiming to enhance the stability and efficacy of agrochemical delivery. Under conducive conditions, proteinoid polymers can self-assemble into hollow proteinoid NPs. These hollow NPs are adept at encapsulating agrochemicals, thereby enabling their controlled release and targeted application. Moreover, a comprehensive review of various methodologies for producing proteinoids, tailored to their specific applications and effectiveness, has been systematically compiled in \cite{sharma-2022-proteinoid-review}. 

\begin{figure}
  \begin{center}
  \includegraphics[width=3.3in,height=2in]{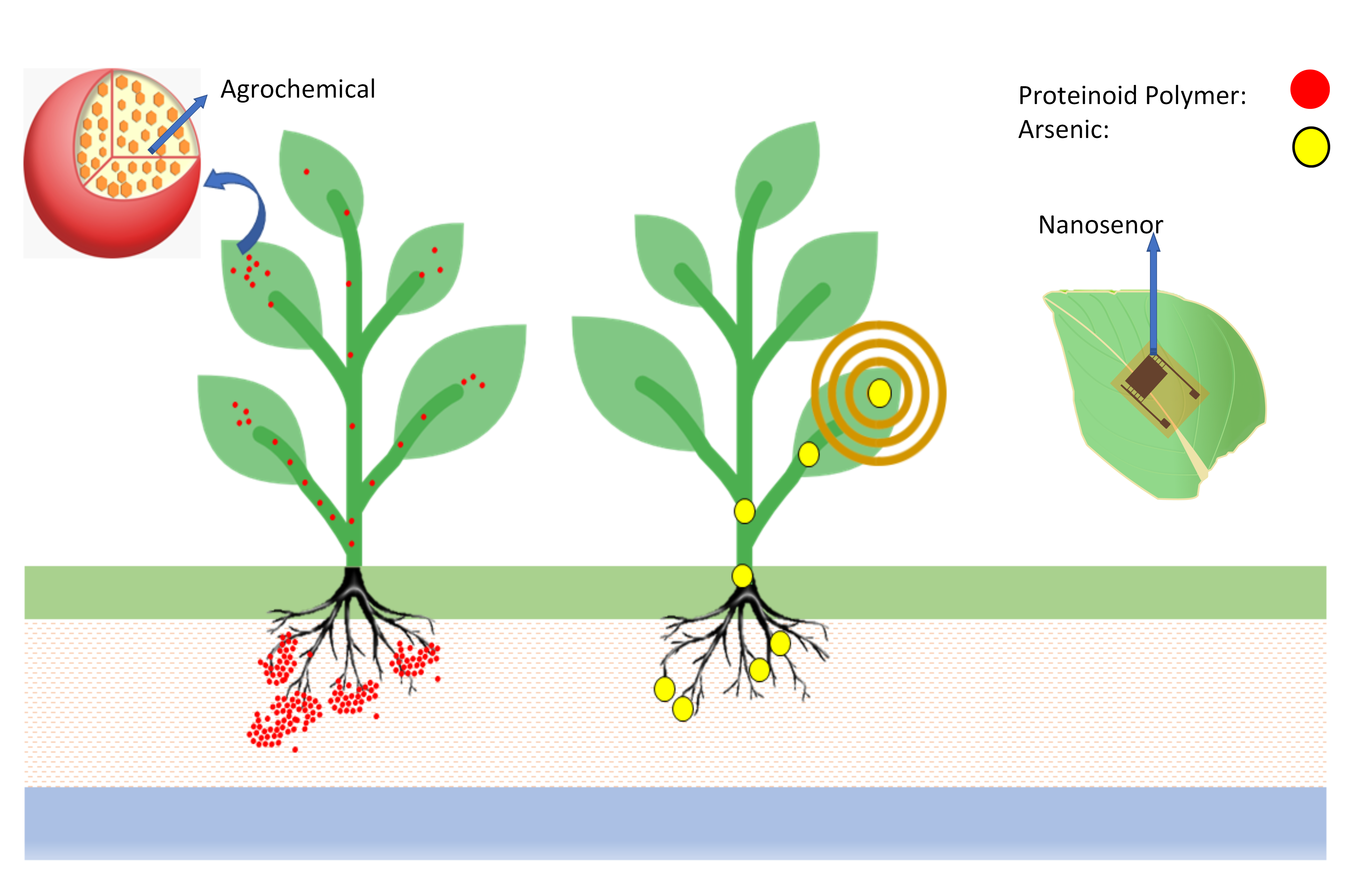}\\
 \caption{Proteinoid
Polymers for delivery of agrochemicals (L), nanobionic optical sensors for monitoring arseic level (M) \& Nanosenor on leaf (R)}\label{Fig-3Nanoparticles}
  \end{center}
\end{figure}

Research in \cite{Subramanian-2022-Nanotech} highlights the use of various types of nano-sensors, made from different nanomaterials, to analyse soil nutrients for increasing crop productivity. Another study \cite{Rojas-2022-Cu-nano}, investigates  cellulose-copper-silica nanoparticles (cellulose-Cu-Si NPs) to enhance antimicrobial properties against Cu tolerant bacteria, offering an alternative to traditional bactericides used in agriculture. Fig. 5 illustrates proteinoid polymers for targeted delivery of agrochemicals, optical nanosensors for arsenic detection in soil and a graphene oxide based humidity sensor attached to a leaf, exemplifying the role of nanotechnology in advancing agricultural practices. Nanomaterials holds potential in several vital areas in agriculture, including:
\begin{itemize}
    \item Improvement of soil health via water retention achieved through the use of nano clays and zeolites (natural geomaterials) \cite{inbook-Manjaiah-2019-Soil}. They facilitate the controlled release of water and agrochemicals and are also effective in removal of toxic pollutants from environment.
    \item  Enhancement of water quality \cite{Wu-water-nano-2010}, thus providing widespread benefits to agriculture.
     \item Delivery of Deoxyribonucleic acid (DNA)  or Ribonucleic acid (RNA) into plant cells \cite{Komarova-2023-DNA-nano}, either for genetic engineering or activation of defense mechanisms against pathogenic threats.
\end{itemize}

\subsection {Enhancing Smart Agriculture through Molecular Communication }
Molecular communication is a thriving multidisciplinary research area which can benefit diverse fields such as health, medicine, agriculture, environment, defense and more. MC notable for its bio-compatibility provides a promising approach for developing nano-networks, with potential applications in smart agriculture, such as precision monitoring of soil \& crop health, targeted drug delivery, disease detection and inter plant communication.

To enable information exchange, within IoNT, researchers are concentrating on MC alongside terahertz (Thz)-band electromagnetic (EM) communication. In their work, the authors in \cite{Akan-2017-MC-IoBNT} delve into the fundamental principles of molecular information and communication, while underscoring the limitations, challenges, open issues and future research directions in this field. Similarly, \cite{Kuscu-2019-MC} presents a profound review regarding design, coding, modulation and detection techniques critical for MC. This research  explores the development of MC transmitter and receiver, focusing on two key strategies: utilizing biological entities- engineered bacteria, virus, protein \& Ca2+ - and creating nanomaterial based  structures such as nanoscale field-effect-transistors biosensor (bioFET).

 Research presented in \cite{Ahmed2022-MC-Plants} establishes a link between MC and plant's MC for detection and comprehension of crops activities at the molecular level. MC takes inspiration from cellular communication, where biochemical processes create, emit, select, and convert information bearing molecules into cellular actions / responses. This approach views biological entities as nanomachines. Utilizing a signal transduction mechanism, MC controls and modifies the unpredictable behavior of these nanomachines by employing information molecules as signals. This process realizes a MC network that mirrors traditional communication technologies. In \cite{Awan-2019-MC-plants}, the authors used the MC concept to expound the mechanism behind information exchange within a plant via a single action potential (AP) pulse (electrochemical signal) to receiver cells. Meanwhile, \cite{Unluturk-2017-MC-Range} applies MC principles to model and analyze pheromones transmissions in plants, investigating its capabilities to extend the range of diffusion-based communication and its potential applications in agriculture.
MC holds great promise for smart agriculture which include following potential benefits and applications.
\begin{itemize}
     \item Precision delivery of bioactive substances for distribution of nutrients, pesticides or growth regulators.
     \item Enabling inter-plant communication among neighbouring crops. If a plant is subject to biotic or abiotic stress, it may release volatile organic compounds (VOCs), which could serve as a signal to nearby plants to activate their defense mechanisms such as discharge of certain drugs or pesticides, thus improving crop health.  
     \item An innovative approach in MC involves using genetically modified bacteria as sensors. This could include bioluminescence feedback communication system, targeted gene expression and the detection of hormone or disease. 
     \item Enabling inter-cellular information sharing within plant cells augmenting the photosynthesis thus improving plant growth and development \cite{Awan-2021-MC-plants}.
     \end{itemize}

\subsection{Internet of Nano Things for Agriculture}
Building on the transformative impact of nanotechnology, IoNT marks a significant advancement. It establishes a network of nano-sized entities, including devices, objects, or organisms, engineered for effective data transmission \cite{akan-2023-IoE}.

These nano-machines each with different functionalities are interconnected to overcome their individual constraints and enhance their operational capabilities. In \cite{Miraz-2015-IoNT-interface} it is pointed out that major portion of work on MC \& IoNT is concentrated within a limited scope, like within in-vivo environments. The transfer of information from implanted IoNT devices to external systems still remains a challenging task. There is a need to build such an interface that these nano devices are able to communicate directly with traditional EM wave- based devices and equipment. Researchers in \cite{Li-2022-IoNT-interface} proposed an innovative approach in which in-vivo IoNT is interfaced with in-vitro system utilizing modified nervous system that transmits signals from nano-machines via nerve fiber to be decoded by receivers on the body's surface. The integration of the IoNT into agriculture is set to make a substantial shift towards  precision farming. 

IoNT can benefit smart agriculture by employing sophisticated monitoring and management systems. Authors in \cite{Balghusoon-IoNT-2020-routing-architecture} presented an IoNT based architecture for plant monitoring system which is depicted in Fig. 6. This architecture encompassed a variety of components like nanonodes (sensors/ actuators), nanorouters, nano micro interface (gateway) and Internet gateway. Furthermore, the study also delves into the development of routing protocols for IoNT network.

Following are some prospective applications where IoNT technology can be used in agriculture.
\begin{itemize}
    \item Real time crop monitoring employing nano-sensors to collect data on plants, soil health, nutrient levels, environmental conditions etc. This data is sent to the farmer for informed decision-making.
    \item Controlled delivery of fertilizers through nanoscale actuators.
    \item Observing the interaction among plant cell organelles and pathogens to enable early detection and preventive measure.
    \item Livestock health tracking and administration of drugs or vaccines.
     \item Enhancement of livestock nutrition and feeding efficiency  through nano-scale digestion-systems.  
\end{itemize}

\subsection {Role of Internet of Bio-Nano Things in Agricultural Innovation}
IoNT has revolutionized the technological landscape through the advancements in nanomaterials like graphene, facilitating the interconnection of nanoscale devices.  Despite its marvellous benefits, IoNT could pose harmful effects on health or contribute to pollution. This challenge has led to the integration of biological entities and nanotechnology giving rise to the IoBNT \cite{Akan-2017-MC-IoBNT}, \cite{Akyildiz-2015-IoBNT}. Along with other critical applications such as healthcare, IoBNT is poised to play a vital role in agriculture with biosensors and actuators to function inside or near plant body.

\begin{figure}
  \begin{center}
  \includegraphics[width=3.5in]{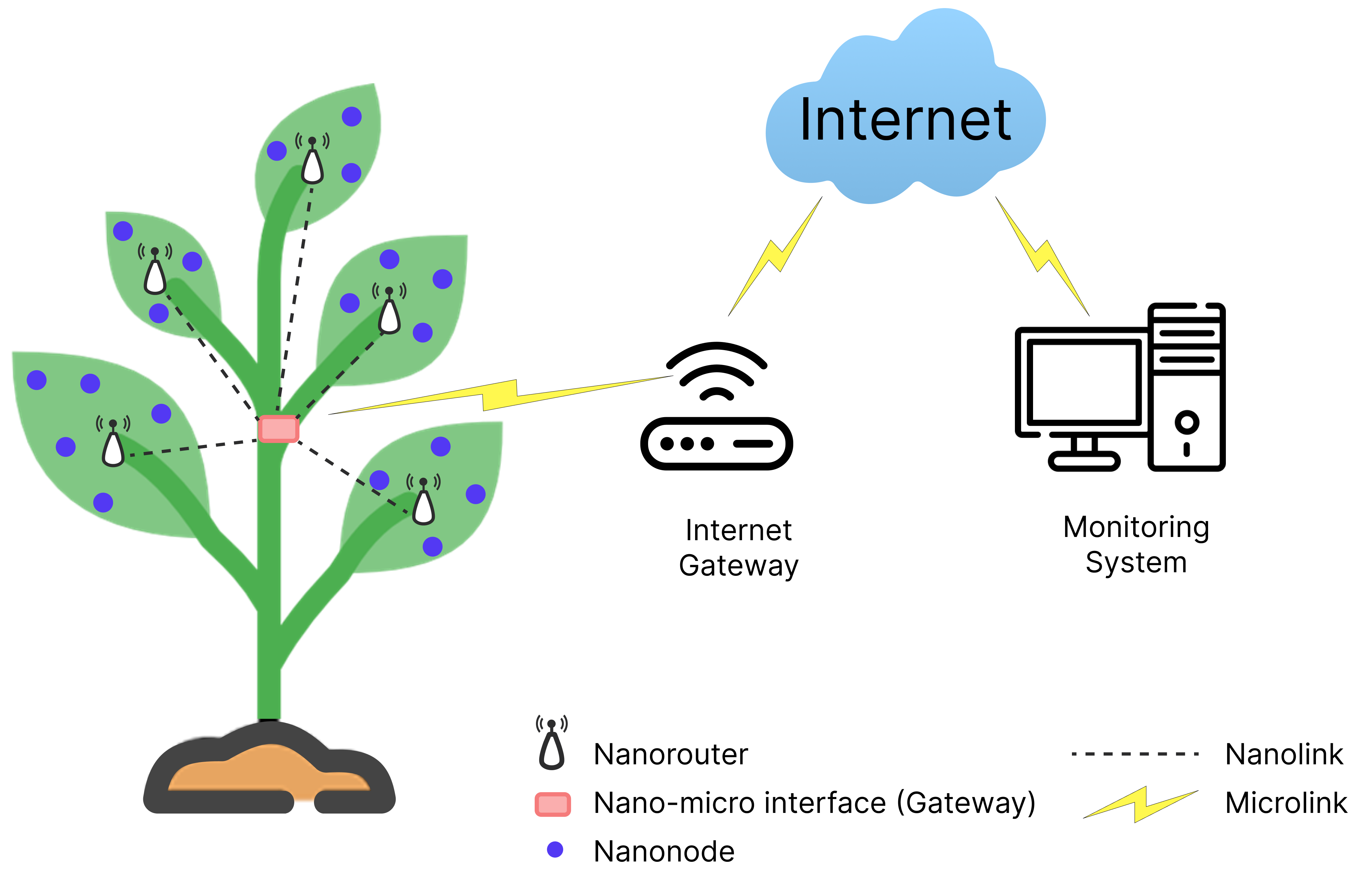}\\
 \caption{IoNT based architecture for plant monitoring system (Adapted from \cite{Balghusoon-IoNT-2020-routing-architecture})}\label{Fig-4 IoNT}
  \end{center}
\end{figure}

In their comprehensive survey, the authors in \cite{Kuscu-IoBNT-Plant} investigated IoBNT framework, exploring its potential applications, challenges and open research directions. This study highlights parallels between IoBNT and plant networks, despite the latter's lack of a physical nervous system. Key similarities between IoBNT and plant networks include:
\begin{itemize}
    \item Electrical communication within plants, from roots to other parts \cite{Awan-2019-plants-elect}, mirrors the data transmission procedures used in IoBNT.
    \item The use of plant capillaries not just for transport of water and nutrients  but also molecular information, akin to IoBNT channels.
    \item Chemical signaling (through pheromones) between plants for growth and defense is similar to MC strategies in IoBNT.
    \item The cooperation between plants, microorganisms and rhizobiome to facilitate resource sharing and growth reflects IoBNT's collaborative nature of bio-nano entities.
\end{itemize}

To sustainably boost agricultural productivity, it's essential to embrace smart farming methods, especially through the adoption of nanobio sensors. This approach facilitates the real-time tracking of vital factors such as soil health, nutrient composition, microbial community structure, pest activity, and crop quality across different growth stages \cite{GhoshMoulick-2020-biosensor}.

\begin{table*}
 \captionsetup{justification=centering}
 \captionsetup{labelsep=newline}
 \renewcommand{\arraystretch}{1.7}
 \setlength{\tabcolsep}{4pt}
 \caption{Applications, Challenges and Research Directions of Nanotechnology and Nanomaterials in Smart Agriculture}\label{table1}
\centering
\resizebox{\textwidth}{!}{
\begin{tabular}{ p{4.5cm} p{4.5cm} p{4.5cm} p{4.5cm}}
 \toprule
\textbf{Technology Domain / Area} & \textbf{Suitability for IoAT} & \textbf{Limitations / Challenges} & \textbf{Potential Research Directions}   \\
\midrule
$\textbf{Nanotechnology}$ & & & \\
\midrule
Nanomaterials \& Nanoparticles
& \mydash To monitor \& enhance soil and crops health.  
& \mydash Adverse affects of some NP on plants: roots shunting, lower germination rates, and reduced transpiration \cite{DUHAN-2017-Nano-11}.
\newline \mydash Lack of nanotech regulations in agriculture.
\newline \mydash Security and privacy.
& \mydash Green Synthesis of NP.
\newline \mydash Development of IoAT regulatory framework.
\newline \mydash Improvement of soil health using nano clays \& zeolites.
\newline \mydash DNA/RNA delivery for plant genetic modification / immunity.
\newline \mydash Robust communication \& energy sustainability for buried sensors.  \\
Proteinoid polymers \cite{sasson-2020-proteinoid-engineering}
& \mydash Nutrient and chemical delivery to crops.
& \mydash To control the degradation \& release rate of NPs encapsulated contents. 
& \mydash Eco-friendly, cost-effective nanoformulations.
\newline \mydash Exploring new amino acids for proteinoid NPs.
\newline \mydash Monitor plant health with proteinoid NP's fluorescence.\\
Molecular Communication
& \mydash Disease detection \& delivery of agrochemicals in crops.
& \mydash MC integration complexity.
& \mydash Utilizing VOCs for plant defense.
\newline \mydash Disease detection using bioluminescent bacteria. \\
IoNT \& IoBNT
& \mydash Real time monitoring \& drug delivery.
& \mydash Integration with existing systems.
\newline \mydash Bio-compatibility and safety assessments.
& \mydash Nano-scale digestion-systems for livestock health improvement.
\newline \mydash Real-time monitoring of plants \& livestock.
\newline \mydash Developing seamless Bio-Nano Interfaces. \\
\bottomrule
\end{tabular}}
\end{table*}

Some prospective areas where IoBNT could be implemented in agriculture include:
\begin{itemize}
    \item Real-time monitoring of livestock to improve their health and quality of their products like milk, meat and eggs.
    \item Improvement of plant health through the application of Bio-Nano Things directly to the plant or soil. This could facilitate in water management, fertility of soil and pest control.
    \item Smart drug delivery focusing to precisely deliver agrochemicals to targeted region of plants.
\end{itemize}

Nevertheless, security and privacy in nano communication, particularly in IoNT, IoBNT and MC is challenging, where conventional cryptographic solutions are not applicable. This necessitates development of robust security mechanisms and protocols to preserve data integrity  \cite{Rawahi-2018-Security-MC},\cite{Dressler-2012-security-MC}, ensuring the reliability of smart farming systems employing these technologies.



\section {Harnessing Engineered and Natural Phages to Augment Animal and Plant Health}
Bacteriophages, also known as phages, are viruses that specifically infect bacteria. They are ubiquitous and offer a wide range of applications. With the emergence of designer phages in microbiome engineering, we are at a crucial juncture poised to unravel a vast array of possibilities. These meticulously crafted biological agents hold great promise in revolutionizing the health of plants and animals, markedly improving agricultural productivity, and offering novel approaches to address the daunting issue of antibiotic resistant bacteria. In this section, the role of phages in plant and animal health, techniques \& technologies used, and some future perspectives will be discussed. Additionally, an overview of their applications, the challenges encountered, and directions for future research are summarized in Table VI.


\subsection {Introduction to Phage Therapy}
The growing challenge of bacterial resistance to antibiotics necessitates innovative strategies. Phages presents a viable solution to this pressing issue by targeting and eliminating mutated bacteria. The concept of \enquote{designer phages} takes this idea further by engineering synthetic bacteriophages tailored for specific applications. It is easy to engineer phages with small genomes utilizing synthetic methods, whereas for larger phages recombination based approaches are in use. The accuracy, effectiveness, and versatility of phages augment their position as a key innovation in the dynamic fields of biotechnology and health sciences \cite{Trivedi-2020-phages}, \cite{Peixoto-2021-phages-animal}. 

 In the study \cite{Salmond-2015-phages-pastpresent}, the authors provide an overview of a century long journey of phage research and its pivotal role in foundational and applied sciences of biology. In \cite{Doss-2017-phages}, the focus is on the recent developments in phage therapy, emphasizing its potential uses in treating bacterial diseases in humans, plants, and aquatic life. \cite{Górski-2018-phages} further underscores the growing recognition and adoption of phage therapy as an effective alternative to counter antibiotic resistance, suggesting potential uses beyond traditional antibacterial treatments. The exploration of phage therapy's potentials and challenges is examined in \cite{Anders-2014-phage-constraints}. Authors in \cite{Ross-2016-phage-range}, examine the concept of host range, spectrum of bacterial strains a specific phage can infect, in bacteriophages. In addition, the role of phages as bio preservatives in various food products has been explored in \cite{Ramos-2021-phage-food}.

Bacteriophages exhibits certain traits conducive to vaccine development \cite{Bates-2022-Phages-vaccine}. They are cost-effective and simple to mass produce, display resilience i.e can withstand harsh environmental conditions, and capable of activating both innate and adaptive immune responses. Most importantly, since they only infect or target bacterial cells, they can be safely used for humans, animals or plants without posing any significant side effects.

Clinical trials across Europe, the UK, and Australia are intensifying to assess the efficacy of diverse phage treatments against multidrug-resistant bacterial infections among varied patient demographics \cite{strathdee-2023-phage}. Trials involving genetically modified and synthetic phages are in their initial stages, undergoing stringent safety evaluations. While phages may not completely supplant antibiotics, given their extensive use in agriculture and animal husbandry, the implementation of phage-based strategies holds significant potential to greatly improve the management and responsible use of antibiotics, aligning with the principles of the One Health approach.

\subsection {Phage Therapy for the Health of Plants \& Animals}
Bacteriophages are ubiquitous and  most eco-friendly antibacterial solutions, as they have no detrimental effects on the environment. Phage therapy holds great potential for applications in agriculture, veterinary medicine and humans healthcare. Regardless recent trials \& experiments, it has yet to be fully embraced by conventional medicine due to safety, efficiency and regulatory challenges.

The damage caused by bacteria to plants is diverse and widespread, primarily because of scarcity of effective bactericides and development of bacterial resistance overtime. This situation poses a great risk to a sustainable agricultural production system. Research indicates that over 200 bacterial species contribute to crop damage from the  preharvest period through to storage and transportation \cite{farooq-2022-croploses}. Phage therapy has emerged as a promising approach for countering various phytobacteria, and has now become commercially accessible for some diseases. However, its efficacy is hampered by challenges like diversity of bacterial pathogens, high likelihood of resistance development, and reduced longevity of phages within the plant ecosystem. \cite{farooq-2022-croploses} underscores various environmental factors affecting the performance of phage cocktails in biocontrol applications for plant diseases. Efficiency and reliability of phages can be enhanced by addressing the bacterial resistance problem, which can be achieved by increasing residual activity through the use of phage mixtures, sunlight avoidance, leveraging the propagation of bacterial strains, and protective formulations. The challenges associated with the implementation of phages in agriculture and the potential solutions to overcome them are discussed in detail in \cite{HALAWA-2023-challenges}.
Furthermore, recent progress in phage biocontrol and its integration into comprehensive plant protection strategies have been noted in \cite{HOLTAPPELS-2021-Phages-60}, leading to an increase in the number of phage products gaining  market approval to fight plant pathogens \cite{Huang-2022-product}.

Mirroring plant ecology, bacterial diseases adversely affect animal's health, with bacteriophages showing great promise in the treatments of these diseases. Research documented in \cite{Sheng-2006-phage-animals}, \cite{Lee-2000-phages-animals} \& \cite{Vangelis-antibiotic-animals} validate the effectiveness of bacteriophages in the treatment of various types of bacterial infections in livestock, with clinical studies demonstrating up to a 99 percent reduction in bacterial pathogens. This reduction includes zoonotic pathogens, which are transmissible to humans, observed in pigs or cattle. Researchers in \cite{Chan-2013-phage-animals}, explore the impact of phage cocktails- mixture of various bacteriophages- on bacterial treatment of cattle and poultry. Beyond their utility in disease management, phages also contribute to growth promotion in animal production \cite{Ikusika-2022-animalgrowth-phage}. Moreover, authors in \cite{Ferriol-2021-animals-phage}, present an overview of recent phage therapy studies conducted on both livestock and pets. Subsequent investigations \cite{ijms-2020-phage-animals-21103715}, \cite{Cieślik-2021-phage-animals-model} employ animal models to assess the efficacy of phage therapy against bacterial infections in vivo. Fig. 7 illustrates application of designer phages in combating microbiome-related diseases within agricultural settings, thus playing a pivotal role in enhancing agricultural productivity.

The benefits of bacteriophage therapies are numerous and include the following.

\begin{itemize}
    \item Phage preparation is both simple and cost-effective \cite{Skurnik-2007-phage-cost}.
\end{itemize}

\begin{itemize}
    \item There is no residual presence in tissues post administration, eliminating the requirement for a withdrawal phase or period.
\end{itemize}

\begin{itemize}
    \item Unlike antibiotics which acts as bacteriostatic agents, obligatorily lytic bacteriophages effectively kills bacteria \cite{Kutter-2010-PhageTherapy}.
\end{itemize}

\begin{itemize}
    \item  They multiply at the infection site, thus eliminating the need for repeated doses \cite{LocCarrillo-2011-phage-advantage}.
\end{itemize}

\begin{itemize}
    \item  Bacteriophages, aside from those possessing a lipid component, are composed of only nucleic acids and proteins, rendering them safe for application in humans, animals, and plants.
\end{itemize}

\begin{itemize}
    \item The likelihood of bacteria developing resistance to a specific phage is low. Nevertheless, should resistance occurs, it is much easier to find or design a new phage than to develop a new antibiotic \cite{Elbreki-2014-Phage-resistance}.
\end{itemize}

\begin{itemize}
    \item Waste generated from the production and treatment of phages is largely biodegradable and only impacts the targeted hosts (bacteria).
\end{itemize}

\begin{figure}
  \begin{center}
  \includegraphics[width=3.5in]{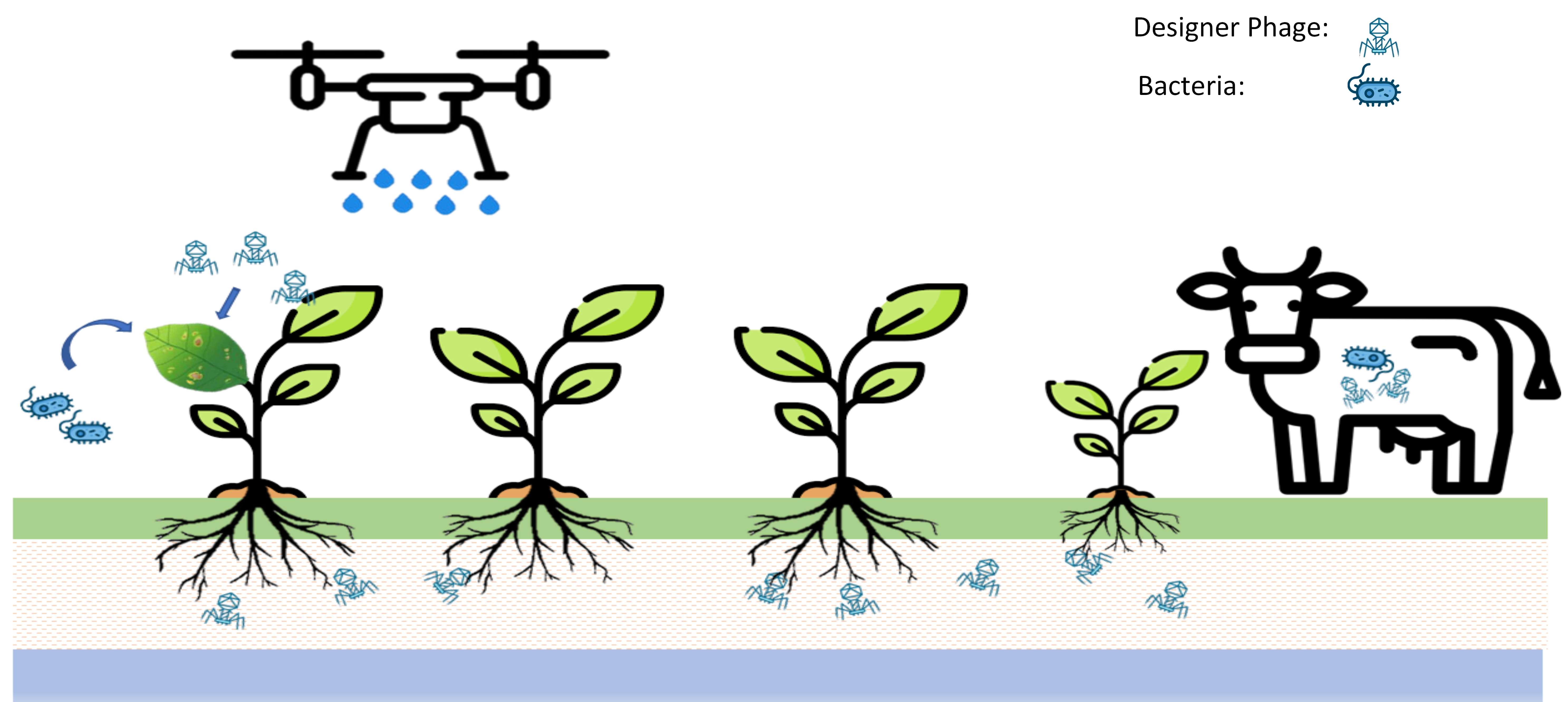}\\
 \caption{Designer phages in treating microbiome-associated diseases in plants and
animals to enhance agricultural productivity}\label{Fig-5 Designer Phages}
  \end{center}
\end{figure}

\subsection {Techniques and Technologies used in Phages}
Traditionally, the use of engineered phages was restricted due to \enquote{phage display} techniques and foundational research. However, advancements in molecular biology and sequencing technologies have enabled the development of modular designer phages aimed at combating antibiotic resistant bacteria. Authors in \cite{HUSSAIN-2023-techniques}, present a comprehensive review of phage engineering techniques, detailing both their advantages and disadvantages in developing phages best suited for contemporary applications. To address the challenge of time consuming bacteria detection, \cite{Chin-2016-phage-biosensor} introduced a method that utilizes wireless phage-coated magnetoelastic biosensors. This innovative approach enables real-time monitoring of the growth of specific types of bacteria. In \cite{López-2019-phage-ML} one-class learning algorithms are investigated to analyze the effectiveness of phage-bacteria interactions. Algorithms for predicting the lifestyle
\begin{table*}
 \captionsetup{justification=centering}
 \captionsetup{labelsep=newline}
 \renewcommand{\arraystretch}{1.7}
 \setlength{\tabcolsep}{4pt}
 \caption{Applications, Challenges and Research Directions of Phages in Smart Agriculture}\label{table1}
\centering
\resizebox{\textwidth}{!}{
\begin{tabular}{ p{4.5cm} p{4.5cm} p{4.5cm} p{4.5cm}}
 \toprule
\textbf{Technology Domain/ Area} & \textbf{Suitability for IoAT} & \textbf{Limitations/ Challenges} & \textbf{Potential Research Directions}   \\


\midrule
$\textbf{Designer Phages}$\\
\midrule
Phage biocontrol \cite{HOLTAPPELS-2021-Phages-60}
& \mydash Mitigating bacterial plant diseases.
\newline \mydash Reduce chemical pesticide usage.
\newline \mydash Mitigating bacterial diseases in livestock.

& \mydash Regulatory hurdles.
\newline \mydash Detailed safety assessments required.
\newline \mydash Phage diversity in agriculture largely unexplored \cite{Fernandez-2018-phages}.

&  \mydash Design stable phages for diverse environments.
\newline \mydash Automating phage dispensing with IoT integration.
\newline \mydash Develop phage cocktails to target multiple bacterial strains \cite{Molina-2021-phages}.
\newline \mydash Augment phages' payload for multifunctionality.
\newline \mydash Integrating phage therapy with antibiotics for improved
outcomes.
\\ 
 \bottomrule
\end{tabular}}
\end{table*}
and traits of phages exists, yet their databases require updates to include a wide range of phage genome sequences \cite{McNair2012-algo}.

Although research on phages has spanned over a hundred years, advanced genome-editing methods have only emerged recently. With public databases now accumulating a vast array of phage genomes \cite{hatfull-2008-bacteriophage-technology}, it is expected that effective engineering strategies and techniques will emerge shortly for them. From technological perspective, it is vital to develop simple and effective protocols that can be used ubiquitously for targeting specific bacteria strains. It is envisioned that synthetic methods will take over recombination based approaches. 
The challenge or limitation of assembling large genomes in synthetic phage engineering is likely to be addressed with reduced costs of gene synthesis and the advent of high-precision \& long range polymerases.

\subsection {Future Directions \& Research Opportunities}
It is imperative to conduct thorough analysis and safety evaluations of phages prior to their therapeutic use. Some future directions for research are mentioned below. 
    \begin{itemize}
    \item Exploring novel strategies to cater for the growing bacterial resistance to phages.
    \end{itemize}
    
\begin{itemize}
    \item Investigating ways to augment the phages' payload capacity for enabling multi-functional capabilities and applications.
\end{itemize}

\begin{itemize}
    \item Improving the stability and delivery of phages to designated locations \cite{Mulani-2019-phage-limitation}.
\end{itemize}

\begin{itemize}
    \item Integrating phage therapy with antibiotics for improved outcomes.
\end{itemize}

\begin{itemize}
    \item Despite the commercialization of certain phage products, their performance in terms of stability and efficacy in agricultural sector is unsatisfactory. To address this, strategies like encapsulation or embedding techniques are necessary to ensure they perform well in adverse conditions.
\end{itemize}

\begin{itemize}
    \item Enhancing the performance of phages in the treatment of plant diseases by developing techniques to protect them against the effects of sunlight.
\end{itemize}

\begin{itemize}
    \item Establishment of a comprehensive regulatory framework enabling large-scale clinical trials for phage therapy.
\end{itemize}


\section {Internet of Fungus for Modern Agriculture}
Humans have developed various platforms and methods to  facilitate information exchange and provide mutual assistance.  Similarly, in the natural world all living organism, from animals  to plants, engage in communication. Among plants this  communication occurs due to the mycelial (fungus) network,  which interconnects plants and helps information and nutrients  exchange between them. The internet of fungus (IoF) is vital in  modern agriculture, particularly for its role in the improvement  of plant health, plant security, nutrients delivery and overall  crops productivity. In this section, we will explore the concept  of IoF, examine their communication mechanisms, applications within agriculture and some potential future research directions and opportunities. In addition, a succinct summary is provided in Table VII.

\subsection {Introduction to the Internet of Fungus}
Plants not merely grow, they interact with each other and exchange vital information and nutrients. Albert Bernard Frank, a 19th century German biologist, discovered that fungi attach themselves to plant roots in symbiotic associations and characterized it as \enquote{mycorrhiza}. In this symbiotic partnership, plants furnish the essential carbohydrates for fungi, and in exchange, the mycelium fungi aid plants in water uptake and delivery of key nutrients like nitrogen and phosphorous. Fig. 8 shows fungal networks facilitating the delivery of nutrients and water to plants. These fungal filaments significantly extend the reach of plants beyond their natural capabilities. Its functions like a natural network that connects a vast array of flora including plants, trees, shrubs, and flowers.

 Mycelial networks not only enhance plants growth but also helps activation of defence chemicals, thus strengthening plant's immune response and making them resilient to diseases \cite{Song-2014-mycelial-defense}. The fungal community is not limited to a single species- it's diverse and have special functionalities. Some specialize in nutrient transport, while others are responsible for breaking down organic matter.

It is also important to recognize the multifaceted nature of mycelial networks, as their functionality transcends beyond beneficial aspects. While it is true that they facilitate nutrient sharing and inter-plant communication, they can also serve as a means for toxic plants to detrimentally affect the network. These networks may act as conduits for plants to pilfer vital nutrients and resources from each other. A notable example includes plants that lack chlorophyll, like Phantom Orchids, which derives essential nutrients from neighboring plants and trees via mycelial networks. Moreover, American Black Walnut Tree and Golden Marigolds have been discovered to release toxins into the network to suppress the growth of near by plants to gain a competitive edge in accessing resources like water and light.

\begin{figure}[H]
  \centering
  \includegraphics[width=3.3in, height=2in]{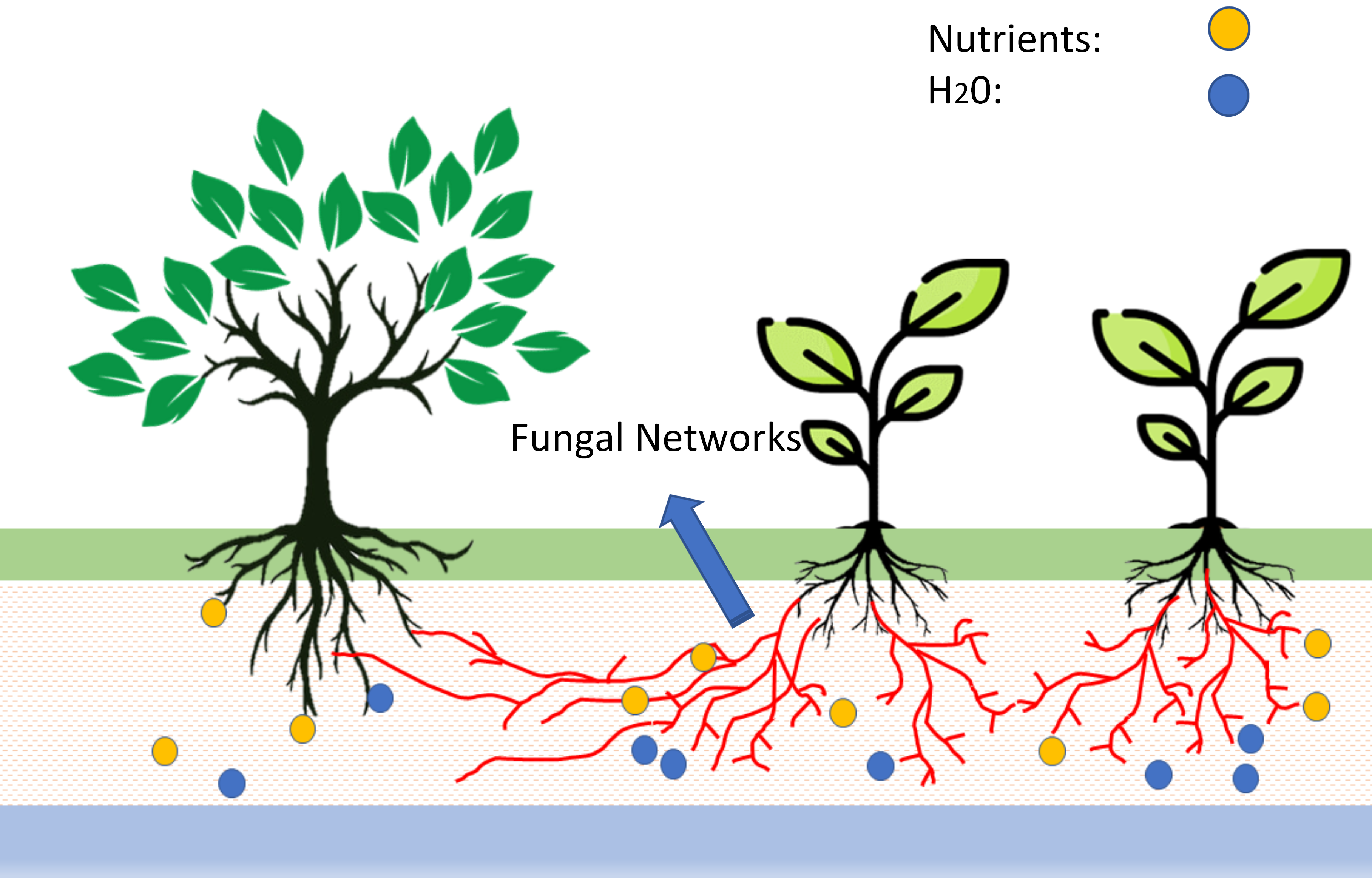}
  \caption{Fungal network for delivery of nutrients and water to plants}\label{Fig-6 Fungal NW}
\end{figure}

\subsection {Communication Mechanism within Fungal Networks}
The mycorrhizal fungal networks \cite{Simard-2018-fungal} enable advanced communication between trees in a forest. These networks allows resource sharing, defensive signaling, and kin recognition. The interactions within these networks reflect cognitive abilities akin to perception, learning, and memory, impacting the trees' well-being and longevity. Remarkably, the structure of these networks bears a resemblance to human neural networks, possessing characteristics that enhance both local and global communication efficiencies, essential for functions akin to intelligence.

This intricate communication network is established through the mutualistic, or symbiotic, relationship known as \enquote{mycorrhiza} occurring between fungi and plants. In this symbiosis, plants supply sugars - glucose \& sucrose formed by photosynthesis- to fungi in exchange of crucial resources like water, nitrogen and phosphorous. Approximately up to 80 percent of all terrestrial vascular  plants have this symbiotic relationship, which is realized through the plant's roots \cite{French-2017-percentage-fungus}. Apart from nutritional and informational exchange, the plants benefits from fungi priming, where the initial interaction between plant roots and fungi enhances plant's immunity response. This increased immunity helps plants in combating diseases, thereby supporting the survival of the entire ecosystem.

Effective communication necessitates three fundamental elements: a sender, a message, and a receiver. The diversity in communication methods and responses is substantial. Vast research has been conducted into both inter-species and intra-species communication exploring information exchange  between fungi , plant cells and bacteria \cite{christensen2011lipid}, \cite{tarkka2009inter}. Several crucial elements of this communication  \cite{Tavanti-2012-fungal} have been identified, such as the message, which can be a protein, alcohol, lipid, or gas, the sender: fungi, and the recipients, which vary from plants and mammals to fungi and bacteria. 

There is critical need to understand the intricate communications \& interactions that are taking place among soil organisms. These complex communications highlights the remarkable intelligence and sophistication of nature, offering promising prospects for the development of novel agricultural techniques. Such methods aim to enhance the productivity and sustainability of the ecosystem.

\subsection {Applications in Agriculture}

Plants and trees utilize mycelium networks to interact with their surroundings \cite{Gorzelak-2015-mycelium}. These networks serve two key functions for plants and trees: conserving their carbon sources and facilitating the exchange of information with the environment. Authors in \cite{Sirimorok-2023-mycelium}, extended this concept to smart farming security. The study presented a proof of concept for enhancing farm security by integrating a mycelium network with IoT infrastructure. This approach introduced a virtualized network tailored for specialized sensor communication, emphasizing the network's adaptability and its ability to be dynamically reconfigured to meet various needs under changing conditions. 

Mycorrhizal fungi play a pivotal role in improving the soil structure and fertility \cite{Miller-2000-fungi-soil}. By leveraging their hyphae, narrow extensions which can effectively penetrate small soil pores, they increase the surface area through which plants can access nutrients. Furthermore, mycorrhizal fungi are also beneficial in promoting soil aggregation, a process which helps soil particles to stick together. This aggregation helps in soil aeration \& water retention, nutrient accessibility, erosion prevention, root growth and decomposition of organic matter.
Among mycorrhizal fungi, Arbuscular Mycorrhizal Fungi (AMFs) standout for their ability to benefit host plants through the production of organic acids \cite{Fall-2022-fungal-soil}. These acids brings significant enhancements to the soil environment, including altering soil pH to improve nutrient solubility, carbon sequestration, detoxification, metal chelation, and enhanced root penetration.

Additionally, fungi can also be employed in agriculture for their natural bio-control attributes. Plant diseases, caused by bacteria, fungi, viruses, and nematodes, significantly impact agricultural yields, posing a critical threat to global food production \cite{savary-2019-fungal-global}. Biological control is an effective countermeasure, employing antagonistic organisms to suppress the pathogenic activity. This method is increasingly favored over chemical pesticides, owing to its wide range of benefits. Contrary to chemical methods, biological control does not contribute to pathogen resistance, environmental contamination, or the proliferation of secondary pests. Moreover, it aligns with organic farming practices \cite{ab-2018-fungal-emerging}. The efficacy of Trichoderma in bio-control is particularly noteworthy, as demonstrated in \cite{thambugala-2020-fungi}, underscoring its potential in sustainable agricultural practices.


\begin{table*}
 \captionsetup{justification=centering}
 \captionsetup{labelsep=newline}
 \renewcommand{\arraystretch}{1.7}
 \setlength{\tabcolsep}{4pt}
 \caption{Applications, Challenges and Research Directions of Internet of Fungus in Smart Agriculture}\label{table1}
\centering
\resizebox{\textwidth}{!}{
\begin{tabular}{ p{4.5cm} p{4.5cm} p{4.5cm} p{4.5cm}}
 \toprule
\textbf{Technology Domain/ Area} & \textbf{Suitability for IoAT} & \textbf{Limitations/ Challenges} & \textbf{Potential Research Directions}   \\

\midrule
$\textbf{Fungal}$\\
\midrule
 Trichoderma 
 & \mydash Enhances disease protection, plant growth/ resistance, nutrient efficiency,and reduces agrochemical pollution.\cite{thambugala-2020-fungi}.  
 & \mydash Most effective on underdeveloped plants. 
 \newline \mydash Slight harm to host cells in some cases \cite{MUKHERJEE-2022-Fungal-15}. 
 & \mydash Developing scalable, cost-effective trichoderma- based NP synthesis for agrochemicals \cite{Alghuthaymi-2022-fungal}.
  \newline \mydash Trichoderma formulations for leaf diseases, enhanced shelf-life \& consistent field performance \cite{kumar-2022-fungal}.
  \newline \mydash Assessing new trichoderma strains under varied conditions \cite{Hariharan-2022-fungal-challenges}.  \\ 
 
Mycorrhizal fungi / Mycelium Networks 
& \mydash Amplify plant/ tree water \& nutrient absorption and immunity \cite{Simard-2004-mycelium-doi:10.1139/b04-116}. 
\newline \mydash Improves soil structure as a binding agent \cite{Sarker-2022-mycelium}.
& \mydash Soil activities like digging/tillage disrupt mycelium. 
\newline \mydash Fertilizers/ fungicides can eliminate mycelium.  
& \mydash Develop security/communication system for smart farming \cite{Sirimorok-2023-mycelium}. 
\newline \mydash Integrating mycelium with IoAT for stress detection (drought/salinity), soil improvement, and plant health monitoring.
\newline \mydash Development of fungal based bio-sensors for detecting pathogens \& nutrients.
\newline \mydash Field studies to examine impact of fungi on plant biodiversity.
\newline \mydash Mycorrhizal fungi's role in plant growth via phytohormones.
\\ 
 \bottomrule
\end{tabular}}
\end{table*}

\subsection {Future Directions \& Research Opportunities}
The IoF is an innovative paradigm that holds immense potential in revolutionizing the agricultural sector. It can improve crop \& soil health, boost productivity and provide a sustainable ecosystem. This concept underscores the importance of investigating fungal networks to advance agricultural practices. Some prospective areas of research are mentioned below.


\begin{itemize}
    \item Development of fungal based bio-sensors for detecting pathogens, nutrients and soil conditions. Research \cite{Adamatzky-2021-fugalCPU} indicates that mycelium exhibit potential for electrical conductivity, particularly notable when dehydrated, as this state increases its resistance. Such characteristic makes mycelium suitable for being utilized as capacitor, or for the conversion of electrical energy to heat. Thereby, underscoring its applicability in bio-electronics. This amazing feature of mycelium could be instrumental in the realization of potential sensor network for monitoring health of trees in a forest or plants in a field.
\end{itemize}

\begin{itemize}
    \item Integrating mycelium networks into the IoT for real time data collection, transmission and monitoring.
\end{itemize}

\begin{itemize}
    \item Exploring the role of mycorrhizal fungi in synthesis and transportation of phytohormones (plant hormones), which are growth and development regulators of plants.
\end{itemize}

\begin{itemize}
    \item Investigating the interaction of AMF with soil microorganisms such as beneficial nematodes and its impact on the structure of soil, specifically under drought conditions.
\end{itemize}

\begin{itemize}
    \item Analysis of glomalin, sticky protein produced by AMF, to understand its capacity for carbon sequestration under varied environmental conditions for addressing soil degradation.
\end{itemize}

\begin{itemize}
    \item Research is required to determine the specific phosphorous \& nitrogen levels in the soil that determines whether AMF will enter a symbiotic relation with plants or remain inactive.
\end{itemize}

\begin{itemize}
    \item Further field experiments are necessary to examine the full impact of mycorrhizal fungi on plant biodiversity.
\end{itemize}


\section {Internet of Energy Harvesting Things: Sustainable Power for Agriculture}

In the IoT, ensuring sustainable power for a vast network of interconnected devices is a critical challenge. Conventional power sources, such as direct electrical connections or bulky batteries, are not viable for maintaining the multitude of small, ubiquitous smart objects continuously operational. Energy harvesting emerges as a vital solution, enabling these devices to operate autonomously and overcome power constraints. In this section, we will delve into the realm of energy harvesting for smart agriculture, including its techniques, current trends, and future directions \& research opportunities. Furthermore, a summary of applications, challenges and future research directions is presented in Table VIII.


\begin{table*}
 \captionsetup{justification=centering}
 \captionsetup{labelsep=newline}
 \renewcommand{\arraystretch}{1.7}
 \setlength{\tabcolsep}{4pt}
 \caption{Applications, Challenges and Research Directions of Internet of Energy in Smart Agriculture}\label{table1}
\centering
\resizebox{\textwidth}{!}{
\begin{tabular}{ p{4.5cm} p{4.5cm} p{4.5cm} p{4.5cm}}
 \toprule
\textbf{Technology Domain/ Area} & \textbf{Suitability for IoAT} & \textbf{Limitations/ Challenges} & \textbf{Potential Research Directions}   \\


\midrule
$\textbf{Internet of Energy Harvesting}$\\
\midrule
Sustainable Energy

& \mydash Powering wireless sensor NW with harvestable energy.
\newline \mydash Automated irrigation systems.
\newline \mydash Energy-independent farm operations.

& \mydash Ensuring consistent energy supply.
\newline \mydash Limited battery/ super-capacitor storage capacity. 
\newline \mydash Durability and maintenance.

&  \mydash Drones using harvested energy to monitor agricultural areas.
\newline \mydash Plant based energy harvesting \cite{Howe-2018-IoEH-Plants}.
\newline \mydash Energy-harvesting wearable devices for livestock health monitoring.
\newline \mydash Energy-harvesting from vibrations \& movement of agricultural machinery/ vehicles.
\newline \mydash Exploring lightweight, robust, and long-lasting energy storage technologies.
\newline \mydash Developing self-powered sensors for IoT systems.
\\ 
 \bottomrule
\end{tabular}}
\end{table*}

\subsection{Overview of Internet of Energy}

 The exponential increase in energy consumption, driven by its extensive use in industries, data centers, transportation systems, home appliances, and smart devices, underscores the need of modernizing and automating energy infrastructure. This effort aims to ensure a reliable power supply and minimizing energy wastage. The Internet of Energy (IoEn) concept introduces a transformative approach to how energy is produced, supplied, and managed, addressing the increasing demand for energy through intelligent automation of both energy producers and consumers.

IoEn signifies a paradigm shift in power grid systems, transitioning from traditional centralized power generation and one-way power flow to more efficient, reliable, flexible and sustainable energy networks. The IoEn can be considered as an evolution of smart grid concept, characterized by a smart electricity distribution infrastructure that depends upon a reliable communication network for efficient monitoring and control.

Mirroring internet networks, IoEn aims to establish a resilient system for exchange of energy among producers \& consumers collectively identified as \enquote{prosumers} \cite{Kafle-2016-IoEn}. Achieving this objective necessitates the development of advanced monitoring and control mechanisms, facilitated through the internet, to manage distributed and intermittent energy generation and storage. The IoEn is poised to enable energy exchange among a variety of sources and loads. This includes renewable energy sources, decentralized energy storage solutions, electric vehicles as well as domestic and industrial prosumers.  All these components will be interconnected and regulated via the internet, thus enhancing the efficacy and sustainability of the energy ecosystem.

IoEn integrates the features of IoT and smart grid to enable an advanced energy network for bidirectional communication \cite{Hannan-2017-IoEn}. A considerable amount of research has been conducted to address the energy challenges encountered in IoT based applications  \cite{Akan-2018-IoEH}, \cite{Cetinkaya-2017-IoEn}, \cite{Cetinkaya-2017-IoEn-2}, \cite{Ozger-2017-IoEn}, \cite{Cetinkaya-2019-IoEn}. The main focus of this research is investigating and utilizing energy harvesting techniques and assessing their potential benefits. IoEn finds application across a diverse array of fields, like building management, networking with devices \& transports, climate monitoring, healthcare, industry and smart grids \cite{Hannan-2017-IoEn}. Specifically, in the agricultural sector, IoEn has the potential to significantly improve efficiency and productivity by incorporating renewable energy sources. This integration minimizes reliance on fossil fuels and hence contributes to the improvement of eco-system.

\subsection{Energy Harvesting Techniques}

Energy sources suitable for harvesting are mainly categorized into four groups: motion, heat, light and  EM radiation \cite{Ku-2015-EnergyHarvest}. The feasibility of utilizing these sources is determined by their characteristics, like availability, manageability and predictability. Key techniques employed in energy harvesting for IoT \cite{Akan-2018-IoEH}  including light, thermal, EM, magnetic-field, electric-field, mechanical and flow-based are discussed as under:

\begin{enumerate}

\item \textit{Light Energy Harvesting}: Energy harvesting from light sources is a widely recognized method for power supply, leveraging ambient light energy from sun or artificial sources. When the light interacts with photo voltaic (PV) cells - which are composed of semiconductor materials - it generates energy through a process called photo-voltaic effect \cite{Kafle-2016-IoEn}. Despite the decreasing costs of PV modules, their increasing efficiency and the ease of installation \& use, their application in mission critical scenarios is constrained by the fluctuations of output power, inoperability at night, and recurring maintenance costs \cite{Ku-2015-EnergyHarvest}. Nevertheless, they are useful in situations where intermittent reporting is sufficient.

\item \textit{Kinetic Energy Harvesting}: This approach involves harnessing electrical power from mechanical stress, movements and vibrations arising from motion variations. The most prominent technique within this category is transforming wind power into electrical energy via AC generators, leveraging air currents or flow. In addition, piezoelectrics, materials specifically designed for this application, are being used to generate energy from erratic kinetic sources in both indoor and outdoor environments \cite{Matiko-2013-piezo}. Although air flow based energy harvesting systems provide significant power conversion efficiency, their performance is considerably influenced by environmental conditions, similar to solar energy technologies. Conversely, piezoelectric materials operates regardless of ambient factors but designing and fabricating a generalized harvesting system pose substantial challenges \cite{Cetinkaya-2017-IoEn-2}.

\item \textit{Thermal Energy Harvesting}: Thermal or heat energy can be harnessed utilizing two distinct characteristics of materials: pyroelectricity and thermo-electricity \cite{Akhtar-2015-IoEn-thermal-9}. Mostly energy generated in this domain is through thermo-electricity which is based on Seeback effect. In this method a thermal difference forces a junction constituted by two conductors to expand in a specific direction. The amount of harvested energy can easily be adjusted by altering the connection of pairs i.e. series and/ or parallel. Although, generating power from temperature gradients is attractive due to its widespread applicability, the maximum efficiency of power extraction is fundamentally constrained by Carnot cycle \cite{Matiko-2013-piezo}. Due to their compact size, thermo-generators are broadly used in low power and time uncritical consumer electronics applications \cite{Cetinkaya-2017-IoEn-2}.

\item \textit{Electromagnetic Energy Harvesting}:  Electromagnetic or Radio Frequency (RF) energy harvesting has garnered significant attention primarily due to the exponential growth in the number of wireless devices \cite{Sudevalayam-2011-EMHarvesting-7}, \cite{Matiko-2013-piezo}. Realization of battery-less IoT devices and networks is possible due to the broadcasting nature wireless communication: it is easily available and efficiently utilizable. The proliferation of RF systems in urban areas is poised to replace the traditional energy harvesting methods to provide remote services essential for smart city infrastructures. For indoor environments, EM waves emitted from routers, modems, tablets, smartphones and laptops are captured and converted to usable power by specialized power rectifying antennae. This process enables the operationalization of low-power actuators installed in smart homes or buildings. However, it is imperative to note that RF energy harvesting is not deemed suitable for mission critical applications due to their erratic nature and variable output.

\item \textit{Magnetic-Field Energy Harvesting}: This technique involves harnessing energy from the ambient magnetic field generated around conductors that carry AC current, utilizing current transformers \cite{Moghe-2009-Magnetic-field-15}, \cite{Moghe-2012-Magneticfiedl-18}. This approach yields sufficient rate of continuous power, provided that current flow in the line is adequate. Therefore, it is suitable for IoT networks that require considerable power. Nevertheless, extracting energy safely from assets carrying high currents in close proximity to the harvester remains a challenge,  which necessitates implementation of sophisticated control and protection measures.

\item \textit{Electric-Field Energy Harvesting}: An electric-field is produced when a conductor is charged to a specific voltage level. In alternating current (AC) systems, this time-varying field results in a displacement current, facilitating the transmission and storage of electric-field induced charges in a storage element. Since the energy is accumulated from the surrounding field, this method is termed as electric-field energy harvesting \cite{Cetinkaya-2017-IoEn-2}, \cite{Moghe-2009-Magnetic-field-15}, \cite{Vendik-2017-E-Field}. Notably, electric-field is the only source that is consistent and does not depend upon the load \cite{Xu-2013-e-Field}. Due to firm regulations of voltage and frequency, the electric-field is stable and predictable in nature. Consequently, it can be labelled as the most promising strategy to provide long lasting and self sustainable  power to IoT networks irrespective of the ambient factors or conditions \cite{Cetinkaya-2017-IoEn}.

\end{enumerate}

\subsection{Emerging Trends in Energy Harvesting}

 The energy harvesting techniques outlined above find application in areas such as wireless networking and remote monitoring. However, when relying on natural resources, like solar \& wind, their operation is profoundly affected by their availability. Considering that all the energy harvesting techniques are influenced by ambient conditions, grid based factors, or other unmanageable parameters, hybrid solutions emerge as pivotal for ensuring sustained operations in critical scenarios.

Historically, energy harvesting within the Internet of Energy Harvesting Things (IoEHT) mainly aimed at augmenting existing power sources. The current trend, however, is shifting towards the development of fully battery-free and self-sustaining systems. The Internet of Hybrid Energy Harvesting Things (IoHEHT) \cite{Akan-2018-IoEH} addresses this shift by integrating various energy harvesting approaches, thus tackling issues of source availability, reliability, and the limitations inherent in single-source solutions. 

Additionally, a transformative shift is underway in agricultural energy dynamics, driven by concerns over environmental impact and the depletion of fossil fuels. Traditionally dependent on fossil fuels for resource production and mechanization, agriculture is now moving towards energy-efficient practices and renewable energy sources to achieve sustainability and economic feasibility. A significant amount of research has been conducted on energy harvesting for precision agriculture  \cite{Thene-2022-IoEH-Agri}, \cite{Gulec-2020-gulec}, \cite{Saxena-2020-energy-agri}, \cite{chel-2011-renewable}, \cite{guha-2011-rf}, \cite{khernane-2024-renewable}. Furthermore, advancements in energy harvesting techniques for IoT are detailed in \cite{Akan-2018-IoEH}, \cite{Cetinkaya-2017-IoEn}, \cite{Cetinkaya-2017-IoEn-2}, \cite{Sanislav-2021-IoEH-Iot}, \cite{Ma-2020-iOEH-Iot}, reflecting the sector's growing emphasis on sustainable and efficient energy use. These advancements signify a progressive step towards a more eco-friendly and energy-efficient future in both IoT and agriculture.

\subsection{Future Directions \& Research Opportunities}
In precision agriculture, efficiency and productivity are optimized through data-driven techniques, wireless sensors, and automation. Energy harvesting can play a critical role in powering various IoT devices deployed in the agricultural sector. Some prospective future research directions are outlined below.

\begin{itemize}

    \item Exploring novel ambient energy sources through the harvesting of energy from the vibrations and movements associated with agricultural machinery (e.g., tractors) and livestock.

    \item Investigating lightweight, robust, and long-lasting energy storage technologies capable of rapid energy transfer and efficient storage.

    \item Utilizing wireless power transfer technology to charge wireless sensors remotely.

    \item Developing self-powered sensors for IoT systems in agricultural, harnessing ambient energies like light, vibrations, thermal gradients and biochemical energy, to facilitate battery less operations.

    \item Exploring energy harvesting energy from living plants \cite{Howe-2018-IoEH-Plants} to powering wireless sensors, thus improving the sustainability of IoT systems in agricultural sector.

     \item Designing nano-scale and robust harvesting circuits tailored to efficiently address the power requirements of IoT sensor nodes \cite{Zeadally-2020-IoEHT}.

\end{itemize}


\section {The Synergy of Internet of Vehicles, Internet of Drones \& Internet of Space Things for Efficient Farming}

The convergence of IoT, Internet of Vehicles (IoV), Internet of Drones (IoD) and Internet of Space Things (IoST) is anticipated to significantly transform the agricultural sector. This integration will provide farmers and researchers with real-time insights, enhanced resource management, and the facilitation of  sustainable operations \& practices. Specifically, internet connected tractors, harvesters and other agriculture machinery will enhance the productivity levels. Similarly, IoD and IoST frameworks are set to empower farmers to make data-driven and informed decisions. Fig. 9 depicts the synergistic impact of IoST, IoD, IoV, and IoEn on agriculture. In this section we will discuss the pivotal role of IoV, IoD and IoST for augmenting the productivity of agriculture sector. Additionally, some prospective future directions will also be explored, with a summary provided in Table IX.

\subsection{Internet of Vehicles: Optimizing Agricultural Operations}

The Internet of Vehicles (IoV), an innovative concept combining Vehicle Ad-hoc Networks (VANET) with the IoT, enables enhanced communication among vehicles \cite{mahmood-2020-IoV-V2V-connected}, sensors \cite{zhang-2019-IoV-Sensor}, people \cite{Sewalkar-2019-IoV-ppl}, and cloud data centers \cite{ji-2020-IoV-survey}. Despite its potential, the IoV's varied and dynamic topology raises concerns about its safety, necessitating robust security measures. IoV is crucial in the advancement of autonomous driving by leveraging sensors and actuators. Though it is set to bring radical changes in transportation, it confronts challenges in data acquisition and optimizing communications. Pivotal to this evolution is Vehicle-to-Vehicle (V2V) and Vehicle-to-Everything (V2X) communication, in establishing a transportation ecosystem that is safer, more sustainable, and more efficient.

In the study \cite{ji-2020-IoV-survey}, authors presents an overview of VANET and delves into IoV, discussing its architecture, advantages, and applications. Furthermore, a thorough analysis on communication security issues within the IoV framework is provided in \cite{abu-2018-IoV-systematic}. Complementing these insights, \cite{Chowdhury-2021-IoV-int} introduced a novel framework for evaluating the reliability of autonomous vehicles within industrial transportation. This framework integrates safety alerts, GPS data, and onboard unit elements to facilitate the adoption of autonomous vehicles in Intelligent Transportation Systems (ITS).

The incorporation of IoV into agriculture will bring significant enhancements in farming efficiency. By leveraging real-time data, IoV has the capability to refine farming operations, especially in the areas of route and resource management, which will contribute to greater fuel efficiency. This technology has the potential in providing practical insights that can help increase agricultural yields and operational effectiveness. As a key element of IoV, predictive maintenance will be instrumental in minimizing equipment downtime and optimizing overall productivity. It also has the capacity to improve traceability from  farm to  consumer, thereby ensuring better food safety and standards. In addition, IoV can augment safety measures on the farm, providing real-time alerts related to workers and machinery safety, thus cultivating a safer agricultural  environment.


\begin{figure}
  \begin{center}
  \includegraphics[width=3.5in]{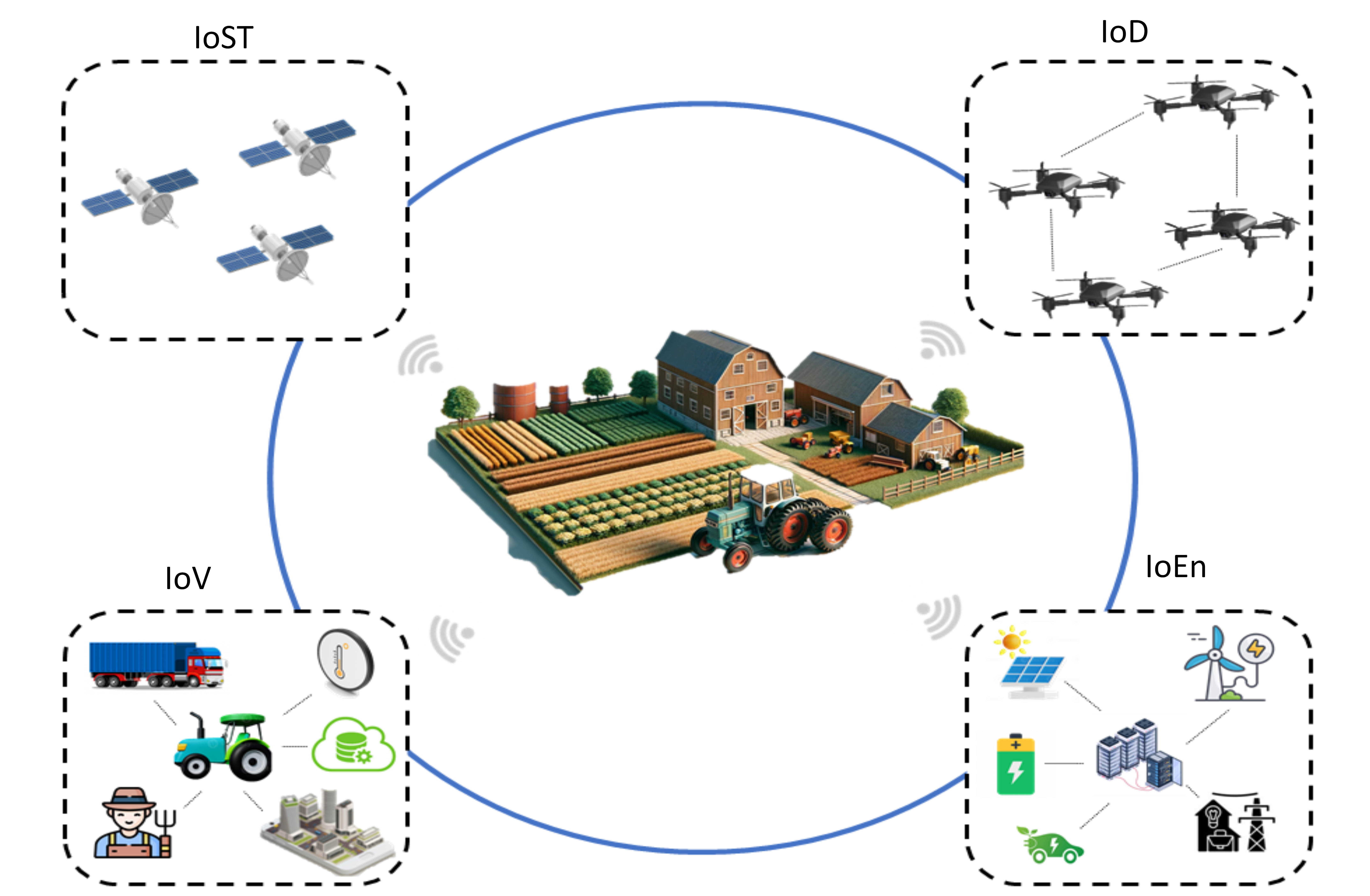}\\
 \caption{Integration of IoXs in Precision Agriculture: The interplay of Internet of Space Things (IoST), Internet of Drones (IoD), Internet of Vehicles (IoV), and Internet of Energy (IoEn) to enhance farm monitoring and management}\label{Fig-4 IoNT}
  \end{center}
\end{figure}

\subsection{Internet of Drones: A new Horizon in Pest Management and Crop Surveillance}

The Internet of Drones (IoD) introduces a sophisticated control and communication mechanism among drones, transcending traditional visual line-of-sight constraints. Powered by advancements in AI, these drones overcome obstacles such as limited processing capacity and the need for direct line-of-sight, enabling them to autonomously collect data, communicate, and make decisions. Such progress opens up  plethora of application areas, including environmental monitoring, disaster response, remote surveillance, and notably, agriculture \cite{BOURSIANIS-2022-IoD-100187}.

In precision agriculture, the extensive network of interconnected sensors is vital for the continuous collection and analysis of field data. This data is essential for driving sophisticated AI prediction models crucial in guiding vital agricultural decisions, optimizing resource allocation, and enhancing production efficiency. The system adeptly gathers, analyzes, and distributes this processed data to relevant parties, facilitating informed decision-making. Additionally, the integration of autonomous drones within the IoT framework offers substantial prospects for enhancing precision agriculture techniques. The IoD revolutionizes precision agriculture through real-time data provision, automation of tasks, and remote monitoring. Such technological advances enable farmers to better utilize resources, improve crop quality and yield, and enhance animal welfare, paving the way for more efficient and sustainable farming practices and contributing to global food security.

In \cite{Caruso-2021-IoD-Data} researchers presented an analytical model for efficient data collection in precision agriculture  using drones. The requirements and associated challenges of using Unmanned Aerial Vehicles (UAVs) in smart agriculture are investigated in detail in \cite{Reddy-2021-IoD}. \cite{Alanezi-2022-IoD-livestock} discuss the role of drones in livestock management, while \cite{Chen-2021-IoD-pests} study their use for pest identification and precision spraying. However, like IoT, the IoD faces significant security and privacy challenges \cite{Yahuza-2021-IoD-Security}, which could hinder its broader application. The implementation of robust security and privacy measures, such as the decentralized blockchain-based security model proposed in \cite{Yazdinejad-2021-IoD-Security}, is crucial for realizing the full potential of IoD. Additionally, \cite{Yang-2022-IoD-security} provide a comprehensive review of the security problems and solutions in IoD, emphasizing the importance of safeguarding these systems to ensure safe and effective operations in agriculture and beyond.


\begin{table*}
 \captionsetup{justification=centering}
 \captionsetup{labelsep=newline}
 \renewcommand{\arraystretch}{1.7}
 \setlength{\tabcolsep}{4pt}
 \caption{Applications, Challenges and Research Directions of IoV, IoD \& IoST in Smart Agriculture}\label{table1}
\centering
\resizebox{\textwidth}{!}{
\begin{tabular}{ p{4.5cm} p{4.5cm} p{4.5cm} p{4.5cm}}
 \toprule
\textbf{Technology Domain/ Area} & \textbf{Suitability for IoAT} & \textbf{Limitations/ Challenges} & \textbf{Potential Research Directions}   \\

\midrule
$\textbf{IoV}$\\
\midrule
Intelligent Vehicles 
&  \mydash Monitor agricultural machinery- tractors \& harvesters in real-time.
\newline \mydash Track agricultural products.
\newline \mydash Ensure Safety \& Security of vehicles \& farm assets.

& \mydash Interoperability \& Standardization issues \cite{Agbaje-2022-IoV-Interopera}.
\newline \mydash Poor NW communication in rural areas.

&  \mydash Leveraging tokens \& blockchain technology \cite{Wang-2021-IoV-token} for data \& transaction security.
\newline \mydash Smart logistics to optimize productivity and risk management.
\newline \mydash Employing vehicular edge intelligence in agricultural vehicles.
\newline \mydash Electrification of agricultural vehicles for ecofriendly operations.
\\ 

\midrule
$\textbf{Internet of Drones}$\\
\midrule
Smart UAVs

& \mydash Sensing \& monitoring.
\newline \mydash Smart spraying.
\newline \mydash Livestock management.

& \mydash Privacy and Security.
\newline \mydash Subpar performance in extreme weather.
\newline \mydash Operational management \cite{Labib-2019-IoD-management}.

&  \mydash Knowledge distillation \& Federated Learning \cite{Yahuza-2021-IoD-Security}.
\newline \mydash Livestock monitoring through wearable sensors.
\newline \mydash Robust communication between IoD \& ground sensors. 
\\ 

\midrule
$\textbf{Internet of Space Things}$\\
\midrule
Satellite Network  

& \mydash Deliver data on weather ,crop health \& soil conditions.
\newline \mydash Accessibility to remote areas.

& \mydash Ensuring data quality \& reliability due to atmospheric conditions.
\newline \mydash High manufacturing \& launch cost.

&  \mydash Optimizing IoT protocols for ground-space NW communication.
\newline \mydash Low-latency ultra-reliable communication by IoST. 
\newline \mydash Integrating terrestrial NW with drones \& satellites for reliable and robust communication.
\\ 
 \bottomrule
\end{tabular}}
\end{table*}


\subsection{Internet of Space Things: Taking Agricultural Data to New Heights}

In recent years, proliferation of IoT devices has expanded across diverse fields, signalling a trend towards omnipresence in the near future. This increase underscores the importance of efficient connectivity with minimal latency. Generally reliant on 4G and 5G networks, these IoT devices frequently face connectivity issues \cite{Islam-2021-IoST-connectivity}, notably in rural areas like agricultural lands, farms, and forests, where network coverage is typically limited. This constraint has impeded the expansion of IoT's applications in such areas.

The Internet of Space Things (IoST) emerges as a pioneering solution to the connectivity challenges faced by IoT devices, especially in isolated areas. Utilizing a network of small satellites in low Earth orbit, direct communication of IoT devices is facilitated from space, thus circumventing the inherent limitations of ground-based networks. This pioneering technique is revolutionizing connectivity, facilitating access to  remote areas previously inaccessible by conventional communication technologies, thereby significantly enhancing the reach and capabilities of IoT. An important development is the use of  CubeSats \cite{Akyildiz-2019-IoST-cubestats}, initially designed as small, cost-efficient satellites for research purposes, which have now evolved as an integral part of the IoST infrastructure.
Their deployment in lower Earth orbit is not only an economically viable solution but also  a significant advancement in telecommunications technology. 

In smart agriculture, IoST is key to guaranteeing constant connectivity for IoT devices in regions of sparse network coverage, thereby enabling deployment of IoT devices in challenging agricultural environments. Addressing these challenges, \cite{Mallick-2023-IoST-agri}  proposed a Blockchain-based IoST framework specifically designed for agricultural IoT systems. This framework consists of a three-tier architecture: a device layer for data collection, a fog layer for processing, and a cloud layer for analysis and storage.This structured approach not only augments data processing but also improves the security and overall performance of IoT applications in agricultural domain. Additionally, to meet future needs, a novel cloud-compatible architecture for IoST has been proposed in \cite{Priyadarshini-2022-IoST-cloud}, increasing its efficacy and versatility.

The introduction of IoST enables broad and efficient deployment of IoT  in various sectors, notably in agriculture. This advancement is not only limited to Earth; with IoST \cite{Priyadarshini-2022-IoST-cloud}, the concept of farming beyond our planet \cite{MONJE-2003-Space-151}, is becoming increasingly a reality. IoST lays the groundwork for more sophisticated, data-centric agricultural practices, greatly advancing smart farming technologies and expanding the horizons of IoT applications.

\subsection{Future Directions \& Research Opportunities}
The preceding discussion underscores the transformative potential of IoV, IoD and IoST within the agricultural domain. Some prospective research directions are discussed below.

\begin{itemize}
    \item Electrification of agricultural vehicles - tractors, harvesters, crop sprayers and delivery trucks - for eco-friendly operations, facilitating seamless sensor integration, reducing maintenance costs and enhancing communication capabilities both among vehicles themselves and with a central server for enabling real-time data analysis and monitoring.
\end{itemize}

\begin{itemize}
    \item Employing vehicular edge intelligence in agricultural vehicles for enabling precision farming, improving efficiency and optimizing resource utilization. By equipping each vehicle with advanced sensors, they will be transformed into intelligent nodes capable of collecting valuable data such as soil moisture levels, weather conditions, crop health, pests infections and more. These intelligent nodes will process data on the move and make smart decisions based on collected information without the constant need to transmit data to the server.
       
\end{itemize}

\begin{itemize}
    \item The deployment of autonomous vehicles to improve agricultural supply chain and logistics. This includes real-time tracking of vehicle fleets for optimizing  routes, monitoring fuel consumption and ensuring timely deliveries.
    
\end{itemize}

\begin{itemize}
    \item Enhancing drones intelligence through knowledge distillation and federated learning. This will ensure privacy and also enable efficient communication.
\end{itemize}

\begin{itemize}
    \item Utilizing blockchain technology to ensure the integrity and security of data exchanges  between IoV , IoD and satellites.
\end{itemize}

\begin{itemize}
    \item Combining the high-resolution \& targeted capabilities of drones with the extensive coverage and consistency of satellites. This synergy will improve both spatial and temporal sensing resolutions of agricultural monitoring. This collaboration will reduce operational costs and address challenges like limited duration of drone flight and data processing constraints thereby offering a comprehensive solution for monitoring and analysis.
\end{itemize} 

\begin{itemize}
    \item Integrating terrestrial networks with drones and satellites to build a unified, reliable and robust communication infrastructure. Such networks would facilitate in enabling agricultural practices across diverse landscapes and on a larger scale.    
    
\end{itemize}


\section {The Nexus of 6G and Precision Agriculture}
The Internet of Things (IoT), a vast network connecting a multitude of objects and devices, has been instrumental in fostering novel services and technological innovations. As the number of IoT devices continues to surge, there is an increasing demand for robust network infrastructure. Alongside the widespread adoption of IoT, several challenges emerge that require resolution. This interconnected ecosystem depends heavily on sophisticated communication technologies, such as 5G and the upcoming 6G network \cite{Bhatia-2023-6G}, which are crucial for enabling advanced hybrid and multi-cloud solutions. In this section we will discuss the critical role of 6G technology  in enhanced connectivity for modern agriculture, key technological enablers of 6G and some prospective future research directions. In addition, a summary of applications, challenges and future research directions is presented in Table X.

\subsection {Enhanced Connectivity for Modern Agriculture}

As discussed in Section I, advancement in communication technology is  critical for realization of modern agriculture. Such advancements facilitate optimal usage of resources, supports informed decisions and promote sustainable practices in agriculture. It is imperative for farmers to leverage these technologies to boost their productivity in the constantly changing agricultural landscape.

With the exponential increase in the quantity of interconnected devices and pressing need for improved services, existing wireless communication systems are required to evolve to meet user demands, such as reduced latency, increased throughput, omnipresent connectivity and enhanced security. Future oriented technologies are poised to revolutionize communication by providing ubiquitous connectivity. This transformation will commence an era of sophisticated and intelligent services, alongside devices that will operate at extremely high frequency bands. To attain these objectives, several vital technology enablers need to be developed and integrated. These include advanced antenna systems, increased spectral efficiency, the application of nanotechnology, and tactile internet.

The anticipated digital ecosystem necessitates complete connectivity, which is not yet fully materialized by 5G networks. Future generations, 6G and beyond, are expected to revolutionize communication technology and meet the high expectation of users. To achieve this goal we can either enhance 5G networks or develop new communication techniques and strategies \cite{Saad-2020-6G}. Although, 5G technology offers increased speed and connectivity over its predecessors, it faces certain constraints, including limited coverage due to the use of higher frequency bands, substantial infrastructure costs, issues with device compatibility \& energy consumption, security and standardization. Furthermore, despite 5G's promise to enable real-time sensing applications - such as autonomous vehicles, virtual reality, remote surgeries, precision agriculture, and smart cities - yet early deployments have revealed some limitations. The relationship between 5G and 6G is symbiotic, with 5G setting the platform for next generation of wireless technology.

\begin{figure}
\begin{center}
  \includegraphics[width=3.5in]{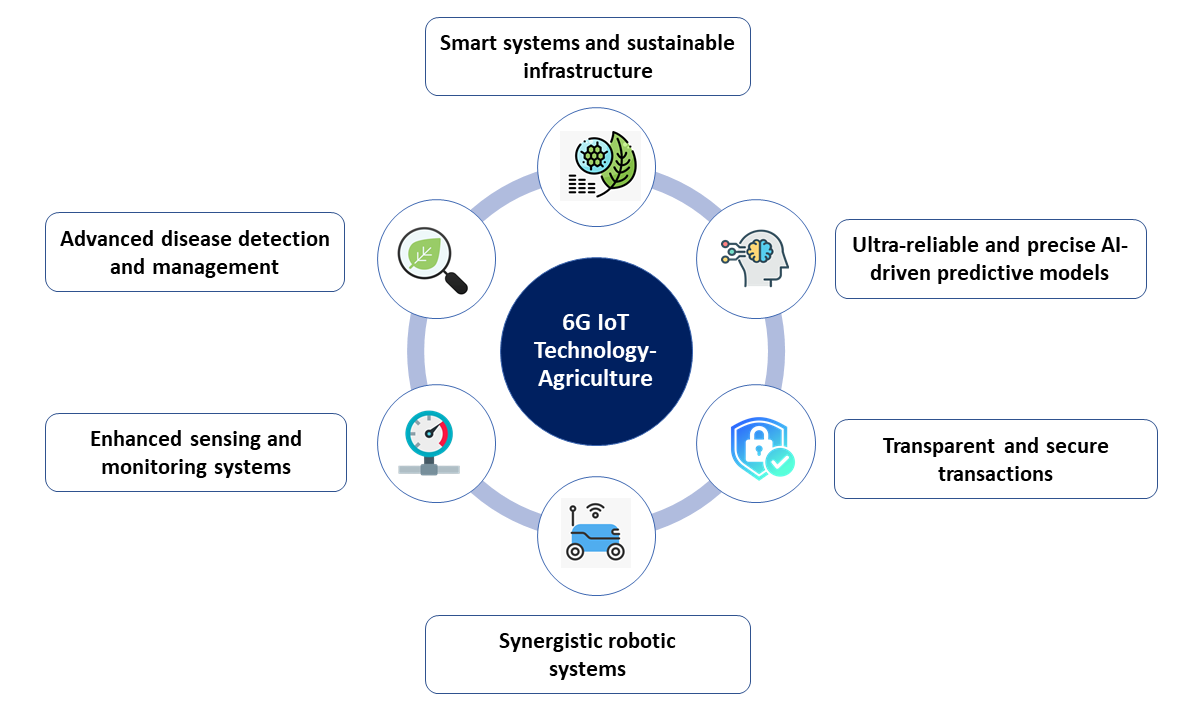}\\
 \caption{Overview of 6G IoT technology applications in agricultural ecosystems}\label{Figure 6G-agriculture}
  \end{center}
\end{figure}

The introduction of 6G is expected to provide high data rates with universal connectivity, enabling the interconnection of everything, everyone and everywhere. Research outlined in \cite{Zhang-2019-6G-req} explore the future perspectives, requirements, framework and leading technologies of 6G systems. In contrast, \cite{Qadir-2022-6G-future} sheds light on current advancements, application scenarios and prevailing challenges in 6G. Moreover, \cite{Hokazono-2022-6G} presents an innovative architecture for 5G and 6G that integrates terrestrial networks with non-terrestrial networks, including high-altitude platform stations (HAPSs), aiming to broaden coverage, increase efficiency, and guarantee the universal availability of mobile services. Additionally, the authors in  \cite{Jiang-2021-6G-req} presents some key performance indicators (KPI) for the upcoming 6G technology, integrating new KPIs with those previously established for 5G networks.

High performance precision agriculture aims to deliver automated irrigation systems, real-time monitoring and data analytics \& insights. Despite the importance of these crucial capabilities, accessing vital data remains a challenging task, primarily due to connectivity problems in rural areas. The integration of smart devices, sensors, automated equipment, and drones holds promise to boost yields and efficiency  within the agricultural sector. However, their deployment requires substantial resources and high data speeds. The advent of 6G technology, promising widespread wireless connectivity, is expected to considerably enhance this sector \cite{Akyildiz-2020-6G-agric}, as illustrated in Fig. 10, which outlines the key 6G IoT technology applications in agricultural ecosystems. It is important to realize that these advancements might primarily benefit large-scale enterprises, or supply-chain players, nevertheless 6G is set to accelerate the speed of progress in agricultural digitization.

\subsection {6G Technological Enablers for Agricultural Advancements}

The introduction of 6G technology is poised to significantly enhance the IoT framework, offering greater stability and network reliability, thereby improving the overall quality of life. It is anticipated to bring considerable improvements in speed, latency, and connection density, opening doors to cutting-edge services like holographic communications, Metaverse and Extended Reality (XR) \cite{Xing-2022-6G}. However, realizing these sophisticated applications necessitates significant advancements in mobile network intelligence, computational capabilities, and sensory technologies, underscoring the imperative for considerable investments in mobile network infrastructure to fully capitalize on 6G's potential.

To achieve the goals of 6G systems, identifying and developing key technological enablers is vital. In \cite{Zhang-2022-6G-review}, researchers delve into key technologies underpinning sixth-generation (6G) mobile communication systems, such as terahertz technology, reconfigurable intelligent surface (RIS), space–air–ground-integrated networks, wireless AI, massive MIMO, integrated sensing and communications (ISAC), and digital twins, examining their current applications and future potential in enhancing smart agricultural practices. Fig. 11 highlights the key technologies anticipated to enable 6G communication, which are briefly described below.

\begin{enumerate}

\item \textit{Quantum Communication \& Computing}: Leveraging quantum mechanics, this technology is poised to significantly impact 6G networks by enabling secure and ultra-fast communication, thereby providing robust protection against cyber security threats such as hacking and eavesdropping \cite{Chowdhury-2020-6G-quant}, \cite{Prateek-2023-6G-quant}. Its applications span diverse fields such as military communications, finance, and any other sensitive data transfer, making it one of the core 6G enabler. Despite its advantages, quantum communication systems face challenges like signal fragility and the need for new and innovative technologies like quantum repeaters and amplifiers to extend their range. Additionally, accurate detection of quantum signals is complicated by challenges like decoherence and noise \cite{Vu-2023-6G-qunatum}.

\begin{figure}
  \begin{center}
  \includegraphics[width=4in, height=2.4in]{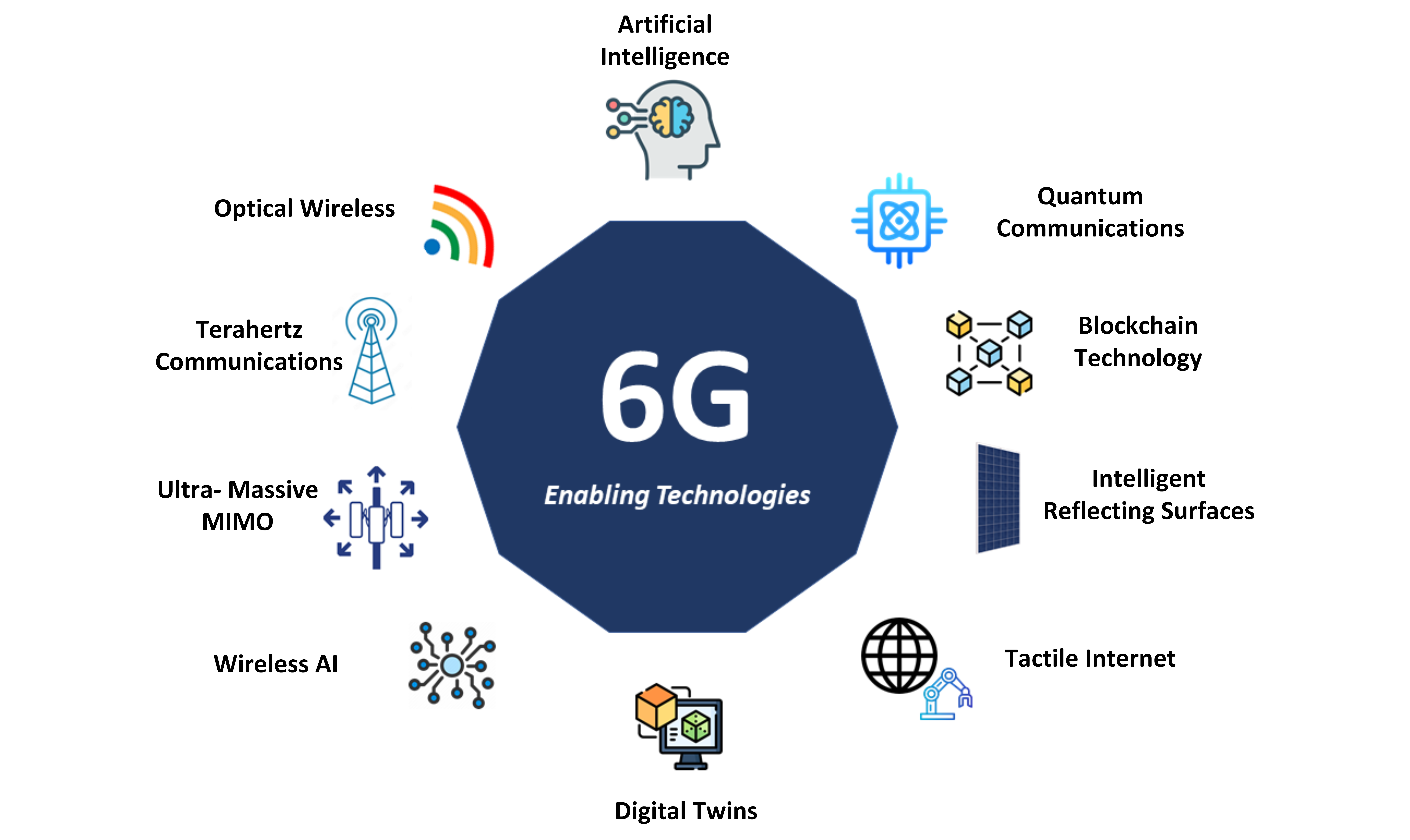}\\
 \caption{Key technologies for enabling 6G communication}\label{Figure 6G Enabling tech}
  \end{center}
\end{figure}
Quantum computing offers remarkable potential in agriculture, promising enhanced crop productivity, precision management, and monitoring plant \& soil health. When integrated with technologies such as IoT, AI and big data analysis, quantum computing can contribute to sustainable farming. By utilizing the quantum entanglement - where the state of one particle instantaneously influences the other - quantum sensors can furnish very accurate measurements of soil moisture and nutrients. The authors in \cite{MARAVEAS-2024-quantum-agri} have conducted a comprehensive survey regarding harnessing quantum computing's capabilities for smart agriculture.

\item \textit{Terahertz Communication}: The rapid advancements in information technology have led to the exponential growth in data traffic, with studies indicating that wireless data traffic doubles every 18 months \cite{Cherry-2004-terahz-data}. As a result, communication technologies have started exploring and embracing the higher frequency bands to achieve higher data rates. 5G, for instance, has officially incorporated millimeter-wave (mm-wave) communications (30-300 GHz) within the sub-100 GHz band. However, there are still limitations and challenges in accommodating billions of devices and achieving data transmission at the terabit per second (Tbps) level.

Addressing these challenges, the terahertz (THz) frequency band, ranging from 0.1 THz to 10 THz, has been proposed due to its potential for extra-low latency, ultra-high bandwidth and exceptionally high data rates. These capabilities render THz waves suitable for micro-sized communications, holographic communications,  and ensuring localized, high-speed, high precision connectivity.

Research on THz technology, which commenced more than a decade ago, has found extensive applications in the agricultural sector. The study \cite{Usman-2022-tHz-plants} presents an integrated communication and sensing system based on THz waves for monitoring plant health. In another work \cite{Zahid-2018-THz-Plants}, the application of THz technology has been shown to assess the water content and detect pesticide residues in plant leaves. Further, \cite{Jiang-2014-THz-quality} discusses the applicability of THz in the detection of agricultural product quality in a non-destructive manner. Moreover, \cite{wedage2021climate} investigated the potential of THz in sensing climate changes for enhancing crop productivity. 

As discussed above THz band can be used for the improvement of agricultural sector. Some of the potential applications will include:

\begin{itemize}
    \item \textbf{Water content \& quality evaluation}: THz waves can detect water content in plant leaves to support irrigation management. Moreover, the interaction of THz waves with water enables the detection of unique spectral imprints from substances within water, thus aiding in identification of contaminants.
\end{itemize}
\begin{itemize}
    \item \textbf{Food quality \& safety}: The application of THz waves allows for the non-destructive assessment of the quality of vegetables, fruits and grains. THz spectroscopy possesses the capability to discern weak inter-molecular interactions which can help in the detection of pesticide residues and heavy metals in food.
\end{itemize}

\begin{itemize}
    \item \textbf{Pest detection}: The sub-millimeter wavelengths of THz radiations have the ability of high penetration and low energy consumption, making it ideal for wireless sensing and detection of pests and diseases in plants.
\end{itemize}

\begin{table*}
 \captionsetup{justification=centering}
 \captionsetup{labelsep=newline}
 \renewcommand{\arraystretch}{1.7}
 \setlength{\tabcolsep}{4pt}
 \caption{Applications, Challenges and Research Directions of 6G in Smart Agriculture}\label{table1}
\centering
\resizebox{\textwidth}{!}{
\begin{tabular}{ p{4.5cm} p{4.5cm} p{4.5cm} p{4.5cm}}
 \toprule
\textbf{Technology Domain/ Area} & \textbf{Suitability for IoAT} & \textbf{Limitations/ Challenges} & \textbf{Potential Research Directions}   \\
\midrule
$\textbf{6G}$ & & & \\
\midrule
Networking \& Communication 
& \mydash Real-time data analysis \& monitoring of fields: leveraging low latency, high bandwidth \& wide spatial reach of 6G.
\newline \mydash Integrating large-scale IoT devices with higher throughput \& efficiency \cite{Zhihan-2021-6G}.

& \mydash High energy consumption due to high frequency bands \cite{mohammed-2021-from5Gto6G}.
\newline \mydash Terahertz band's high vulnerability to obstacles \& atmospheric effects.
\newline \mydash Expensive 6G infrastructure.
\newline \mydash Lack of legal \& policy frameworks.

& \mydash Integrating AI with 6G for NW management in IoAT.
\newline \mydash Zero-Energy sensors for IoAT.
\newline \mydash THz spectroscopy for metal contaminant detection in food.
\newline \mydash Integrating RIS with wireless power transfer systems for charging IoT devices.
\newline \mydash Digital twin for real-time virtual monitoring and predictive analysis of farms.
\newline \mydash Integrated sensing \& communication with space-air-ground-integrated NW.
\\
\bottomrule
\end{tabular}}
\end{table*}

\item \textit{Optical Wireless Communications (OWC)}: Complementing the RF based traditional mobile communication, 6G will significantly incorporate OWC. OWC frequency range encompasses infrared (IR), visible light (VL) and ultraviolet (UV) spectrum \cite{Chowdhury-2018-OWC}. The spectrum of OWC is ultra-wide and remains unregulated. It can provide both indoor and outdoor services with communication distances ranging from a few nm to over 10,000 Km. Some of the notable and well established OWC technologies include visible light communication (VLC), optical camera communication, light fidelity, and optical band based FSO (Free Space Optical) systems.

OWC can introduce transformative changes in the agricultural sector. The reliance of current smart farming practices on RF communication networks could adversely affect  plant health and growth rates. To address these concerns, authors in \cite{Javed-2021-OWC-agri} proposed a solution that involves either replacement or supplementing the existing RF network with LEDs as grow lights. These LEDs not only assist in plant growth but also acts as communication carriers utilizing technologies such as VLC and optical  camera communication for enhanced smart farming. Furthermore, in \cite{Matus-2021-owc}, a scalable and cost-effective optical camera communication  based solution for WSN in precision farming is introduced. In \cite{Putra-2022-FSO}, it is demonstrated that OWC, in comparison to traditional RF, offers high capacity, low latency and reduced interference. Simulation results validate that FSO is suitable for outdoor IoT applications under various weather conditions, maintaining robust performance.

\item \textit{Ultra-massive MIMO}:  Massive multiple-input, multiple output (MIMO) technology is anticipated as a core 6G network enabler. Various terms are employed to express Massive MIMO systems including very large MIMO, advanced antenna systems, large scale antenna systems, hyper MIMO and full dimension MIMO. These systems utilize a large number of antennas to enable high diversity and mitigate path loss in high frequency communication systems \cite{Sakai-2020-MIMO}. Massive MIMO is realized when the number of antennas serving each user is more than 10, escalating to hundreds or even thousands of antennas for what is termed as ultra-massive MIMO \cite{Wang-2023-MIMO}. This ultra-massive MIMO capability enables the generation of spatial beams suitable for THz frequency bands \cite{Dilli-2021-MIMO}. This technology could be beneficial in high-density urban areas and other environments where the demand for wireless data is huge. Ultra-massive MIMO systems have been explored by many researchers as a vital technology for future networks \cite{Faisal-2020-MIMO}, \cite{Akyildiz-2016-MIMO}, \cite{Busari-2017-MIMO}, \cite{wang-2024-tutorial}.

MIMO technology, a precursor of ultra-massive MIMO, has found applications in various agricultural contexts, such as enhancing the resolution of agricultural sensors for crop growth monitoring \cite{Liu-2022-MIMO-irri} and improving underground sensing for precising irrigation \cite{Chaudhary-2011-MIMO-pre}. The use of massive MIMO for agricultural IoT has been investigated in \cite{BANA-2019-MIMO}, illustrating its capability to significantly advance the field. Furthermore, MIMO technology has been employed to ameliorate the performance of integrated air-space-ground communication networks involving  UAV and satellite networks, thereby enabling better communication services for agricultural activities \cite{Zhao-2018-MIMO}.

\item \textit{Intelligent Reflecting Surfaces (IRS)}: Spectral efficiency stands as crucial key performance indicator (KPI) for 6G networks. IRS is poised to bring revolutionary advancements in spectral efficiency. An IRS consists of passive reflecting elements that can be programmed to reflect the incoming signals towards a specific direction. In the context of 6G, reconfigurable intelligent surfaces (RIS) are considered as MIMO 2.0 \cite{Chowdhury-2020-6G-quant}.

IRS can augment 6G network's capacity by enabling multiple users to simultaneously send and receive signals on the same frequency band while consuming less power. Additionally, IRS helps to enhance signal coverage and reduced interference, thereby improving signal strength by evading obstacles. This capability ensures enhanced network performance in challenging environments like urban or indoor settings. When deployed on surfaces such as walls, ceilings, glass etc., IRS transforms the wireless environment into intelligent reconfigurable reflectors known as smart radio environment (SRE) \cite{Di-2020-IRS}. This facilitates passive beam forming, that can notably improve the channel gain with lower costs and energy consumption compared to massive MIMO antenna arrays.

In smart farming, UAVs play a vital role in monitoring large areas and providing valuable insights to the farmers. IRS can significantly reduce power consumption in the UAV systems while also enhancing the line of sight probability \cite{Alsarayreh-2023-IRS} \cite{Hashida-2020-IRS}. Authors in \cite{Liu-2022-irs-agri}, proposed a novel power supply solution called PowerEdge, that incorporates energy harvesting, storage, wireless power transfer, and IRS to attain sustainable smart agricultural operations. Smart agriculture relies on IoT devices, which can be adversely affected by harsh environmental conditions. To overcome these communication challenges, the application of IRS in various contexts, including underwater IoT, underground IoT and in industry 4.0, has been investigated to improve the performance of system \cite{Kisseleff-2021-IRS-agri}.

\item \textit{Digital Twins}: The concept of digital twin was introduced a decade ago and was aimed to provide optimized performance, real-time monitoring, simulations, and precise predictions. Despite its early conceptualization, the theoretical framework and practical implementations of this technology have not yet been fully realized. Currently, digital twin is gaining unprecedented traction and is poised to optimize quality control, process design, and decision making. It achieves this by transforming a physical entity into a virtual counterpart utilizing various techniques \cite{JONES-2020-digital-twin}. A digital twin consists of a physical object, its virtual clone and the data that flows between them \cite{Barricelli-2019-Digital-twin}. The proliferation of AI and big data has further propelled the interest in digital twin technology establishing it as a versatile tool that has the potential for widespread applications in every domain.

In the agricultural sector, researchers have begun to harness the capabilities of digital twin technology. A cost-effective \& highly accurate digital twin framework for farmland has been proposed in \cite{VERDOUW-2021-DT-agri}, utilizing WSN and cloud servers. This model facilitates in near real-time virtual monitoring of farm  conditions and is scalable for large-scale farm deployment. A novel approach that combines DL and digital twin has been applied in aquaponics - symbiosis of fish \& plants - resulting in significant resource savings, such as soil, water and lighting \cite{Ghandar-2021-Digitaltwin-agri}. Additionally, a model that integrates digital twins, bigdata and IoT has been introduced in \cite{Howard-2020-DT-agri-127} for greenhouse agriculture, enabling a conducive greenhouse environment. In underground farming, digital twin technology has been leveraged to achieve low electrical energy consumption \cite{Jans-Singh_Leeming_Choudhary_Girolami_2020-128}. Furthermore, the role of digital twins in agriculture has been investigated in \cite{Christos-2021-DT-agri}, highlighting their current adoption level and outlining a roadmap for their widespread implementation. 

Overall, the agricultural sector stands to gain immensely from digital twin technology, with applications ranging from livestock, field and agricultural machinery monitoring to supplychain and logistics optimization, weather prediction \& risk management, sustainability and resource management.

\item \textit{Tactile Internet (TI)}: 5G-enabled IoT systems  primarily concentrate on enhancing awareness and connectivity. However, with the advent of 6G, a shift towards more sophisticated and real-time human-machine communication is expected, especially for controlling and managing IoT devices through the tactile internet (internet of skills). As outlined in \cite{ABDELHAKEEM-2022-TI} TI is envisioned as a wireless communication system specifically designed to enable the real-time exchange of touch, control and sensory data between humans and machines. The foundational requirements for TI are ultra-low latency, short response times, robust reliability \& resilience, edge clouds network and stringent security measures. Furthermore, TI is anticipated to pioneer applications such as haptic interface that will significantly enhance the interaction between humans and digital environments.

The integration of TI, characterized by its ultra-reliability and responsiveness in haptic experiences, holds the potential to bring transformative advancements in the field of agriculture. TI enhanced robots can be utilized for planting, harvesting and weed control. Furthermore, the application of haptic interfaces could notably improve the navigation and control of robots within complex agricultural environments. Additionally, the incorporation of TI technology in agricultural machinery and vehicles is poised to enhance safety, productivity and efficiency in this domain.

\item \textit{Blockchain}: Blockchain technology emerges as a critical enabler for 6G networks \cite{Hewa-2020-Blockchain-32}. It functions as a distributed ledger system, operated over a peer-to-peer network, thus eliminating the requirement for a centralized control or authority. The ledger spans over a network of computers rather than being confined to a single server. Security in transactions is achieved via a series of blocks linked together, where each block contains a cryptographic hash of the previous one, ensuring the tamper-proof and immutable nature of the system \cite{L.B.-Blockchain-agriculture12091333},\cite{Chowdhury-2020-6G-quant}. Consequently, blockchain is poised to facilitate secure and transparent data communication in 6G networks, proving invaluable in applications where high level of privacy, trust and security are major concerns.

Furthermore, the application of blockchain transcends communication networks. As a ledger system where every participant can store and validate transactions, it offers a viable approach for smart agriculture. It enables the secure management of inventories and contracts, thus can contribute to the establishment of reliable food supply chains. Additionally, it can also facilitate in processing payments between farmers, retailers and consumers in a timely and secure manner, showcasing its versatility and potential in multitude of applications.

\item \textit{Wireless Artificial Intelligence}: The advancements in sensing and communication capabilities of wireless systems have led to swift expansion of IoT and mobile computing networks. The integration of AI with wireless communication, termed as wireless AI, is poised to bring evolutionary advancements in intelligent network systems. AI's significance is evident across various aspects of wireless communication, including the modeling, learning and forecasting of intricate wireless propagation environments, tracking network states, smart scheduling, signal processing and network deployment optimization. In \cite{nguyen-2020-wirelessAI}, the authors conducted a comprehensive survey on the role of wireless AI in communication networks, underscoring its capability to transform the landscape of wireless communication through enhanced intelligence and efficiency.

AI plays a huge role in smart agriculture applications, such as digital soil mapping \cite{Dong-2018-AI-agri-101}, health monitoring of agricultural machinery \cite{Gupta-2020-AI-agri-100}, disease detection in plants \cite{Torai-2020-AI-agri-99} and resource management \cite{Nasir-2023-resourcemgmt}. The research progress regarding AI enabled smart farming has been reviewed in detail in \cite{Godwin-2021-AI-agri-102}. Furthermore, the integration of AI with wireless communication and sensor networks, facilitating sophisticated sensing and monitoring capabilities is discussed in \cite{Vijayakumar-2021-AI-agri-103}, \cite{Vincent-2019-AI-104}, \cite{Dasgupta-2020-AI-105}, \cite{Somov-2018-AI-106}.

\item \textit{Integration of Wireless Information and Energy Transfer}: WIET is poised to be one the most cutting-edge technologies in 6G. Future wireless systems will feature the capability of wireless power transfer, utilizing radio waves for the dual purpose of transmitting both energy and information. This dual purpose optimizes the use of RF spectrum, enabling a network that can power trillion of low-powered devices ubiquitously and on demand. Thereby, WIET will enhance the longevity of battery-powered wireless sensor networks \cite{Wang-2014-WIET} ,\cite{Clerckx-WIET-agri}.

In the context of smart farming, thousands of wireless senors are deployed in various environments such as soil, water, plants and on machinery, all of which require periodic battery charging. WIET technology promises to extend battery life of these sensors thereby enhancing operational efficiency.

\end{enumerate}

\subsection {Future Directions \& Research Opportunities}
IoT is steering the agricultural sector towards a new era of smart farming. This approach aims to blend modern scientific and technological advancements with traditional farming methods, leading to automated and intelligent cultivation and production processes \cite{Yang-2021-6G}. Current wireless technologies, such as ZigBee and Wi-Fi \cite{Hidayat-2020-6G}, while crucial, fall short in meeting the evolving needs of smart agriculture. This gap underscores the necessity of integrating the emerging 6G technology to enable more advanced applications in the agricultural sector. 
Several promising research directions are highlighted below.


\begin{itemize}
    \item The utilization of THz waves for pest detection and disease control emerges as a promising research direction. THz waves' high penetration, high sensitivity and minimal energy consumption characteristics make them suitable for non-invasive monitoring of water content in plant leaves. 
    Furthermore, THz spectroscopy, could be instrumental in identifying metal substances in food products, that may be contaminated by pesticides, thus ensuring food safety.
\end{itemize}

\begin{itemize}
    \item Establishment of network infrastructure in rural areas is costly. To address this issue IRS can be used to improve the coverage, quality and accuracy of communication service. Additionally, integration of IRS with wireless power transfer systems can offer solution to charging battery-powered IoT devices deployed in remote areas. 
\end{itemize}

\begin{itemize}
    \item The development of haptic-driven robotic systems for executing agricultural tasks like planting, harvesting and weed control.
\end{itemize}

\begin{itemize}
    \item Utilizing digital twin technology for real-time virtual monitoring and predictive analysis of farms.
\end{itemize}

\begin{itemize}
    \item Leveraging integrated sensing and communication with space-air-ground-integrated networks for ubiquitous monitoring of farms and forests.
\end{itemize}

\begin{itemize}
    \item Using the high data rates, ultra-low latency, reliability and security features of 6G for data analytics and repository management in agricultural. Moreover, this technology could enable autonomous agricultural machinery realization.
\end{itemize}

\begin{itemize}
    \item The employment of quantum sensors for accurate measurement of photosynthetically active radiation (PAR) \cite{coffin-2021-PAR}, \cite{AKITSU-2017-quantum}, leaf water potential (LWP) \cite{LACERDA-2022-quantum-water}, and soil ions \& pH levels.
\end{itemize}


\section {Blockchain Technology for Smart \& Sustainable Agriculture Sector}
Blockchain technology is a digital system that enables electronic documentation, validation and verification thereby eliminating the need for intermediaries. It ensures data and information visibility for all participants, with all records being immutable that are impervious to tampering or erasure. The latest trends and improvements in blockchain technology offer some exciting features that are well suitable for agriculture. This section provides an overview of the blockchain technology, explores its applications \& challenges within the agricultural sector, and highlights potential future research directions and opportunities, with a summary presented in Table XI.

\subsection{Overview of Blockchain Technology}
Blockchain technology operates as a decentralized and immutable ledger that facilitates in documenting transactions and tracking assets. Assets encompass both tangible items including vehicles, land, and currency, as well as intangible items like brand identity, copyrights, and patents. Thus anything of value can be exchanged and monitored on a blockchain network, effectively minimizing risks and reducing costs for all the participants.

At its core, blockchain is revolutionizing information management by furnishing a secure, decentralized system for record keeping. It logs transactions across multiple independent nodes through cryptographically linked blocks, thus eliminating the need for centralized control and trust establishment. According to IBM key elements of a blockchain are as under:

\begin{itemize}
    \item \textbf{Distributed ledger technology}: Every participant in the network is granted access to the distributed ledger along with its immutable transaction logs. This system ensures that transactions are logged once, thereby avoiding unnecessary duplication. 
    
\end{itemize}

\begin{itemize}
    \item \textbf{Immutable records}: Once a transaction is logged in the shared ledger, it becomes immutable, preventing any participant from modifying or tampering with it. To correct any transaction errors, it is imperative to add a new transaction to reverse its effect. Both transactions are visible to ensure integrity.
    
\end{itemize}

\begin{itemize}
\item \textbf{Smart contracts}: These are digital contracts or set of rules stored on a blockchain. They execute automatically when some predetermined conditions are fulfilled. Smart contracts facilitate the automation of agreements, workflows and transactions, thereby eliminating the need for manual intervention in the execution process.

\end{itemize}

Blockchain's contributions to high security, enhanced trust and improved efficiency enable its application beyond mere financial transactions. It extends into areas such as supply chain management, voting systems, and data protection, thus reshaping our digital interactions with a new level of transparency and immutability \cite{Pahontu-2020-blockchain-industries}.  Recent studies further highlight the role of distributed architectures in blockchain systems for mitigating single-point failures and improving scalability, efficiency, and security—key features for agricultural applications such as fog radio access networks \cite{wang2022blockchain-reviwer} and other scalable frameworks \cite{cao2022blockchain-reviwer}.

\subsection{Blockchain Technology in Agriculture}
The recent advancements in blockchain technology are demonstrating significant benefits in its adoption in various areas. Especially, this technology's deployment in agriculture has resulted in numerous positive and evolutionary advancements. According to research \cite{L.B.-Blockchain-agriculture12091333}, blockchain can be utilized in agriculture focusing on areas such as integration with IoT and sensors \cite{uddin-Blockchain-IoT-2021survey}, supply chain management \cite{D.S-2021-Blockchain-supply}, agricultural finance \cite{rocha-2021-blockchain-agribusiness}, production management \cite{FERRANDEZPASTOR-Blockchain-tracibility-2022-100381}, and livestock management\cite{alshehri-2023-blockchain-livestock}, among others. This integration is enhancing efficiency \cite{alsharari-2021-Blockchain-eff-integrating}, transparency \cite{iqbal-2020-Blockchain-transp-safe}, and traceability \cite{FERRANDEZPASTOR-Blockchain-tracibility-2022-100381} in agricultural processes, signifying a major advancement in the sector's technological evolution. For a thorough insight, comprehensive surveys exploring blockchain's role in agriculture can be found in \cite{L.B.-Blockchain-agriculture12091333}, \cite{LIU-2021-Blockchain-review-126763}, \cite{DEY-2021-Blockchain-review128254}, and \cite{Ferrag-2020-blockchain-review}. Fig. 12 highlights some major applications of blockchain technology in agricultural sector which are outlined below.

\begin{enumerate}

\item \textit{Blockchain \& IoT Fusion for Smart Agriculture}:
IoT based systems operating on inefficient centralized models, encounters difficulties in merging data from disparate sources across multiple platforms \cite{Pahl-2018-IoT-Blockchain}. Additionally, data access issues are exacerbated by the structure of data ownership. By implementing the distributed architecture via blockchain, security is improved and a consolidated data repository is created for all stakeholders in the agricultural supply chain. This approach mitigates the risk of single point of failure and enhances the management and governance of farms.

\begin{figure}
  \begin{center}
  \includegraphics[width=3in, height=3.2in]{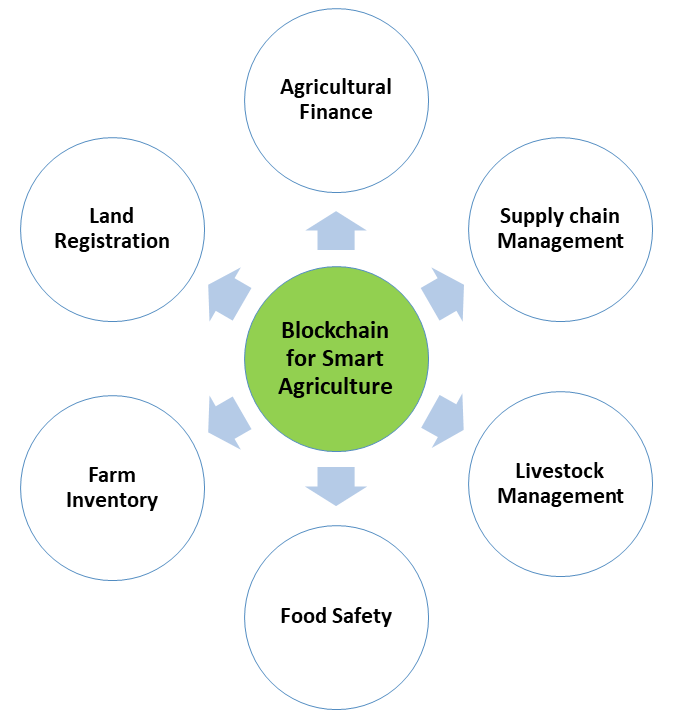}\\
 \caption{Blockchain applications in smart agriculture}\label{BC}
  \end{center}
\end{figure}

Centralized storage, typical of IoT-based systems, makes data susceptible to loss and manipulation. A distributed database can mitigate this issue by facilitating a decentralized storage.
Blockchain, in this context, provides a secure, scalable and robust framework for IoT and ICTs in the realm of smart farming. Several smart farming models based on blockchain \& IoT have been successfully implemented. In \cite{BORDEL-2019-Blockchain} authors introduced an Ethereum blockchain network designed for  automating water control in irrigation systems. A generalized blockchain based security architecture for smart sensing in IoT driven agriculture is presented in \cite{Vangala-2021-Blockchain}. \cite{UrRahman-2020-Blockchain} explores smart contracts for enabling scalable data sharing in the context of IoT for smart agriculture. Comprehensive literature reviews on blockchain \& IoT are provided in \cite{torky2020integrating}, \cite{ferrag2020security}, \cite{dey2021blockchain}, \cite{alkhateeb2022hybrid}, \cite{mathur2023survey}, underscoring the importance of these technologies in advancing precision agriculture.

\item \textit{Application of Blockchain for Livestock Management}:
Blockchain has the potential to improve the precision and effectiveness of livestock management. The complete process can be digitized and rendered transparent with the added benefits of security and integrity of stock data. This approach will streamline the entire stock management process through the automation of recording and tracking of transactions.

In \cite{Shoufeng-2021-Livestock-blockchain}, the authors proposed a blockchain based reconciliation mechanism aimed at boosting consumer trust in the traceability of the beef supply chain. Further, \cite{Kampan-2022-livestock} investigates the impact of blockchain on improving the traceability and integrity of live-stock based products. A novel cloud based livestock monitoring system that utilizes RFID and blockchain technology is detailed in \cite{Yang-2020-livestock}. Additionally, potential blockchain applications in the animal production and health sector are examined in \cite{Makkar-2020-blockchain-animal}.

\item \textit{Advancing Agricultural Finance through Blockchain}:
Blockchain is transforming the finance industry, holding great promise to reduce fraudulent activities and enable secure \& fast transactions. It is poised to streamline financial transactions, thereby reducing costs, enhancing security and improving efficiency. Like its impact on other sectors it has the potential to revolutionize agriculture by enabling autonomous financial transactions, reconciliation mechanisms and auditing processes with increased transparency. Consequently, it offers a secure, robust and resilient payment and transaction system in agriculture.

In \cite{Sadayapillai-2022-blockchain-finance}, authors investigate blockchain smart contracts for enhancing the income of farmers within the agricultural supply chain. \cite{Hiren-2023-Finance} introduces \enquote{AgriOnBlock} a blockchain based approach to tackle challenges like financial losses and agricultural produce degradation. \cite{Kalaiselvi-2021-loan-blockchain} explores blockchain technology for streamlining loan distribution process in agricultural sector. Whereas, an IoT and blockchain based novel framework for food supply chain finance has been presented in \cite{Hong-2021-finance-blockchain}.

\begin{table*}
 \captionsetup{justification=centering}
 \captionsetup{labelsep=newline}
 \renewcommand{\arraystretch}{1.7}
 \setlength{\tabcolsep}{4pt}
\caption{Applications, Challenges and Research Directions of Blockchain in Smart Agriculture}
\label{tableBlockchainContinued} 
\centering
\resizebox{\textwidth}{!}{
\begin{tabular}{ p{4.5cm} p{4.5cm} p{4.5cm} p{4.5cm} }
 \toprule
\textbf{Technology Domain/ Area} & \textbf{Suitability for IoAT} & \textbf{Limitations/ Challenges} & \textbf{Potential Research Directions} \\
\midrule
Blockchain Technology & & & \\
\midrule
Transparency \& Security
& \mydash Supply chain management.
\newline \mydash Secure data sharing.
\newline \mydash Agriculture finance.
& \mydash Interoperability across blockchain networks.
\newline \mydash High energy consumption by blockchains.
\newline \mydash Lack of regulatory policies.
& \mydash Addressing scalability issues in large scale deployments.
\newline \mydash Integrating energy harvesting to reduce costs.
\newline \mydash Tokenizing agricultural assets for new opportunities.
\newline \mydash Develop quantum blockchain hybrid models.
\newline \mydash Agricultural insurance to facilitate compensations for losses.
\\
\bottomrule
\end{tabular}}
\end{table*}

\item \textit{Improving Supply Chain Management using Blockchain}:

Blockchain is set to revolutionize agricultural supply chain by establishing a secure, transparent and efficient system to trace provenance of agricultural products from farm to consumers, thereby guaranteeing food safety. In the context of agriculture, blockchain facilitates traceability by assigning a distinct digital identifier to each product, that is updated with a new digital signature and tamper-proof timestamp at every point of transfer. This creates an immutable record of transactions that can be used for precise location tracking and verification of its authenticity. Consequently, agribusinesses can ensure the delivery of premium-quality products and services to their customers, with the additional benefit of mitigating fraud and counterfeiting risks.

In \cite{Katsikouli-2020-supplychain}, the authors examines the advantages and associated challenges of using blockchain in food supply chain management. \cite{Niu-2022-supplychain} delves into the application of blockchain across transnational agricultural supply chains. The research presented in \cite{HU-2021-supplycahin} proposes a blockchain and edge computing-based trust framework for the organic agricultural supply chain (OASC), aiming to overcome issues related to trust and transparency. \cite{KAMBLE-2020-Supplychain} investigates the adoption of blockchain in agriculture supply chains, by focusing on key enablers like traceability and auditability. Meanwhile, \cite{Mukherjee-2021-supply}  assesses the potential of blockchain technology to promote sustainable practices within agricultural supply chains.

\item \textit{Leveraging Blockchain for enhancing Agricultural production, Management and Governance}:
The integration of blockchain technology into the agricultural sector is in its early stage, yet significant applications have already been exhibited to improve aspects such as food quality, production, management, and governance. It enables a secure and robust system for precise tracking of agricultural goods, thereby ensuring food quality and minimizing fraud risks. The decentralization feature of blockchain promotes an effective and transparent approach to food production. Furthermore, blockchain technology contributes to the enhancement of food system governance by developing an accountable mechanism for overseeing food subsidies and various other food-centric governmental initiatives.

In \cite{Shih-2019-blockchain}, authors propose a blockchain based traceability system to improve agricultural productivity, increase sales and mitigate food wastage. \cite{harshavardhan-2019-blockchain} delves into how blockchain fosters sustainable agricultural practices and boosts economic efficiency. The study \cite{Makkar-2020-blockchain-animal} explores blockchain's utility to enhance the tracking of animals and their by-products. Research presented in \cite{G-2021-blockchain} investigates how blockchain can secure commercial transactions in agriculture, thus safeguarding farmer and business operations. \cite{Anand-2021-blockchIN} details a decentralized marketplace powered by blockchain, allowing farmers to transact directly with consumers. Furthermore, the capability of blockchain for better governance of agricultural products in a big-data context is examined in \cite{guo-2022supply}, showcasing the technology's extensive relevance to the agricultural sector.

\item \textit{Challenges Associated with Blockchain in Agriculture}:
Potential benefits of blockchain technology in the agricultural domain have been discussed above in detail. Nonetheless, there are specific challenges associated with this technology that require attention, as outlined below.

\begin{itemize}
    \item Accurate initial data entry is crucial for maintaining the integrity and reliability of the system throughout its life cycle.
\end{itemize}

\begin{itemize}
    \item A large number of network nodes or users are required to make blockchain robust.
\end{itemize}

\begin{itemize}
    \item The impact of blockchain implementation encompass significant energy consumption costs and considerable financial expenditures.
\end{itemize}

\begin{itemize}
    \item Additional processes and compliance requirements for data management escalate operational overhead.
\end{itemize}

\begin{itemize}
    \item Issues related to interoperability and smooth data transfer across blockchain and legacy systems needs to be resolved.
\end{itemize}

\begin{itemize}
    \item Scalability issues with large scale deployment.
\end{itemize}

\begin{itemize}
    \item The agricultural supply chain is affected by unpredictable natural events such as droughts, floods, and pest infestations, complicating decision-making based on specific attributes and consequently leading to reluctance in the adoption of blockchain technology.
\end{itemize}

\begin{itemize}
    \item Considerable transformations in associated applications and systems, such as the digitization of land records, are imperative.
\end{itemize}

\begin{itemize}
    \item While blockchain mitigates the risk of fraud, threats posed by malicious entities such as hackers, intruders, and viruses require in-depth investigation.
\end{itemize}

\end{enumerate}

\subsection{Future Directions \& Research Opportunities}
The transformative impact of blockchain technology in agriculture is undeniable, opening up potential future research directions and opportunities, as discussed below.

\begin{itemize}
    \item Employing energy harvesting techniques to reduce the high energy consumption costs associated with blockchain technology and minimize environmental impacts.
\end{itemize}

\begin{itemize}
    \item Development of scalable blockchain frameworks or hybrid models designed to handle high volume of transactions in the agricultural supply chain.
\end{itemize}

\begin{itemize}
    \item Research is needed on devising intelligent mechanisms, such as integration of IoT and AI, to validate the accuracy and authenticity of agricultural data prior to its blockchain recording.
\end{itemize}

\begin{itemize}
    \item Tokenization of agricultural assets to enhance liquidity and offer new financing opportunities for farmers and agricultural enterprises.
\end{itemize}

\begin{itemize}
    \item Exploring quantum technology to develop quantum-blockchain hybrid models, focusing on the development of enhanced consensus protocols and the use of quantum keys for secure data encryption.
\end{itemize}

\begin{itemize}
    \item Harnessing the decentralized systems and smart contracts for agricultural insurance to facilitate compensations for losses due to natural disasters, thus eliminating reliance on third-party data evaluation \cite{Polymeni-2023-6G-Blockchain}.
\end{itemize}

\section {Machine learning in Precision Agriculture}

The incorporation of IoT into agriculture has significantly increased this sector's efficiency and productivity through the goal of full process automation.  There is a growing emphasis among researchers on the use of intelligent computational methods to further augment agricultural practices. Numerous ML algorithms have been developed and applied for purposes like monitoring, controlling, predicting, and smart decision making.  ML applications span across various sectors including agriculture, education, healthcare, military, finance, transportation, marketing, manufacturing, and cybersecurity. In this section, we will explore the basics of ML, investigate its utilization in precision agriculture, and identify prospective directions for future research. In addition, a summary of applications, challenges, and future research directions is also presented in Table XII.

\begin{figure*}
  \begin{center}
  \includegraphics[width=6.5in]{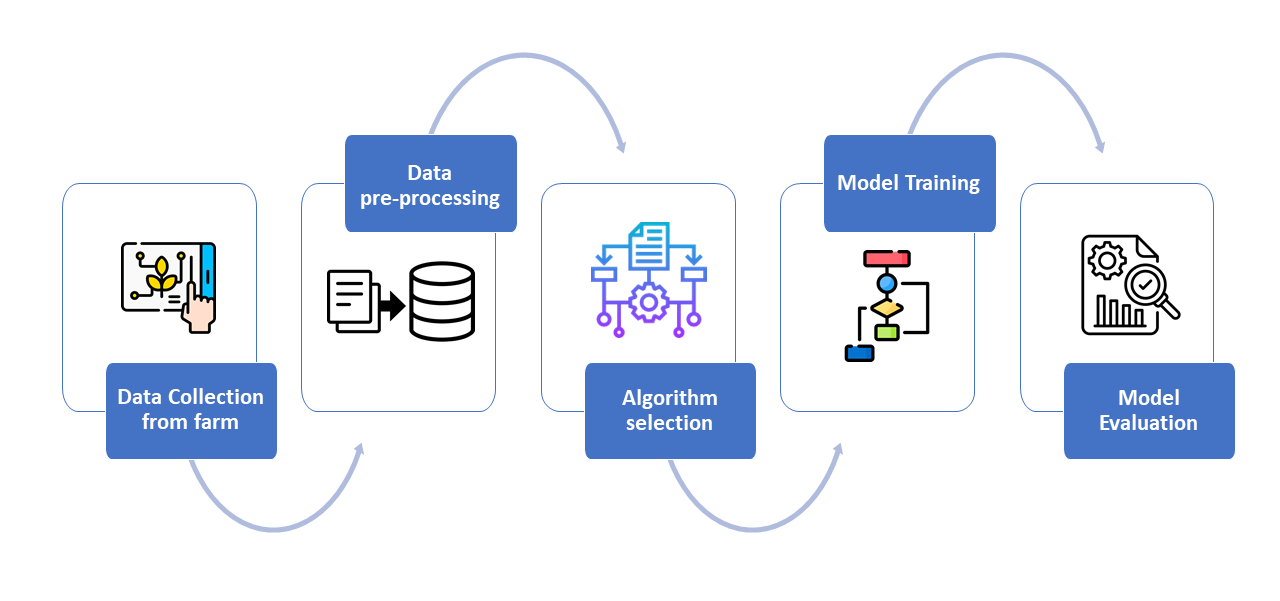}\\
 \caption{Machine Learning (ML) process in agriculture: From data collection to model evaluation}\label{Figure ML - process}
  \end{center}
\end{figure*}

\subsection{Overview of Machine Learning}
As an important subset of AI, ML is reshaping industries by digitizing and analyzing extensive data sets. This process involves the use of advanced algorithms to extract meaningful insights from diverse data, where ML and data mining practices are often carried out using Python. This methodology facilitates the efficient gathering and analysis of real-time data across various systems, thereby enabling the extraction of actionable information.

\begin{enumerate}

\item \textit{Learning System of a ML algorithm}:
The evolution of ML is attributed to advancements in computational learning theory and pattern recognition, employing specialized algorithms and models to analyze data and make informed predictions \cite{Angra-2017-ML}. Utilizing these models, data scientists, researchers, engineers, and analysts produce valid and reliable results for informed decision-making. The stages of the ML process in agriculture, which include data collection, data pre-processing, algorithm selection, model training and model evaluation, are represented in Fig. 13. ML algorithms excel at identifying patterns and relationships within historical data sets, progressively improving their efficacy through the assimilation of extensive datasets.

As outlined in \cite{IBM_ML}, the learning mechanism of a ML algorithm can be categorized into three main elements:
\begin{itemize}
    \item \textbf{A decision process}: ML algorithms are typically used for either prediction or classification tasks. These algorithms derive estimates about data patterns based on input data, which can be either labeled or unlabeled.
    
\end{itemize}

\begin{itemize}
    \item \textbf{An error function}: The model's predictions are being evaluated by this function. The error function can make comparisons to well known examples to validate the accuracy of the model.
    
\end{itemize}

\begin{itemize}
    \item \textbf{An updating or optimization process}: The model adjusts its weights to align more closely with the training data, thus reducing the discrepancy between actual examples and model predictions. This iterative process of evaluation and optimization continues autonomously until the model reaches a predefined accuracy level.
    
\end{itemize}

\item \textit{Types of ML models}:
ML models generally fall under three main types based on how they learn and predict outcomes \cite{ayodele-2010-machine}. The categories are summarized in Fig. 14 and are briefly outlined as follows.

\begin{figure*}
  \begin{center}
  \includegraphics[width=6.5in]{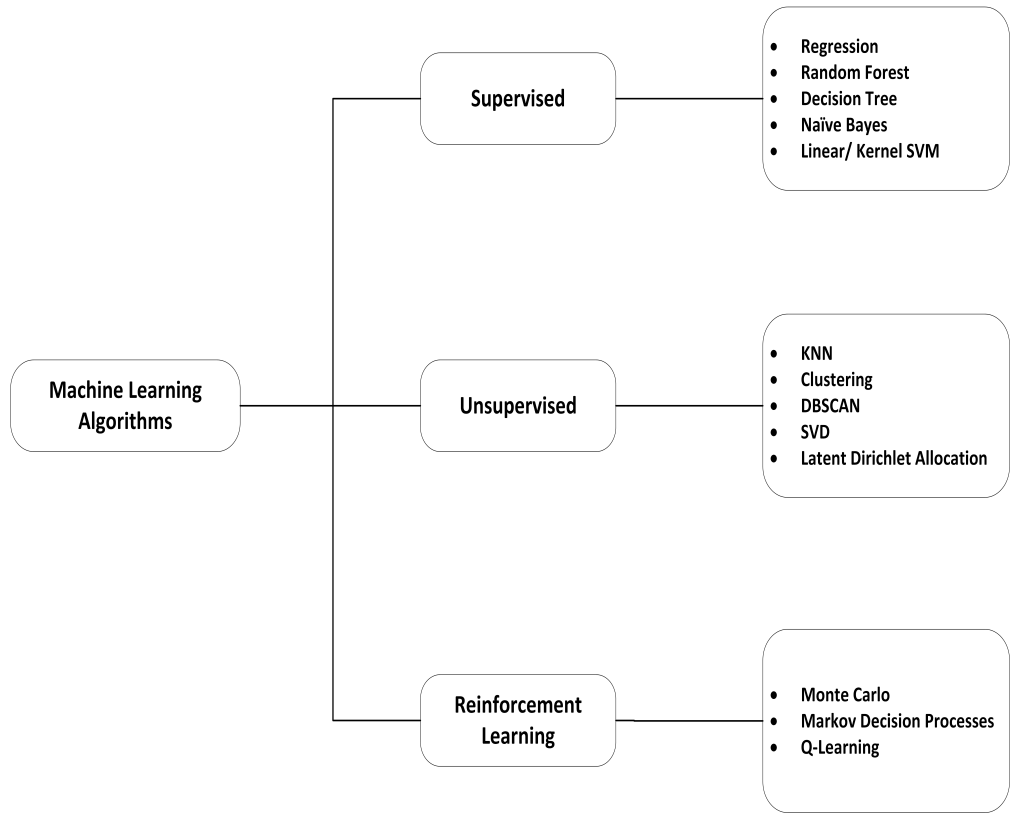}\\
 \caption{Categorization of Machine Learning (ML) algorithms: Supervised, Unsupervised, and Reinforcement Learning}\label{Figure ML -Flow}
  \end{center}
\end{figure*}

\begin{itemize}
    \item \textbf{Supervised learning}: This approach employs labeled data sets to train algorithms for data classification or prediction of outcomes accurately. Upon data injection into the model, it adjusts its weights to achieve optimal fit. Supervised learning model include algorithms such as linear regression, neural networks, naive bayes, random forest, logistic regression, and support vector machine (SVM). 
    
\end{itemize}

\begin{itemize}
    \item \textbf{Unsupervised learning}:Unlike supervised learning, in this technique the model is trained without any labeled data sets. The objective is to discover patterns and relationships in the input data without the need of human intervention. Examples of unsupervised learning algorithms include principal component analysis (PCA), singular value decomposition (SVD), k-means clustering, neural networks and probabilistic clustering methods.
    
\end{itemize}

\begin{itemize}
    \item \textbf{Reinforcement learning}: This model utilizes a trial and error learning strategy to force the system to encourage autonomous problem-solving. Its aim is to take the best actions in a given scenario or environment to maximize rewards. Algorithms used in this method include Markov decision process (MDP), Q-learning, Deep Q Network (DQN), and policy gradient methods.
    
\end{itemize}

\item \textit{Challenges Associated with ML}:
Machine learning has introduced transformative advancements across all facets of human life. However, some major challenges in this domain are discussed below.

\begin{itemize}
    \item \textbf{Underfitting of training data}: When the model fails to establish a precise correlation between input and output patterns.
\end{itemize}

\begin{itemize}
    \item \textbf{Overfitting of training data}: When the model is trained on excessive training data, it leads to subpar performance.
\end{itemize}

\begin{itemize}
    \item \textbf{Scalability}: The challenge of processing large data sets and complex models, which requires substantial computational resources like extensive memory, high processing power, and ample storage capacity.
    
\end{itemize}

\begin{itemize}
    \item \textbf{Model selection}: The need to select the optimal model from a diverse pool, as there is no one-size-fits-all solution. 
    
\end{itemize}

\begin{itemize}
    \item \textbf{Hyperparameter tuning}: The complex and time-consuming process of adjusting the model parameters to optimize performance.
    
\end{itemize}

\begin{itemize}
    \item \textbf{Data availability}: The scarcity of sufficient and relevant data presents another bottleneck in ML projects.
    
\end{itemize}

\end{enumerate}

\subsection{Machine Learning Applications in Precision Agriculture}
Recent advancements in ML techniques have found significant applications in the agricultural sector, with the aim of reducing crop costs and maximizing yield. ML enabled precision agriculture addresses unpredictable weather conditions by eliminating randomness and optimizing every phase of the farming process. Some state-of-the-art ML techniques proposed or implemented to boost efficiency and productivity in this field are discussed below.   
\begin{enumerate}
\item \textit{Disease Detection}: Agricultural diseases can be effectively detected and managed using ML methods. They can be used to guide targeted application of pesticides to protect crops from infections, thus reducing labor requirements. This approach facilitates producers by providing important statistics and enabling strategic planning for the use of fertilizers, pesticides and irrigation accordingly.

Research efforts includes investigating DL algorithms for image based disease detection in plants \cite{mohanty-2016-ML-disease}, and the development of mobile disease diagnostics of plants using ML techniques \cite{hughes-2015-ML-plant}.  Additionally, \cite{amara-2017-deep} delves into employing DL to automatically classify banana leaf diseases, whereas \cite{waheed-2020-ML-plant} explores dense CNN model for detecting diseases in corn leaves. These studies underscore the critical role of ML in enhancing disease detection capabilities in agriculture.

\item \textit{Soil Characteristics \& Weather Prediction}: Soil properties and geographic conditions of climate play critical role in determining crop \& seed selection, land preparation and the choice of fertilizer. Predictions regarding soil characteristics encompass nutrients, surface humidity, porosity, aggregate stability, and temperature. ML can be effectively utilized in prediction of soil properties and weather conditions, thus enhancing agricultural practices.

 Authors in \cite{wang-2018-modeling-ML-soil}, proposed a ML approach to measure nutrient solution components in soilless agriculture. The study \cite{park2015-soil} demonstrates the capability of ML to enhance the spatial resolution of soil moisture data. Further, \cite{morellos2016machine-soil} explores ML based prediction of the nitrogen, organic carbon and moisture content of soil utilizing VIS-NIR spectroscopy. A ML-derived pedotransfer function is used to investigate the soil electrical conductivity and organic carbon content in \cite{benke2020development-soil}. Additionally, the impact of ML in seasonal forecasting of daily mean air temperatures is presented in \cite{shin2020seasonal-ml}, while the significance of ML in accurately predicting rainfall across diverse climates has been demonstrated in \cite{cramer2017extensive-ml-climate}. These studies underscores the broad utility of ML in enhancing agricultural and environmental assessments.

\item \textit{Crop yield Prediction}:
Predicting crop yields and subsequently enhancing these yields is a crucial aspect of agriculture. Several factors such as soil type, pH value, nutrients, fertilizers, rainfall, temperature, sunlight and harvesting schedules, significantly influence the accuracy of crop yield predictions. Yield prediction necessitates the utilization of historical agricultural data alongside contemporary technologies like ML, with the goal to accurately project crop outputs and optimize yields.

In \cite{kuwata2015estimating-yield}, authors use DL, a subset of ML, to estimate crop yields. \cite{nevavuori2019crop-yield} utilizes deep CNN to predict crop yields. The study in \cite{khaki2019crop-yield} conducts crop yield prediction using deep neural networks. \cite{cai2019integrating-yield} explores ML approaches to predict wheat yield incorporating satellite and climate data. DL approach is used in \cite{koirala2019deep-yield} for fruit detection and yield forecasting. Furthermore, \cite{maimaitijiang2020soybean-yield} discusses improvement of soybean yield predictions using drones equipped with  sensors and deep neural network analysis.


\begin{table*}
 \captionsetup{justification=centering}
 \captionsetup{labelsep=newline}
 \renewcommand{\arraystretch}{1.7}
 \setlength{\tabcolsep}{4pt}
 \caption{Applications, Challenges and Research Directions of Machine Learning in Smart Agriculture}\label{table1}
\centering
\resizebox{\textwidth}{!}{
\begin{tabular}{ p{4.5cm} p{4.5cm} p{4.5cm} p{4.5cm}}
 \toprule
\textbf{Technology Domain/ Area} & \textbf{Suitability for IoAT} & \textbf{Limitations/ Challenges} & \textbf{Potential Research Directions}   \\

\midrule
$\textbf{Machine learning (ML) }$\\
\midrule
Analytics \& Automation

& \mydash Farming equipment's predictive maintenance.
\newline \mydash Automation of agriculture processes like irrigation, spraying etc.
\newline \mydash Enhanced crop management.

& \mydash High computational power.
\newline \mydash Inaccurate predictions from flawed data.

&  \mydash Development of scalable \& efficient algorithms for IoT \& drones.
\newline \mydash Predictive analytics for climate resilience.
\newline \mydash Metaheuristic algorithms for the precise localization of nodes.
\newline \mydash Integrating ML with genomic selection techniques to develop robust crops.
\\ 
\bottomrule
\end{tabular}}
\end{table*}

\item \textit{Weed Management}:
Weed detection is vital for crop management due to its competition with crops for space, light, soil nutrients and water, significantly reducing  yield. Consequently, the timely identification and appropriate management of weeds are required for maintaining optimal crop health and enhancing productivity. Leveraging ML techniques, it is possible to accurately detect various weed types and their growth stages, facilitating targeted actions. Proper weed management ensures that crops can utilize pesticides and fertilizers more effectively, thus improving agricultural outcomes.

In study \cite{espejo2020towards-weed-ML}, a novel weed identification system is introduced, combining fine-tuned convolutional networks with traditional classifiers. \cite{chechlinski2019system-weed-ML} presents a novel agro-robotics system based on CNN for automated weeding. A comprehensive review of weed detection using various ML techniques has been discussed in \cite{liu2020weed-ML}. Furthermore, \cite{wang2019review-weed-ML} provides a detailed review of the progress in weed detection facilitated by machine vision and image processing techniques.

\item \textit{Water Management}:
 Water management is a key aspect of agricultural practices, substantially affecting the results. Optimization of water usage can be achieved through the application ML techniques. ML enables the implementation of optimal water scheduling systems - be it daily, weekly, or monthly - tailored to the specific requirements of the crop and soil parameters. Consequently, the integration of ML with water management can improve agricultural productivity while simultaneously conserving water resources.

A vast body of literature exists where ML methods are applied to historical data, offering real-time forecasting and decision making capabilities to farmers in the context of smart irrigation \cite{janani2019study-water}, \cite{sharma2020machine-water}, \cite{vij2020iot-water}, \cite{kondaveti2019smart-water}. This data is gathered through sensors and IoT devices.

\item \textit{Livestock Production \& Management}:
Similar to other technologies such as IoT \& Blockchain, ML has been extensively researched for its application in the improvement of livestock production and management. This research can be classified into two broad categories: animal welfare and livestock production. In animal welfare, ML techniques are applied for health monitoring of animals, facilitating early disease detection through the  analysis of animal behavior, vital signs and environmental conditions. While, in livestock production ML aids in achieving a balanced production, enabling farmers and producers to realize financial benefits.

Researchers in \cite{alonso2020intelligent-livestock}, \cite{berckmans2017general-livestock}, \cite{halachmi2016precision-livestock}, \cite{garcia2020systematic-livestock}, \cite{morota2018big-livestock} focus on the use of ML in precision livestock farming . Whereas,  ML applications in addressing animal diseases can found in studies \cite{zhang2021application-disease}, \cite{machado2015variables-disease}, \cite{hyde2020automated-disease}.

\item \textit{Smart Harvesting}:
Smart harvesting allows farmers to reduce human intervention in the harvesting process through the use sensors, IoT devices, ML, UAVs and robots. Robots equipped with ML and AI capabilities have been developed for the harvesting of vegetables and fruits. This technological advancement offers several benefits to farmers including reduction in labor requirements, better crop yield, decreased costs and improved insights into the harvesting process.

Authors in \cite{horng2019smart-harvesting}, introduce a system that IoT with smart image recognition, employing neural networks to enhance the efficiency of harvesting operations. Further \cite{zhang2020technology-harvest}, reviews advances in mechanical harvesting technologies using robotic harvesters integrated with ML. A novel  grading system for mangoes using ML methods to achieve higher accuracy in harvesting, is proposed in \cite{pise2018grading-harvest}. \cite{maponya2020pre-harvest} evaluates the efficacy of ML classifiers, SVM and RF, for accurate pre-harvest crop classification. Lastly,  \cite{purandare2016analysis-harvest} presents an integrated system of IoT and ML to minimize post harvest losses by recommending optimal harvesting times, disease management and storage conditions.

\end{enumerate}

\subsection{Future Directions \& Research Opportunities}
In precision agriculture, the application of AI and ML is increasingly prominent. ML is improving agricultural practices, providing cutting-edge solutions for more efficient and effective farming operations. Some promising research directions are discussed as under.

\begin{itemize}
    \item Investigating metaheuristic algorithms for the precise localization of nodes, thus optimizing sensor deployment in the agricultural fields while reducing costs. 
    
\end{itemize}

\begin{itemize}
    \item Exploring the application of autonomous swarm drones that utilize ML methods, to monitor crops and livestock. Additionally, the drones can execute operations like spraying of fertilizers and pesticides in large agricultural farms. 
    
\end{itemize}

\begin{itemize}
    \item Designing an evaluation framework to benchmark classifier performance in the diverse agricultural environment.
    
\end{itemize}

\begin{itemize}
    \item Addressing the communication gap between domain experts is crucial to facilitate the adoption of best practices from ML within agricultural sector.
    
\end{itemize}

\begin{itemize}
    \item Integrating ML with renewable energy sources deployed at the farms, to ensure uninterrupted power supply under unpredictable conditions.
    
\end{itemize}

\begin{itemize}
    \item Leveraging ML to develop optimal permaculture designs for sustainable agriculture by analysing critical factors like climate, geographic conditions, soil parameters and water resources.
    
\end{itemize}

\begin{itemize}
    \item Integrating ML with genomic selection techniques to develop  climate-resilient crops. 
    
\end{itemize}


\section{Discussion}
This section examines key factors for successful IoE adoption in agriculture, offering insights to support the sustainable integration of IoE technologies across diverse agricultural contexts.

\subsection{Importance of User-Centric Design for IoE Adoption}
The effective integration IoE in agriculture fundamentally relies on a user-centered design approach, particularly to facilitate non-technical users, such as farmers. A CISCO survey indicates that 60\% of IoT initiatives encounter adoption challenges when usability and training requirements are overlooked. Addressing these challenges requires attention to the following key aspects as outlined in Table XIII:

\begin{enumerate}

    \item \textit{Understanding End-Users Needs}: Direct engagement with end-users / farmers is essential to comprehend their daily challenges, workflows, and specific requirements \cite{wong2018designing}. This understanding is crucial for developing solutions that are both relevant and beneficial to their needs. 
     
    \item \textit{Ease of Use and Accessibility}: The broad scope of IoE in agriculture, spanning biological, molecular, and ML-driven processes, often produces complex data that may be difficult for farmers to interpret. Simplified dashboards, mobile alerts, and visual indicators (e.g., color codes) will help farmers to make informed decisions. Integrating advanced IoE data (e.g., from molecular sensors and designer phages) into familiar IoT platforms will enhance accessibility and reduces complexity.

\begin{figure}
  \begin{center}
  \includegraphics[width=3.5in]{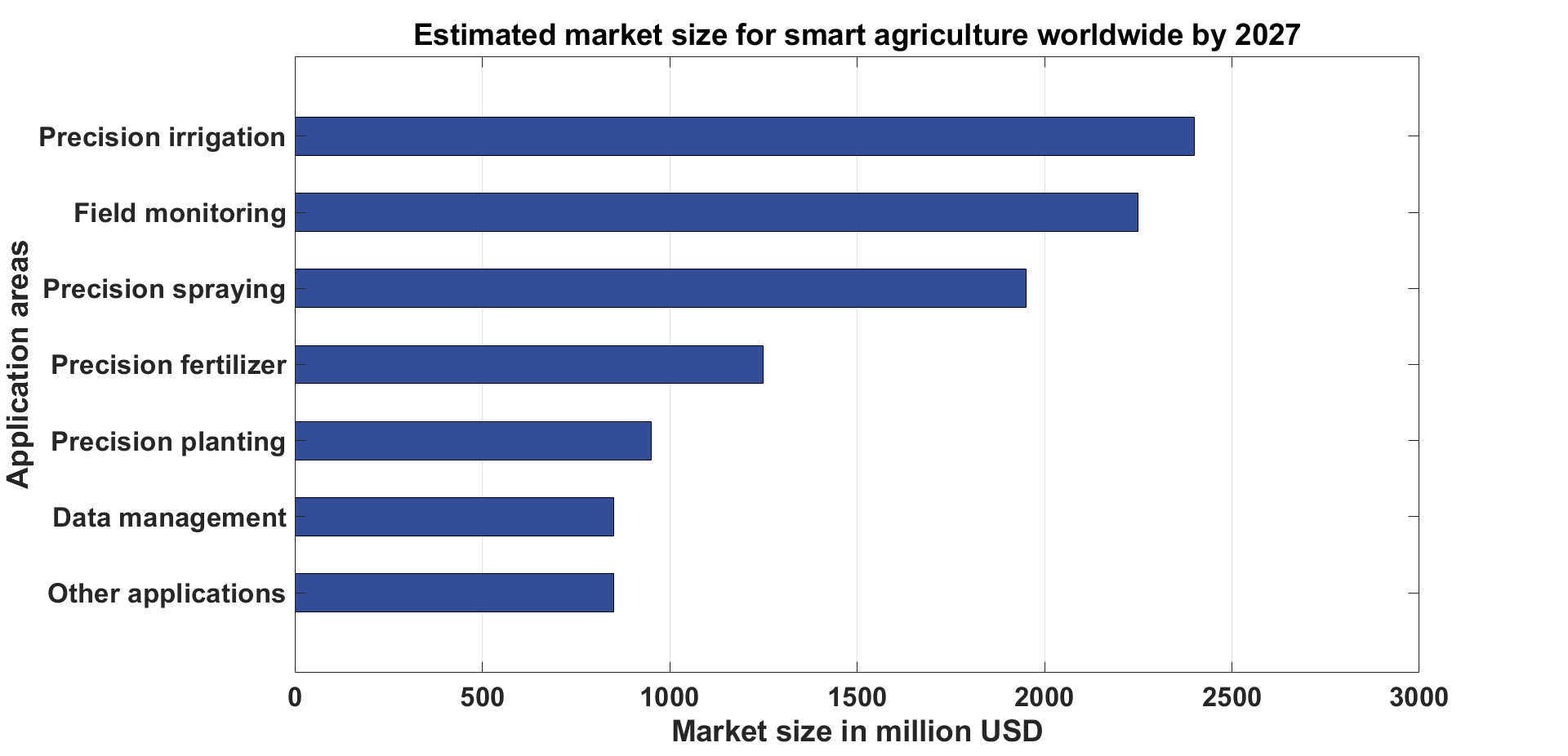}\\
 \caption{Projected market size for smart agriculture by 2027\cite{thilakarathne2023towards}}\label{BC}
  \end{center}
\end{figure}

    \begin{figure*}[h]
  \centering
    \begin{subfigure}{0.48\textwidth}
      \centering
      \includegraphics[width=\linewidth]{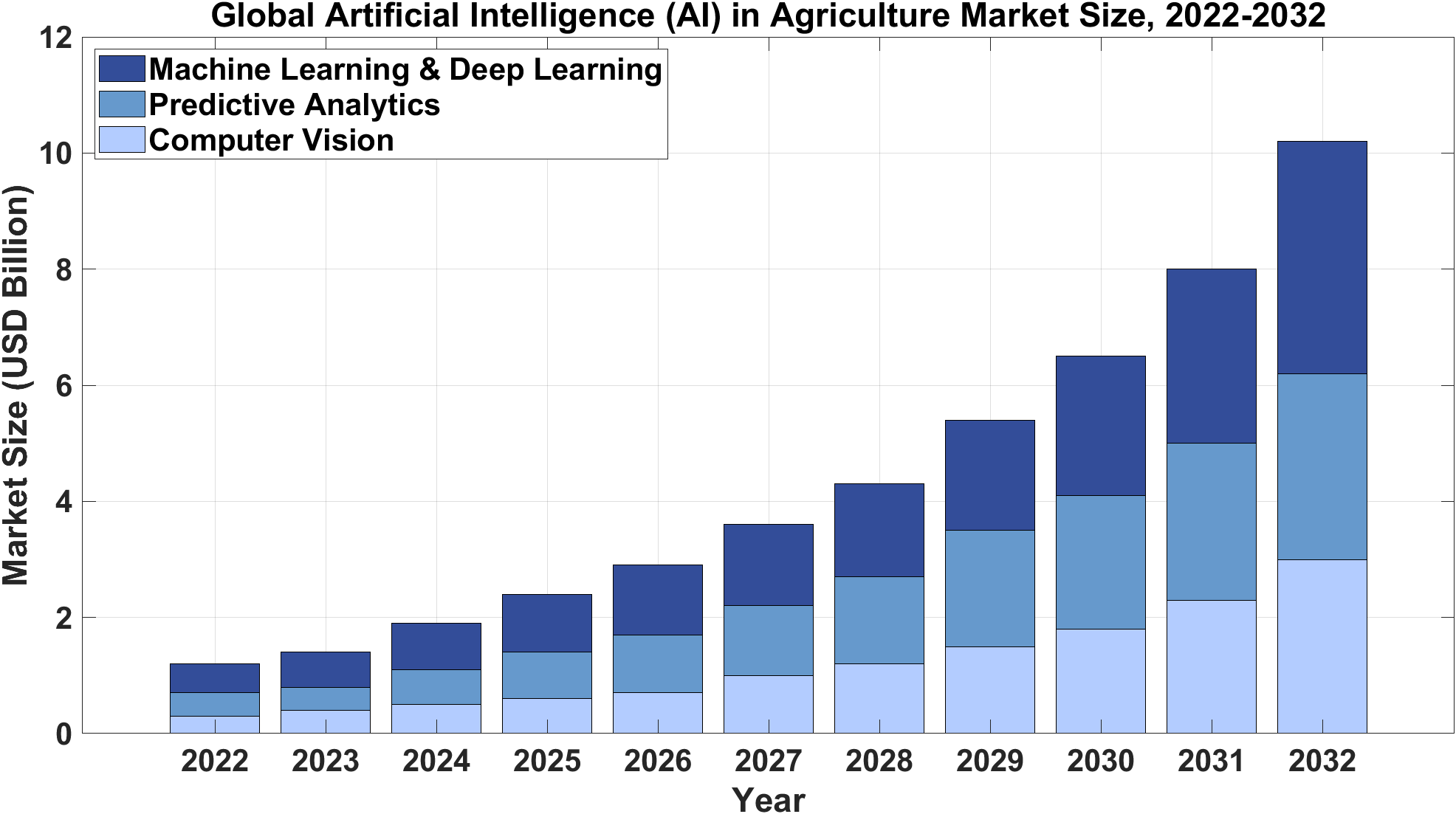}
      \caption{Projected market size for AI in smart agriculture}
      \label{fig:AI_market_size}
    \end{subfigure}
    \hfill
    \begin{subfigure}{0.48\textwidth}
      \centering
      \includegraphics[width=\linewidth]{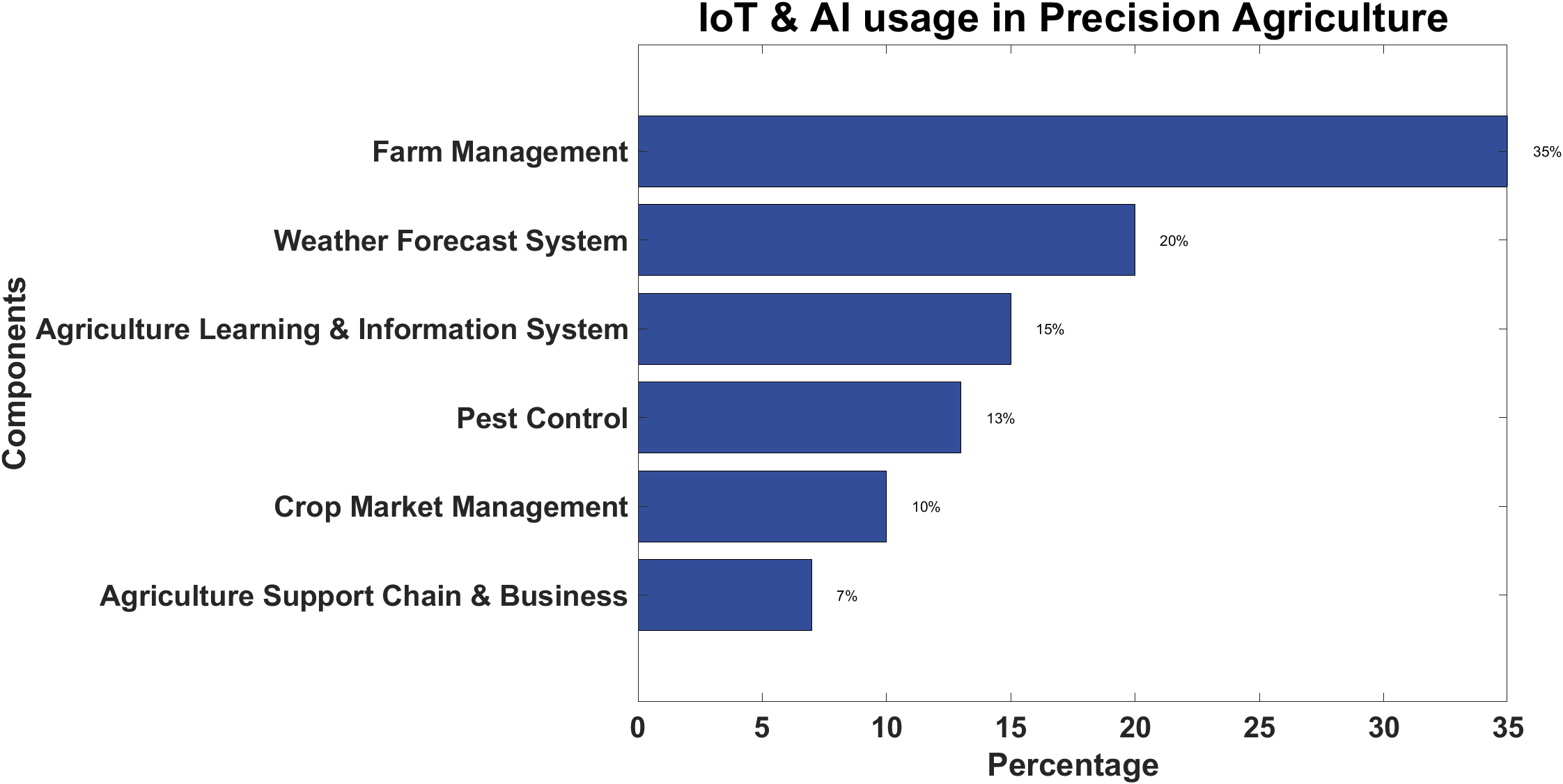}
      \caption{Percent usage of AI \& IoT in smart agriculture}
      \label{fig:AI_IoT_usage}
    \end{subfigure}
    \caption{AI Market Projections and Usage in Smart Agriculture}
    \label{fig:AI_IoT_comparison}
\end{figure*}

    \item \textit{Affordability}: Making IoE solutions more affordable for small and medium-scale farmers is essential for promoting widespread adoption. Governments can support affordability through subsidies, and public-private partnerships can provide accessible financing and deployment models.

     \item \textit{Localization and Training}: Robust IoE implementation in agriculture requires solutions tailored to specific local contexts which addresses language, cultural nuances, and region-specific requirements.

\end{enumerate}

\subsection{Quantitative Impact of IoE Technologies on Agricultural Productivity}

The integration of specialized IoE domains in agriculture offers substantial benefits across critical areas, such as yield enhancement, water efficiency, and cost reduction. Leveraging advanced IoT, AI, and nano-enabled devices / entities allows agricultural practices to become more data-driven and precise, significantly boosting productivity and promoting sustainability. Table XIV presents an overview of these key benefits and their impact on modern agriculture.

Projections indicate strong growth in the global smart agriculture market, with estimates from the European Agricultural Sector suggesting it will reach USD 34.1 billion by 2026 \cite{thilakarathne2023towards}. Fig.15 further illustrates the projected market size of various smart agriculture applications by 2027, underscoring the increasing demand and adoption of IoE technologies in agriculture.

Fig. 16 illustrate the significant growth and varied applications of AI and IoT within precision agriculture \cite{forextv2024ai}. Fig. 16(a) presents the projected expansion of the global AI in agriculture market, anticipated to reach USD 10.2 billion by 2032. This growth is largely driven by advances in ML, predictive analytics, and computer vision—technologies essential for crop yield forecasting, disease detection, and resource optimization. In Fig. 16(b), the distribution of IoT and AI applications reveals that 35\% of usage focuses on farm management activities (e.g., irrigation and crop monitoring), while 20\% is dedicated to weather forecasting for informed scheduling, and 15\% supports agricultural information systems for training and knowledge dissemination.

These findings highlight the increasing role of IoT and AI in promoting sustainable, efficient, and resilient agricultural practices, underscoring their potential to meet global food security and environmental goals.

\begin{table*}[h!]
\captionsetup{justification=centering}
\captionsetup{labelsep=newline}
\renewcommand{\arraystretch}{1.3} 
\setlength{\tabcolsep}{3pt} 
\caption{Key Aspects of User-Centric Design for IoE Adoption in Agriculture}
\label{table:user-centric-design}
\centering
\scriptsize 
\resizebox{\textwidth}{!}{
\begin{tabular}{p{2.5cm}|p{4.5cm}|p{5cm}|p{4.5cm}}
\hline\hline
\multicolumn{1}{c|}{\textbf{Aspect}} & \multicolumn{1}{c|}{\textbf{Objective}} & \multicolumn{1}{c|}{\textbf{Solution}} & \multicolumn{1}{c}{\textbf{Impact}} \\
\hline
Understanding End-User Needs & Align IoE solutions with farmers' workflows & Direct farmer engagement & Increases relevance and usability \\
\hline
Ease of Use and Accessibility & Simplify complex data for non-technical users & User-friendly dashboards, alerts, visual indicators & Enhances decision-making for farmers \\
\hline
Affordability & Make IoE accessible to small/medium-scale farmers & Government subsidies, public-private partnerships & Expands technology adoption \\
\hline
Localization and Training & Adapt IoE to local language and cultural contexts & Regional language support, practical training programs & Boosts adoption and reduces training time by 30\% \\
\hline
Data Security and Privacy & Protect sensitive agricultural data & Blockchain frameworks, decentralized architectures & Strengthens data integrity \\
\hline\hline
\end{tabular}}
\end{table*}

\begin{table*}[h!]
\captionsetup{justification=centering}
\captionsetup{labelsep=newline}
\renewcommand{\arraystretch}{1.5} 
\setlength{\tabcolsep}{6pt} 
\caption{Quantitative Impact of IoE Technologies in Precision Agriculture}
\label{table:IoE_agriculture_livestock}
\centering
\scriptsize 
\resizebox{\textwidth}{!}{
\begin{tabular}{>{\centering\arraybackslash}p{2.5cm}|p{5cm}|>{\centering\arraybackslash}p{2.5cm}|>{\centering\arraybackslash}p{2.5cm}|>{\centering\arraybackslash}p{2.5cm}}
\hline\hline
\multicolumn{1}{c|}{\textbf{Technology}} & \multicolumn{1}{c|}{\textbf{Application}} & \multicolumn{1}{c|}{\textbf{Cost Reduction}} & \multicolumn{1}{c|}{\textbf{Yield Improvement}} & \multicolumn{1}{c}{\textbf{Water Savings}} \\
\hline
\textbf{IoT} & Crop monitoring, automated irrigation, soil management & 15-20\% reduction in input and 60\% labor costs \cite{assimakopoulos2024implementation-cost-table} & 20-30\% \cite{duguma2024contribution} & upto 30\% \cite{friha2021internet} \\
\hline
\textbf{5G/ 6G} & Real-time remote monitoring & 20-40\% reduction in operational costs \cite{farmonaut2024-6G-op} & 30-40\% (Est. similar tech) & 40\% \cite{GSMA5GHub-water} \\
\hline
\textbf{IoD} & Aerial crop health monitoring, targeted spraying  & 30\% general cost reduction \cite{DJI2024-drones} & 20\% \cite{zhichkin2023efficiency-drones} & Typically upto 30\%, while upto 90\%  using ULV spraying \cite{upadhyaya2022efficacy-drones-water}\\
\hline
\textbf{IoV} & Autonomous farming tasks, supplychain, logistics & 20\% fuel saving \cite{MITNews2022-vehicle}, 30\% op cost \cite{padhiary2024navigating} & 10-15\% \cite{nasa_spinoff_john_deere_2017}, \cite{padhiary2024navigating} & - \\
\hline
\textbf{IoST} & Satellite-based macro monitoring, remote sensing & 5\% reduction in operational costs \cite{khlystov_satellite_agriculture_2023} & 10-12\% \cite{segarra2024satellite} 100\% \cite{MIT-sunday_african_farmers_2024} using private satellite data & 5-10\% \cite{khlystov_satellite_agriculture_2023} \\
\hline
\textbf{ML} & Predictive analytics and forecasting & 22\% reduction in operational costs \cite{futurefarming_ai_precision_agriculture} & 30\% \cite{shaikh2022machine} & 15-30\% \cite{shaikh2022machine} \\
\hline
\textbf{Blockchain} & Supply chain traceability, transactions & 30\% reduction in transaction costs \cite{ibm_blockchain} & Indirect, 10-12\% reduction in post-harvest losses  (Est. similar
tech) & up to 30\% integrating Blockchain \& IoT \cite{naqash2023blockchain} \\

\hline
\textbf{IoEn} & Smart energy management, Renewable energy integration &  30\% reduction in cost \cite{friha2021internet} & 10-15\%  \cite{conklin2023renewable} & 20-30\% using agrivoltaics \cite{enel2023water} \\

\hline
\textbf{Nanoparticles} & Targeted delivery  & upto 50\% \cite{vishwakarma2023nanoengineered} & 25-30\% \cite{vishwakarma2023nanoengineered} & upto 60\% through water retention \cite{omar2024improving} \\

\hline
\textbf{Designer Phages} & Disease control & 40\% \cite{phages_on_the_farm} & Upto 10\% \cite{nakayinga2021xanthomona} & Indirect, 10-12\% reduction in diseases (Est. similar tech) \\
\hline
\textbf{IoF} & Soil health and nutrient sharing & 20-40\% reduction in fertilizer costs \cite{qian2024effect} & 10-40\% \cite{qian2024effect} \cite{phys2023mycorrhizal} & 12-17\% \cite{wu2024amount} \\
\hline\hline
\end{tabular}}
\end{table*}

\subsection{Long-Term Maintenance and Scalability Challenges}
While IoE technologies offer substantial benefits, they also introduce challenges related to long-term maintenance and scalability. Key challenges include:

\begin{enumerate}

    \item \textit{Device Maintenance and Durability}:IoT devices utilized in agriculture are exposed to harsh environmental conditions, including extreme temperatures, humidity, and dust, which can accelerate wear and tear. Mitigation strategies include:

    \begin{itemize}
        \item Developing devices with durable materials and protective casings.
        \item Implementing proactive maintenance techniques, such as predictive maintenance, to identify and address potential issues before failure. By understanding different types of predictive maintenance solutions, farmers can make more informed decisions, resulting in substantial cost savings.
    \end{itemize}

    \item \textit{Energy and Connectivity Constraints}:
     Many IoE devices operate in remote agricultural areas where reliable power sources and internet connectivity are limited.
    For instance, global internet access stands at 81\% in urban areas and 50\% in rural areas, with rural regions of least developed countries at just 26\% \cite{statista_internet_access_2023}. This lack of connectivity poses challenges for real-time data transmission. Suggested solutions include:
    \begin{itemize}
        \item Utilizing low-power sensors and communication modules to extend battery life.
        \item Integrating energy-harvesting technologies to provide continuous, low-cost power.
        \item Deploying LPWAN technologies, such as LoRaWAN, which offer extended range and low power consumption, ideal for remote areas.
    \end{itemize}

    \item \textit{Scalability Across Diverse Farm Types}:
    Adoption of agricultural technology varies significantly by region and farm type. For example, 61\% of farmers in Europe and North America plan to adopt at least one technology within the next two years, compared to only 9\% in Asia \cite{mckinsey_agtech_adoption_2024}. Larger farms are more likely to adopt these technologies due to greater resources, highlighting the need for scalable solutions suitable for smaller farms. Suggested strategies include:

    \begin{itemize}
        \item Designing modular IoE systems that can be customized for different farm sizes.
        \item Utilizing cloud platforms to handle large volumes of data, ensuring scalability and accessibility.
        \item Adhering to industry standards for interoperability between devices and systems, which facilitates easier scaling.
    \end{itemize}

\end{enumerate}

\subsection{Data Security and Privacy Challenges}
Security is essential in IoE-enabled agriculture due to the sensitive nature of agricultural data. In smart farming systems, vulnerabilities can expose critical data related to crop health, soil conditions, and financial details, impacting both farm management and market dynamics. Studies show that 84\% of IoT adopters have experienced security breaches, underscoring the importance of robust security measures [330]. Key risks and mitigations include:
    \begin{enumerate}
          \item \textit{Unauthorized Access and Data Breaches}: IoT devices often collect sensitive data. Implementing multi-factor authentication, role-based access control, and encryption for data in transit and at rest can prevent unauthorized access.
        \item \textit{Data Tampering and Integrity Attacks}: Attackers can alter IoT sensor data, leading to incorrect farming decisions. Cryptographic hash functions and blockchain technology can ensure data integrity and create immutable records, enhancing traceability
        \item \textit{Inference Attacks}: Even anonymized data can be vulnerable to inference attacks when combined with other data sources. Data aggregation, anonymization, and access controls are essential to safeguard privacy.
        \item \textit{Denial of Service (DoS) Attacks}: DoS attacks on IoT devices can disrupt critical operations like automated irrigation. Network security measures, such as firewalls and intrusion detection systems, along with system redundancy, can mitigate this risk.
        \item \textit{Supply Chain Vulnerabilities}: Compromised components in the supply chain can introduce vulnerabilities. Vendor assessments and secure boot mechanisms can verify the integrity of device firmware and prevent tampering.
        
    \end{enumerate}
    A comprehensive strategy that combines technical safeguards, policy development, and continuous monitoring is needed to address these risks. Blockchain-based security frameworks and decentralized architectures are recommended to further enhance data integrity and resilience in IoE-enabled agriculture.

%


\section{Conclusion}
This paper presents an in-depth analysis of agriculture's role in addressing global challenges such as food security and environmental sustainability. Through a structured research methodology, we examine and emphasize the transformative impact of IoT technologies - including remote sensing, wireless sensor networks, and big data analytics - on smart agriculture, we extend these concepts into the IoE paradigm for further improving the agricultural sector.

We underscore the necessity of high data rates, broad bandwidth, and low latency for effective IoAT deployment, achievable through 5G and 6G networks. This paper categorizes key IoAT applications in precision agriculture - spanning water management, disease management, monitoring, harvesting, and supply chain management - to optimize agricultural outcomes while addressing associated challenges and outlining future research directions.

Our aim was to explore the application of novel technologies like nanotechnology, IoF and designer phages within the agricultural domain. Additionally, the importance of various IoX's such as IoEn, IoV, IoD, IoST is discussed for fostering more sustainable, efficient, and productive farming methods. Finally,  the paper investigates the potential applications of innovative and emerging technologies like 6G, blockchain and ML to address the critical challenges of connectivity, data integrity, food security, resource management, and sustainability in agriculture. The discussion further addresses user-centric design, scalability, and security as critical factors for successful IoE adoption. 

The objective of this paper is to inspire researchers to explore a wide range of innovative and cutting-edge technologies for enhancing agricultural practices. We aim to provide comprehensive insights across various agricultural domains, paving the way for future research that is well-informed, promising, and impactful.

\bibliographystyle{IEEEtran}
\bibliography{IEEEabrv,Bibliography}

\vspace{-45mm}

\begin{IEEEbiography}[{\includegraphics[width=1in,height=1.25in,clip,keepaspectratio]{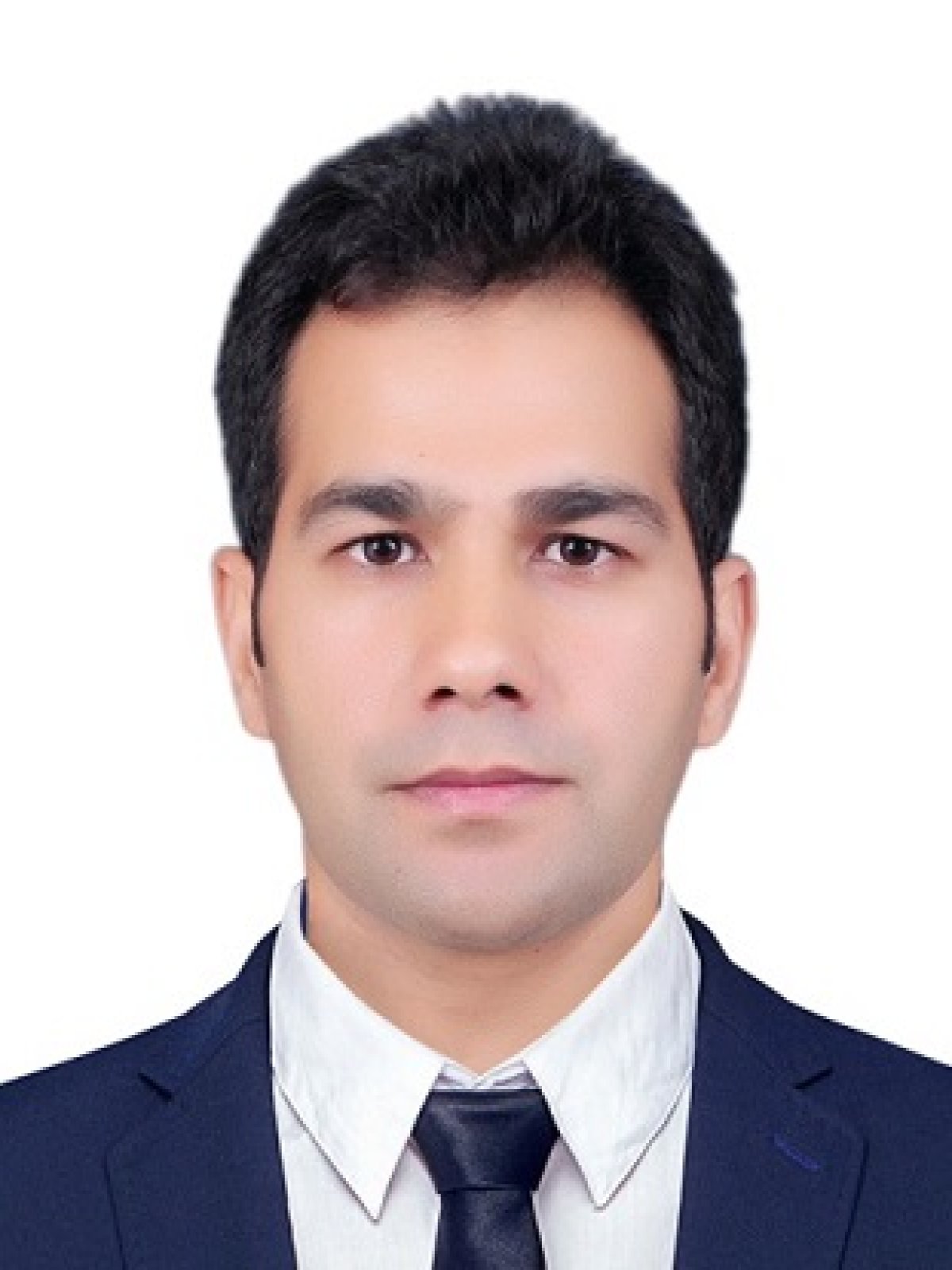}}]{Adil Zaman Babar}
(Student Member, IEEE) received his B.E. degree in Electrical Engineering (Telecommunication) from Air University, Islamabad, Pakistan, and his M.Sc. degree in Electrical Engineering (Communication and Electronics) from UET Peshawar, Pakistan. He is currently a Research Assistant at the Center for NeXt-Generation Communications (CXC), Koç University, Istanbul, Turkey. His research interests include the Internet of Everything (IoE), Wireless Communication, and Networks.
\end{IEEEbiography}

\vspace{-55mm} 

\begin{IEEEbiography}[{\includegraphics[width=1in,height=1.25in,clip,keepaspectratio]{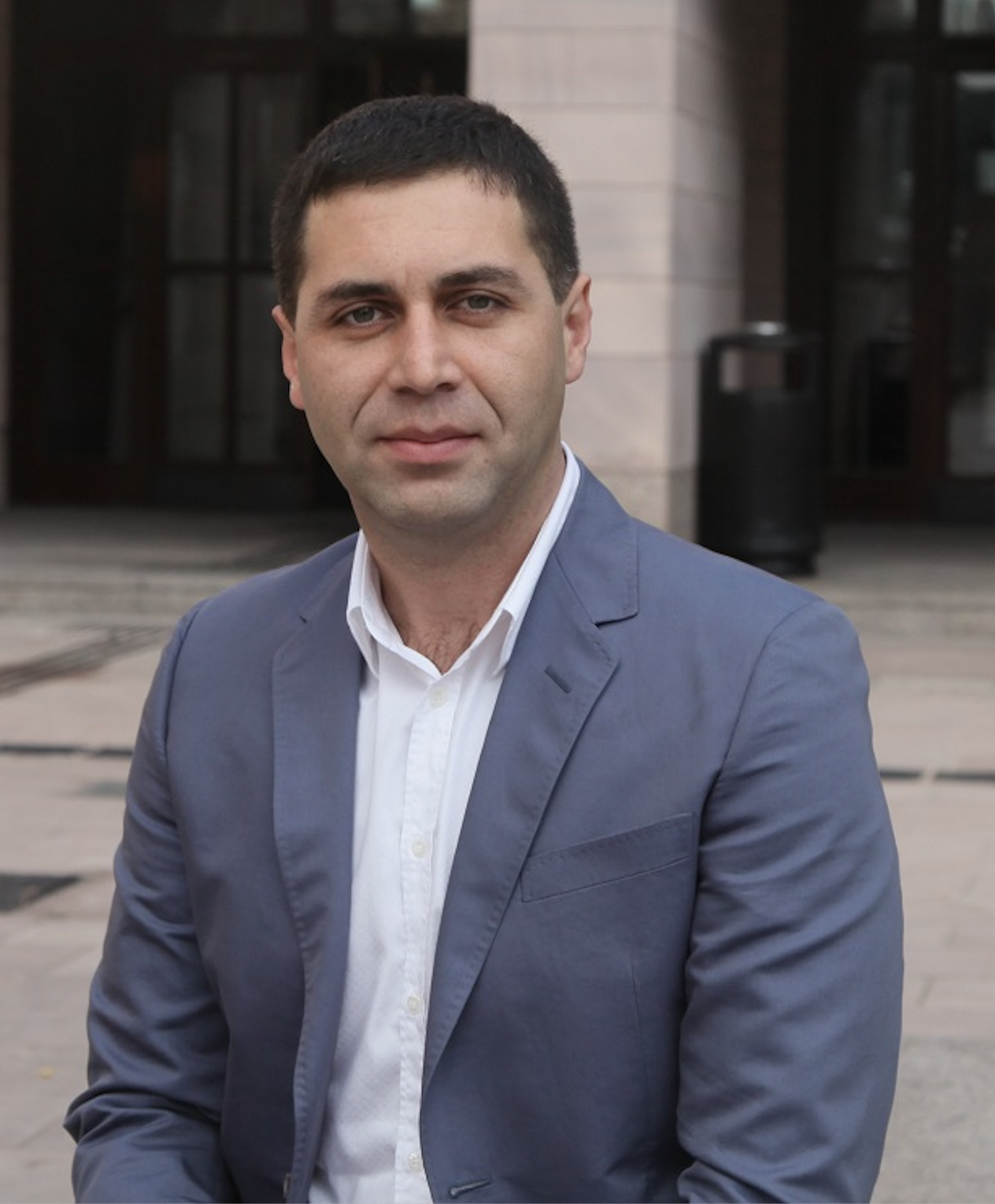}}]{Özgür B. Akan}
(Fellow, IEEE) received his Ph.D. degree from the School of Electrical and Computer Engineering, Georgia Institute of Technology, Atlanta, in 2004. He is currently the Head of the Internet of Everything (IoE) Group at the Department of Engineering, University of Cambridge, UK, and the Director of the Centre for NeXt-Generation Communications (CXC), Koç University, Istanbul, Turkey. His research interests include wireless communication, nano and molecular communications, and the Internet of Everything.
\end{IEEEbiography}

\end{document}